\documentclass[preprintnumbers,nofootinbib,
superscriptaddress,amsmath,floatfix,prd]{revtex4}

\usepackage{amssymb}  
\usepackage{graphicx,subfigure}
\usepackage{exscale}
\usepackage{textcomp}
\usepackage{enumerate}

\usepackage{color}

\usepackage[normalem]{ulem}  

\newcommand{\non}{\nonumber\\}

\newcommand{\be}{\begin{equation}}
\newcommand{\ee}{\end{equation}}
\newcommand{\bea}{\begin{eqnarray}}
\newcommand{\eea}{\end{eqnarray}}
\newcommand{\ba}[1]{\begin{array}{#1}}
\newcommand{\ea}{\end{array}}

\newcommand{\Tr}{{\rm Tr}}

\begin{document}

\title{Instabilities in relativistic two-component (super)fluids}

\author{Alexander Haber}
\email{ahaber@hep.itp.tuwien.ac.at}
\affiliation{Institut f\"{u}r Theoretische Physik, Technische Universit\"{a}t Wien, 1040 Vienna, Austria}

\author{Andreas Schmitt}
\email{a.schmitt@soton.ac.uk}
\affiliation{Institut f\"{u}r Theoretische Physik, Technische Universit\"{a}t Wien, 1040 Vienna, Austria}
\affiliation{School of Mathematics and STAG Research Centre, University of Southampton, Southampton SO17 1BJ, UK}

\author{Stephan Stetina}
\email{stetina@uw.edu}
\affiliation{Institut f\"{u}r Theoretische Physik, Technische Universit\"{a}t Wien, 1040 Vienna, Austria}
\affiliation{Departament d'Estructura i Constituents de la Materia and Institut de Ciencies del Cosmos,
Universitat de Barcelona, Diagonal 647, E-08028 Barcelona, Catalonia, Spain}
\affiliation{Institute for Nuclear Theory, University of Washington, Seattle, WA 98195, USA}

\date{25 January 2016}

\begin{abstract}

We study two-fluid systems with nonzero fluid velocities and compute their sound modes, which indicate various instabilities.
For the case of two zero-temperature superfluids we employ a microscopic field-theoretical model of two coupled bosonic fields, including an entrainment
coupling and a non-entrainment coupling. We analyse the onset of the various instabilities
systematically and point out that the dynamical two-stream instability can only occur beyond Landau's critical velocity, i.e., in an already energetically unstable regime. 
A qualitative difference is found for the case of two normal fluids, where 
certain transverse modes suffer a two-stream instability in an energetically 
stable regime if there is entrainment between the fluids. Since we work in a fully relativistic setup, our results are very general and 
of potential relevance for (super)fluids in neutron stars and, in the non-relativistic limit of our results, in the laboratory. 
   
\end{abstract}

\maketitle

\section{Introduction}

Two-fluid systems where at least one component is a superfluid are realized in different contexts. Any superfluid at nonzero temperature is such a 
system because it can be described in terms of a superfluid and a normal fluid \cite{tisza38,landau41}. Systems with two superfluid components can be 
realized by mixtures of two different species at sufficiently low temperatures. Examples are $^3$He-$^4$He mixtures, where experimental attempts towards simultaneous superfluidity 
of both components have been made \cite{2002JLTP..129..531T,PhysRevB.85.134529} and superfluid Bose-Fermi mixtures of ultra-cold atomic gases, which have been realized recently
in the laboratory \cite{2014Sci...345.1035F,2015arXiv151006709D}. In the interior of neutron stars, nuclear matter and/or quark matter are likely to become superfluid. In nuclear matter, 
neutrons as well as protons can form Cooper pair condensates, giving rise to a two-fluid system of a superfluid and a superconductor. If hyperons are present and form Cooper pairs, 
even more superfluid components may exist \cite{Gusakov:2009kc}.
In quark matter, the color-flavor locked (CFL) phase 
\cite{Alford:1998mk,Alford:2007xm} and the color-spin locked phase \cite{Schafer:2000tw,Schmitt:2004et} are superfluids, and kaon condensation in CFL may lead to a two-component 
superfluid. In each case, temperature effects will add an additional fluid, which renders dense neutron star matter a complicated multi-fluid system. 

In all mentioned systems, a counterflow between the fluids can be created experimentally or, in the case of neutron stars, will necessarily occur. It is well known from plasma 
physics that this may lead to certain dynamical instabilities, called ``two-stream instabilities'' (or sometimes ``counterflow instabilities''). Such an instability manifests itself 
in a nonzero imaginary part of a sound velocity, where the magnitude of the imaginary part determines the time scale on which the given mode becomes unstable. In this paper, 
we will
compute the critical velocity of two-fluid systems at which the two-stream instability sets in. For the case of two superfluids we will start from a $U(1)\times U(1)$ symmetric 
Lagrangian for two complex scalar fields. We include two different coupling terms between the fields: a non-derivative coupling and a derivative coupling, the latter giving rise to 
entrainment between the two fluids (also called Andreev-Bashkin effect \cite{1976JETP...42..164A,2007JETP..105..135S}). 
We restrict ourselves to uniform superfluid velocities, but will allow for arbitrary angles between the directions of the counterflow and the sound mode, thus being able to 
analyse the full angular dependence of the instability. In our zero-temperature approximation, the sound modes are identical to the two Goldstone modes that arise from 
spontaneous breaking of the underlying global symmetry group, and we study them through the bosonic propagator in the condensed phase and through linearized two-fluid
hydrodynamics. 

The two-stream instability can occur in a single superfluid at nonzero temperature \cite{Schmitt:2013nva}, in mixtures of two superfluids 
\cite{PhysRevA.63.063612,2004MNRAS.354..101A,2011PhRvA..83f3602I,2015EPJD...69..126A,2015arXiv151006709D}, 
and in a superfluid immersed in a lattice \cite{PhysRevA.64.061603}. In each case, it is interesting to address 
the relation between this dynamical instability and Landau's critical velocity, where the quasiparticle energy of the Goldstone mode becomes negative (for studies of Landau's critical 
velocity in a two-fluid system see Refs.\ \cite{2003JETPL..78..574A,2006JETP..103..944A,2008JLTP..150..612K}). We shall thus also compute the 
onset of this energetic instability and in particular ask the question whether an {\it energetic} instability is a necessary condition for the two-fluid system to 
become {\it dynamically} unstable. 

Finally, we shall compare our results for the two-component superfluid with the case where one or both of the superfluids is replaced by a normal, ideal fluid. Even though we
always neglect dissipation, there is an important difference between a superfluid and a normal fluid. In a superfluid, density and velocity oscillations are not completely independent 
because they are both related to the phase of the condensate. As a consequence, there is a constraint to the hydrodynamic equations, and 
only longitudinal modes are allowed. 
We will discuss additional solutions that occur in the presence of one or two normal fluids and point out an interesting manifestation of the two-stream instability 
in the presence of entrainment for the case of two normal fluids, which is completely absent if at least one of the fluids is a superfluid.

Our study is very general, and we can in principle extrapolate all results from the ultra-relativistic to the non-relativistic limit. While liquid helium 
and ultra-cold gases are of course most conveniently described in a non-relativistic framework, a relativistic treatment is desirable in the astrophysical context.
On the microscopic level, this is mandatory for quark matter and for sufficiently dense nuclear matter in the core of the star. Under some circumstances, 
for instance in rapidly rotating neutron stars, fluid velocities can assume sizable fractions of the speed of light, such that also on the hydrodynamic level relativistic
corrections may become important. The non-relativistic limit can always be taken straightforwardly by increasing 
the mass of the constituent fluid particles and/or by decreasing the fluid velocities, such that our results can also be applied to superfluids in the laboratory.

Besides a possible realization in the laboratory, the two-stream instability may be of phenomenological relevance for neutron stars: 
pulsar glitches, i.e., sudden jumps in the rotation frequency of the star, are commonly explained by a collective unpinning of superfluid vortices from the ion 
lattice in the inner crust of the neutron star, and hydrodynamic instabilities are one candidate for triggering such a collective effect 
\cite{2004MNRAS.354..101A,Peralta:2006um,Haskell:2015jra}. In this scenario the role of the second (normal) fluid, besides the neutron superfluid, is played by the 
lattice of ions, not unlike the above mentioned atomic superfluid in an optical lattice \cite{PhysRevA.64.061603}. Recently it has been argued that also the superfluids in the core 
of the star might be important for the glitch mechanism, because entrainment effects between the superfluid neutrons and the lattice in the inner crust \cite{pethick2010superfluid,Chamel:2012pk}
reduce the efficiency of the transfer of angular momentum from the superfluid to the crust \cite{Andersson:2012iu}, and thus an analysis of hydrodynamic instabilities in a two-superfluid system of 
neutrons and protons is of phenomenological interest. We emphasize that our results apply to an idealized situation and are thus not directly applicable to the actual 
physics inside a compact star. For instance, we ignore any effects of electromagnetism, which would be necessary to describe the coupled system 
of neutrons, protons and electrons, unless protons and electrons can be viewed as a single, neutral fluid.
Also, we start from a bosonic effective theory while the superfluids in a neutron star are mostly of fermionic nature. And we do not include any effects of 
rotation or a magnetic field, i.e., superfluid vortices or superconducting flux tubes.

Our paper is organized as follows. In Sec.\ II, we present a general derivation of the sound modes from the hydrodynamic equations, 
without reference to any microscopic 
model. In particular, we discuss the differences between a system of two superfluids, a superfluid and a normal fluid, and two 
normal fluids. In Sec.\ \ref{sec:micro} we present a microscopic model for a two-component superfluid, discuss its phase diagram, first 
in the absence of any superfluid velocities,
and derive the quasiparticle excitations. We 
present our main results in Sec.\ \ref{sec:instabilities}, where we compute and interpret various kinds of instabilities in the two-fluid system. As far as possible, the results in this section are formulated in a self-contained way, such that readers only interested in the main results may almost exclusively focus on Sec.\ \ref{sec:instabilities}. 
We summarize and give an outlook in Sec.\ \ref{sec:summary}.

\section{Sound modes from hydrodynamic equations}
\label{sec:hydro}

In this section we derive the sound modes in the presence of nonzero fluid velocities from the hydrodynamic equations. 
As a warm-up exercise, and in order to establish our notation, we will discuss a
single fluid before turning to two coupled fluids. Our derivation is general in the sense that it holds for normal fluids as well as for superfluids, i.e., from the 
final result we will be able to discuss the three cases of two superfluids, one superfluid and one normal fluid, and two normal fluids. Since we neglect 
any dissipative effects, our normal fluid is an ideal fluid.

\subsection{Single fluid}

We start from the conservation equations 
\be \label{conserve}
\partial_\mu j^\mu = 0 \, , \qquad \partial_\mu T^{\mu\nu} = 0   \, , 
\ee
where $j^\mu$ is a conserved current, and 
\be\label{Tmunu}
T^{\mu\nu} = j^\mu p^\nu -g^{\mu\nu}\Psi  
\ee
is the stress-energy tensor with the ``generalized pressure'' $\Psi$ and the space-time metric $g^{\mu\nu}={\rm diag}\,(1,-1,-1,-1)$. 
The conjugate momentum $p^\mu$ is related to the current via
\be \label{jdef}
j^\mu = \frac{\partial \Psi}{\partial p_\mu} = 2\frac{\partial \Psi}{\partial p^2} p^\mu \, , 
\ee
where, on the right-hand side, we have assumed that $\Psi$ is a function only of the Lorentz scalar $p^2=p_\mu p^\mu$. With this relation between the conjugate momentum and the current it 
is obvious that the stress-energy tensor is symmetric. The Lorentz scalar $p$ is the chemical potential measured in the 
rest frame of the fluid, which can be seen as follows. We write $j^\mu = nv^\mu$, where $n$ is the charge density 
measured in the rest frame of the fluid and $v^\mu$ the four-velocity of the fluid with $v_\mu v^\mu =1$. (We work in units where the speed of light is set to one.)
This implies $n^2=j^2$ and thus 
\be \label{n}
n = 2\frac{\partial \Psi}{\partial p^2} p \, . 
\ee
The ``generalized energy density'' is defined as $\Lambda = T^\mu_{\;\;\mu}+3\Psi$. 
Using Eqs.\ (\ref{Tmunu}) and (\ref{n}), we obtain $\Lambda + \Psi = p_\mu j^\mu = pn$, which is the usual 
thermodynamic relation at zero temperature if $p$ is identified with the chemical potential in the rest frame of the fluid. 

The current can now be written as $j^\mu = n p^\mu/p$, from which we conclude that the fluid velocity is 
\be \label{vmu}
v^\mu = \frac{p^\mu}{p}  \, .
\ee
With $v^\mu = \gamma(1,\vec{v})$, where $\gamma=1/\sqrt{1-\vec{v}^2}$ is the usual Lorentz factor, the three-velocity becomes $\vec{v}=\vec{p}/\mu$, with  
$\mu = p_0$ being the chemical potential measured in the frame in which the fluid moves with velocity $\vec{v}$. 
In that frame, the charge density is $j^0 = n\mu/p$. With the help of Eqs.\ (\ref{jdef}), (\ref{n}), (\ref{vmu}) and the relation $pn=\Lambda + \Psi$, we see that the stress-energy tensor
(\ref{Tmunu}) can be written in the familiar form $T^{\mu\nu}=(\Lambda+\Psi)v^\mu v^\nu-g^{\mu\nu}\Psi$.  

We write the hydrodynamic equations (\ref{conserve}) as 
\begin{subequations} \label{jT}
\bea
0&=&\partial_\mu j^\mu = \frac{n}{p}\left[g_{\mu\nu}+\left(\frac{1}{c^2}-1\right)\frac{p_\mu p_\nu}{p^2}\right]\partial^\mu p^\nu \, , \\[2ex]
0&=&\partial_\mu T^{\mu\nu} = j^\mu(\partial_\mu p^\nu -\partial^\nu p_\mu) \, , \label{jT2}
\eea
\end{subequations} 
where, in the second relation, we have used $\partial_\mu j^\mu=0$, and we have introduced the speed of sound $c$ in the rest frame of the fluid, 
\be
c^2 = \frac{n}{p}\left(\frac{\partial n}{\partial p}\right)^{-1} \, . 
\ee

To compute the sound modes, we need to treat the temporal and spatial components of the momentum $p^\mu = (\mu, \vec{p})$ separately. Each component is assumed to fluctuate
harmonically about its equilibrium value with frequency $\omega$ and wave vector $\vec{k}$,
\be \label{osc}
\mu(\vec{x},t) = \mu +\delta\mu \,e^{i(\omega t-\vec{k}\cdot\vec{x})} \, , \qquad  \vec{p}(\vec{x},t) = \vec{p} +\delta\vec{p} \,e^{i(\omega t-\vec{k}\cdot\vec{x})} \, , 
\ee
where $\delta\mu$ and $\delta\vec{p}$ are the fluctuations which will be kept to linear order. Then, Eqs.\ (\ref{jT}) become 
\begin{subequations} \label{hydrolin}
\bea
0 &=& \frac{n}{p}\left[\omega\delta\mu-\vec{k}\cdot\delta\vec{p}+\frac{\mu\omega_v}{p^2}\left(\frac{1}{c^2}-1\right)(\mu\delta\mu-\vec{p}\cdot\delta\vec{p})\right]   \, , \label{hydrolin1} \\[2ex]
0 &=& \frac{n}{p}\vec{p}\cdot(\omega\delta\vec{p}-\vec{k}\delta\mu)  \, , \label{hydrolin2}\\[2ex]
0 &=& \frac{n}{p}\left[\mu\omega_v\delta\vec{p}-\vec{k}(\mu\delta\mu - \vec{p}\cdot\delta\vec{p})\right] \, ,\label{hydrolin3}
\eea
\end{subequations}
where $\omega_v \equiv \omega - \vec{v}\cdot\vec{k}$. We have decomposed 
Eq.\ (\ref{jT2}) into its temporal component (\ref{hydrolin2}) and its spatial components (\ref{hydrolin3}).

In the case of a superfluid, the conjugate momentum can be written as the gradient of 
a phase $\psi\in [0,2\pi]$, $p^\mu = \partial^\mu \psi$. As a consequence, we have
\be \label{super}
\mu = \partial_0\psi \, , \qquad \vec{v} = -\frac{\nabla\psi}{\mu} \, .
\ee
(The first equation, which relates the time evolution of the phase to the chemical potential, is sometimes called Josephson-Anderson relation.)
In a microscopic model, $\psi$ is the phase of the expectation value of the complex scalar field (the Bose-Einstein condensate), see Sec.\ \ref{sec:micro}.  
With $\partial_0\vec{p}=-\nabla\partial_0\psi=-\nabla\mu$ and $\vec{p}=\mu\vec{v}$ 
we see that the fluctuations in the chemical potential and the superfluid velocity are not independent because they are both 
related to the phase. As a consequence, only longitudinal modes are allowed where $\delta\vec{p}$ oscillates in the direction of the wave vector $\vec{k}$, 
$\omega\delta\vec{p}=\vec{k}\delta\mu$, and Eqs.\ (\ref{hydrolin2}) and (\ref{hydrolin3}) are 
automatically fulfilled. [This is also obvious from Eq.\ (\ref{jT2}) with $p^\mu = \partial^\mu\psi$.] It only remains to solve the continuity equation which yields 
a quadratic polynomial in $\omega$. Since in the given approximation all modes are linear in momentum, we can write $\omega = uk$ with the angular-dependent sound velocity 
\be \label{usingle}
u = \frac{(1-c^2)v\cos\theta \pm c\sqrt{1-v^2}\sqrt{1-c^2v^2-(1-c^2)v^2\cos^2\theta}}{1-c^2v^2} \, .
\ee
Here, $v\equiv |\vec{v}|$ is the modulus of the three-velocity and $\theta$ the angle between the directions of the fluid velocity and the propagation of the 
sound mode, $\vec{k}\cdot\vec{p}=\mu vk\cos\theta$. For $v=0$, these modes reduce to $u = \pm c$, as it should be. 
For a given angle $\theta \in [0,\pi]$, we consider only the upper sign
in Eq.\ (\ref{usingle}), such that the sound speed is positive for small velocities $v$. Since we shall later be interested in instabilities, we may already ask at this
point when the speed of sound becomes negative. This happens at the critical velocity $v=c$, where $u$ starts to become negative in the upstream direction $\theta=\pi$.
We also see that $u$ never becomes complex because the arguments of either of the square roots in the numerator only become negative for unphysical
velocities larger than the speed of light, $v>1$.   

In the case of a normal fluid, the same longitudinal modes are found.
But, $\vec{p}$ is now allowed to oscillate in transverse directions with respect to $\vec{k}$, 
which yields the additional mode $u = \vec{v}\cdot\hat{\vec{k}}$ with the conditions for the fluctuations  
$\vec{p}\cdot\delta\vec{p} = \mu\delta\mu$ and $\mu \vec{k}\cdot\delta\vec{p} = \vec{k}\cdot\vec{p}\,\delta\mu$.

\subsection{Two fluids}

We now generalize the results to the case of two coupled fluids. In this case, we write the stress-energy tensor as
\be
T^{\mu\nu} = j_1^\mu p_1^\nu +j_2^\mu p_2^\nu -g^{\mu\nu} \Psi \, . 
\ee
Now $\Psi$ is a function of all Lorentz scalars that can be constructed from the two conjugate momenta $p_1^\mu$, $p_2^\mu$, i.e., $\Psi = \Psi(p_1^2,p_2^2,p_{12}^2)$, where 
we have abbreviated $p_{12}^2 \equiv p_1^\mu p_{2\mu} = \mu_1\mu_2-\vec{p}_1\cdot\vec{p}_2$. The currents are defined as in Eq.\ (\ref{jdef}) and can be written as  
\begin{subequations} \label{j1j2}
\bea
j_1^\mu &=& {\cal B}_1 p_1^\mu + {\cal A}\, p_2^\mu \, , \\[2ex]
j_2^\mu &=& {\cal A}\, p_1^\mu + {\cal B}_2 p_2^\mu \, , 
\eea
\end{subequations}
where
\be
{\cal B}_i \equiv 2\frac{\partial \Psi}{\partial p_i^2} \, , \qquad 
{\cal A} \equiv \frac{\partial \Psi}{\partial p_{12}^2} \, ,
\ee
with the fluid index $i=1,2$.
In general, the currents are not four-parallel to their own conjugate momentum, but, if $\Psi$ depends on $p_{12}^2$, receive a contribution from the 
conjugate momentum of the other fluid. This effect is called entrainment and ${\cal A}$ is called entrainment coefficient. To guarantee the symmetry of the stress-energy tensor,
there can only be one single entrainment coefficient, appearing
in both currents. The conservation equations now read
\begin{subequations}
\bea
\partial_\mu j_1^\mu &=& \Big[{\cal B}_1 g_{\mu\nu}+b_1 p_{1\mu}p_{1\nu}+a_1(p_{1\mu} p_{2\nu}+p_{2\mu}p_{1\nu})+a_{12} p_{2\mu}p_{2\nu}\Big]\partial^\mu p_1^\nu \non[2ex]
&&+\Big[{\cal A} g_{\mu\nu}+a_1p_{1\mu}p_{1\nu}+d\,p_{1\mu} p_{2\nu}+a_{12}p_{2\mu}p_{1\nu}+a_2p_{2\mu}p_{2\nu}\Big]\partial^\mu p_2^\nu\, , \label{cons1}\\[2ex]
\partial_\mu j_2^\mu &=& (1\leftrightarrow 2) \, , \label{cons2} \\[2ex]
\partial_\mu T^{\mu\nu} &=& j_1^\mu(\partial_\mu p_1^\nu-\partial^\nu p_{1\mu})+j_2^\mu(\partial_\mu p_2^\nu-\partial^\nu p_{2\mu}) \, ,
\eea
\end{subequations}
where Eq.\ (\ref{cons2}) is obtained from Eq.\ (\ref{cons1}) by exchanging the indices $1\leftrightarrow 2$ ($a_{21}=a_{12}$), and we have abbreviated 
\bea
b_i &\equiv& 4\frac{\partial^2 \Psi}{\partial (p_i^2)^2}\, , \qquad a_i \equiv 2\frac{\partial^2 \Psi}{\partial p_i^2\partial p_{12}^2} \, , 
\qquad a_{12} \equiv \frac{\partial^2 \Psi}{\partial (p_{12}^2)^2} \, , \qquad d\equiv 4\frac{\partial^2 \Psi}{\partial p_1^2\partial p_2^2} \, .
\eea
Here, $b_1$, $b_2$ are susceptibilities that are given by the properties of each fluid separately, while $d$ describes a non-entrainment coupling between the two fluids, 
and the coefficients $a_1$, $a_2$, $a_{12}$ are only nonzero in the presence of an entrainment coupling.
Again we introduce fluctuations in the temporal and spatial components of the conjugate momenta, as given in Eq.\ (\ref{osc}). The conservation equations then become
\begin{subequations}
\bea
0&=& \Big[\omega {\cal B}_1+\omega_{v_1}\mu_1(\mu_1b_1+\mu_2a_1)+\omega_{v_2}\mu_2(\mu_1a_1+\mu_2a_{12})\Big]\delta\mu_1
+\Big[\omega {\cal A}+\omega_{v_1}\mu_1(\mu_1a_1+\mu_2d)+\omega_{v_2}\mu_2(\mu_1 a_{12}+\mu_2 a_2)\Big]\delta\mu_2 \non[2ex]
&&-(\omega_{v_1}\mu_1 b_1+\omega_{v_2}\mu_2 a_1)\vec{p}_1\cdot\delta\vec{p}_1 
-(\omega_{v_1}\mu_1 d +\omega_{v_2}\mu_2 a_2)\vec{p}_2\cdot\delta\vec{p}_2 -{\cal B}_1\vec{k}\cdot\delta\vec{p}_1-{\cal A}\vec{k}\cdot\delta\vec{p}_2\non[2ex]
&&-(\omega_{v_1}\mu_1 a_1+\omega_{v_2}\mu_2 a_{12})\vec{p}_2\cdot\delta\vec{p}_1
-(\omega_{v_1}\mu_1 a_1+\omega_{v_2}\mu_2 a_{12})\vec{p}_1\cdot\delta\vec{p}_2  \, , \label{one}\\[2ex]
0&=& (1\leftrightarrow 2) \, , \label{two}\\[2ex]
0&=& (\vec{p}_1{\cal B}_1+\vec{p}_2 {\cal A})\cdot(\omega\delta\vec{p}_1-\vec{k}\delta\mu_1) + (\vec{p}_2{\cal B}_2+\vec{p}_1 {\cal A})\cdot(\omega\delta\vec{p}_2-\vec{k}\delta\mu_2) 
\, , \label{2fl3}\\[2ex]
0&=& (\omega_{v_1}\mu_1 {\cal B}_1+\omega_{v_2}\mu_2{\cal A})(\omega\delta\vec{p}_1-\vec{k}\delta\mu_1)+
(\omega_{v_2}\mu_2 {\cal B}_2+\omega_{v_1}\mu_1{\cal A})(\omega\delta\vec{p}_2-\vec{k}\delta\mu_2) \, , \label{2fl4}
\eea
\end{subequations}
where we have inserted $\partial_\mu T^{\mu 0}=0$ (\ref{2fl3}) into the spatial components $\partial_\mu T^{\mu \ell}$ ($\ell=1,2,3$) to obtain Eq.\ (\ref{2fl4}).

\subsubsection{Super-Super}

If both fluids are superfluids, we have $\omega\delta\vec{p}_i=\vec{k}\delta\mu_i$. The conservation of energy and momentum 
is automatically fulfilled in this case, and the two continuity equations read
\begin{subequations} \label{dmu1dmu2}
\bea
0&=& \left[{\cal B}_1(\omega^2-k^2)+\mu_1^2b_1\omega_{v_1}^2+2\mu_1\mu_2a_1\omega_{v_1}\omega_{v_2}+\mu_2^2a_{12}\omega_{v_2}^2\right]\delta\mu_1 \non[2ex]
&&+\left[{\cal A}(\omega^2-k^2)+\mu_1^2a_1\omega_{v_1}^2+\mu_1\mu_2(a_{12}+d)\omega_{v_1}\omega_{v_2}+\mu_2^2a_2\omega_{v_2}^2\right]\delta\mu_2 \, , \\[2ex]
0&=& \left[{\cal A}(\omega^2-k^2)+\mu_2^2a_2\omega_{v_2}^2+\mu_1\mu_2(a_{12}+d)\omega_{v_1}\omega_{v_2}+\mu_1^2a_1\omega_{v_1}^2\right]\delta\mu_1\non[2ex]
&&+\left[{\cal B}_2(\omega^2-k^2)+\mu_2^2b_2\omega_{v_2}^2+2\mu_1\mu_2a_2\omega_{v_1}\omega_{v_2}+\mu_1^2a_{12}\omega_{v_1}^2\right]\delta\mu_2 \, .
\eea
\end{subequations}
After some rearrangements this can be compactly written as
\be
(u^2\chi_2 + u\chi_1+\chi_0)\delta\mu = 0  \, , 
\ee
with $u=\omega/k$, the vector $\delta \mu = (\delta\mu_1,\delta\mu_2)$, and the $2\times 2$ matrices $\chi_2$, $\chi_1$, $\chi_0$ whose entries are given by 
\be
(\chi_2)_{ij} = \frac{\partial^2\Psi}{\partial\mu_i\partial\mu_j} \, , \qquad  
(\chi_1)_{ij} = \left(\frac{\partial^2\Psi}{\partial\mu_i\partial p_{j\ell}}+\frac{\partial^2\Psi}{\partial\mu_j\partial p_{i\ell}}\right)\hat{k}_\ell \, , \qquad 
(\chi_0)_{ij} = \frac{\partial^2\Psi}{\partial p_{i\ell}\partial p_{jm}} \hat{k}_\ell\hat{k}_m \, , 
\ee
with $i,j = 1,2$, and spatial indices $\ell,m=1,2,3$. (We have reserved $i,j$ for the two different 
fluid species, which should not lead to any confusion since spatial indices will not appear explicitly from here on.) 
The sound velocity $u$ is then determined from 
\be \label{detu}
{\rm det}\,(u^2\chi_2 + u\chi_1+\chi_0) = 0 \, , 
\ee
which is a quartic polynomial in $u$ with analytical, but very complicated, solutions. We shall discuss and interpret these solutions in Sec.\ \ref{sec:instabilities},
after we have introduced a microscopic model for the two-component superfluid in Sec.\ \ref{sec:micro}.

\subsubsection{Super-Normal}

Let us now assume that one of the fluids is a normal fluid. Say fluid 1 is a superfluid, $\omega\delta\vec{p}_1=\vec{k}\delta\mu_1$, while we make no assumptions about 
$\delta\vec{p}_2$. Of course, we still find the same modes as for the two superfluids since the normal fluid can also accommodate the longitudinal oscillations of a superfluid. 
An additional mode is found if we enforce a transverse mode by requiring $\omega\delta\vec{p}_2\neq\vec{k}\delta\mu_2$. Then, Eq.\ (\ref{2fl4}) yields the mode 
\be \label{vnk1}
u = \frac{\vec{v}_2\mu_2{\cal B}_2+\vec{v}_1\mu_1{\cal A}}{\mu_2{\cal B}_2+\mu_1{\cal A}}\cdot\hat{\vec{k}} \, .
\ee
This is the generalization of the mode $u = \vec{v}\cdot\hat{\vec{k}}$ 
mentioned in the discussion of the single normal fluid. We may apply this expression to a single superfluid at nonzero temperature.
In this case, there are two currents, the conserved charge current $j_1^\mu = j^\mu$ and the entropy current $j_2^\mu= s^\mu$ 
(which is also conserved if we neglect dissipation). Their conjugate momenta 
are $p_1^\mu=\partial^\mu \psi$, where $\psi$ is the phase of the condensate, and $p_2^\mu=\Theta^\mu$, whose temporal component is the temperature, $\Theta_0=T$, measured in the 
normal-fluid rest frame, where $\vec{s}=0$. Analogously to Eq.\ (\ref{j1j2}) we can write \cite{Carter:1995if,2013PhRvD..87f5001A,Schmitt:2014eka} 
\begin{subequations} 
\bea
j^\mu &=& {\cal B} \partial^\mu \psi  + {\cal A} \Theta^\mu  \, , \\[2ex]
s^\mu &=& {\cal A}\partial^\mu \psi + {\cal C} \Theta^\mu \, . 
\eea
\end{subequations}
Consequently, we can identify $\vec{v}_2\mu_2{\cal B}_2+\vec{v}_1\mu_1{\cal A}\to \vec{s}$ and $\mu_2{\cal B}_2+\mu_1{\cal A}\to s_0$. Note that $\mu_2$, the temporal component
of the conjugate momentum $p_2^\mu$ corresponds to the temperature $T$, and the velocity $\vec{v}_2$ corresponds to $\vec{\Theta}/T$. The four-velocity of the normal fluid is defined 
by $v_n^\mu = s^\mu/s$, which yields the three-velocity $\vec{v}_n = \vec{s}/s_0$. (If normal fluid and superfluid velocities are used as 
independent hydrodynamic variables, one works in a ``mixed'' representation with respect to currents and momenta:
 while the superfluid velocity corresponds to the {\it momentum} of one fluid, $v_s^\mu = \partial^\mu\psi$, the normal fluid 
velocity corresponds to the {\it current} of the other fluid, $v_n^\mu = s^\mu/s$.) Inserting all this into Eq.\ (\ref{vnk1}) yields
\be
u = \vec{v}_n\cdot\hat{\vec{k}} \, . 
\ee
This is in exact agreement with Ref.\  \cite{PhysRevB.77.144515}, where this mode has been discussed in the non-relativistic context. (In Refs.\ \cite{2013PhRvD..87f5001A,Alford:2013koa},
where sound modes in a relativistic superfluid at nonzero temperatures were studied within a field-theoretical setup, this mode was not mentioned because the calculation 
was performed in the rest frame of the normal fluid.)

\subsubsection{Normal-Normal}

In the case of two normal fluids, we make no assumptions about the fluctuations $\delta\vec{p}_i$, $\delta\mu_i$. Let us first suppose there was no coupling between the 
two fluids at all, i.e., ${\cal A}=a_1=a_2=a_{12}=d=0$. Without loss of generality, we can work in the rest frame of one of the fluids, 
say $\vec{p}_2=0$ and thus $\omega_{v_2}=\omega$. Then, 
we can express $\vec{p}_1\cdot \delta\vec{p}_1$, $\vec{k}\cdot\delta\vec{p}_1$, $\vec{k}\cdot\delta\vec{p}_2$ in terms of $\delta\mu_1$ and $\delta\mu_2$ to 
obtain 
\be
0=\mu_1\omega_{v_1}[{\cal B}_1(\omega^2-k^2)+\omega_{v_1}^2\mu_1^2b_1]\delta\mu_1 + \mu_2\omega[{\cal B}_2(\omega^2-k^2)+\omega^2\mu_2^2b_2]\delta\mu_2 \, .
\ee
This equation yields the separate modes of the two fluids: by setting $\delta\mu_2=0$ we find the modes for fluid 1, and by setting $\delta\mu_1=0$ we find the modes for fluid 2. With 
\be
{\cal B}_i=\frac{n_i}{p_i} \, , \qquad b_i=\frac{n_i}{p_i^3}\left(\frac{1}{c_i^2}-1\right) \qquad (i=1,2) \, , 
\ee
we recover the modes discussed above for the single fluid. 

Now let us switch on the coupling between the fluids. The simplest 
case is to neglect any entrainment, $a_1=a_2=a_{12}=0$, but keep a nonzero non-entrainment coupling, $d\neq 0$. For a compact notation we define the  
``mixed susceptibilities''
\be 
\Delta_{1/2} \equiv \frac{p_{2/1}}{n_{1/2}}\frac{\partial n_{1/2}}{\partial p_{2/1}}\, , 
\ee
(such that $d = \Delta_1B_1/p_2^2=\Delta_2B_2/p_1^2$), which are only nonvanishing for nonzero coupling $d$. 
Then, for $\delta\mu_2=0$ we find the modes $u=v_1\cos\theta$, and 
\be \label{dmu20}
\delta\mu_2=0:\qquad u = \frac{(1-c_1^2)v_1\cos\theta\pm c_1\sqrt{1-v_1^2}\sqrt{1-(1+\Delta_1)v_1^2[c_1^2+(1-c_1^2)\cos^2\theta]+\Delta_1c_1^2}}{1-c_1^2[v_1^2(1+\Delta_1)-\Delta_1]} 
\, . 
\ee
As for the case of a single fluid, we may ask whether and when the speed of sound turns negative. And, in contrast to the single fluid, $u$ may even become complex. The critical 
velocities for these two instabilities (to be discussed in detail in Sec.\ \ref{sec:instabilities}) are, respectively,
\be \label{dmu20vcr}
\delta\mu_2=0:\qquad  v_{{\rm c}}^< = \frac{c_1}{\sqrt{c_1^2\sin^2\theta+\cos^2\theta}} \, , \qquad v_{{\rm c}}^> = v_{{\rm c}}^<\frac{\sqrt{1+\Delta_1 c_1^2}}{c_1\sqrt{1+\Delta_1}}
\ge v_{{\rm c}}^<
 \, ,
\ee
For $\delta\mu_1=0$ we have 
\be \label{dmu10}
\delta\mu_1=0:\qquad u = \frac{\Delta_2 c_2^2v_1\cos\theta\pm c_2\sqrt{1-v_1^2}\sqrt{1-v_1^2+\Delta_2(c_2^2-v_1^2\cos^2\theta)}}{1-v_1^2+\Delta_2c_2^2} \, .
\ee
Again, we can easily compute the critical velocities,
\be \label{dmu10vcr}
\delta\mu_1=0:\qquad v_{{\rm c}}^< = \frac{1}{\sqrt{1+\Delta_2\cos^2\theta}} \, , \qquad v_{{\rm c}}^> = v_{{\rm c}}^<\sqrt{1+\Delta_2c_2^2} \ge v_{{\rm c}}^<
\, .
\ee
In both cases, it is obvious (and we have indicated it by our choice of notation), that the critical velocity for a negative sound velocity is smaller than or equal to that for a 
complex sound velocity. We shall come back to this observation and also discuss the general case with entrainment 
at the end of Sec.\ \ref{sec:instabilities}.

\section{Microscopic model and elementary excitations}
\label{sec:micro}

In this section we present a microscopic bosonic model for two complex scalar fields $\varphi_1$, $\varphi_2$, which 
we shall evaluate at zero temperature in order to provide 
the input for the hydrodynamic results of the previous section. We shall also compute the Goldstone modes, which, for small momenta and in the given approximation, 
are identical to the 
sound modes computed from linearized hydrodynamics. Our starting point is the Lagrangian
\be \label{L}
{\cal L} = \sum_{i=1,2}\left(\partial_\mu\varphi_i\partial^\mu\varphi^*_i-m_i^2|\varphi_i|^2-\lambda_i|\varphi_i|^4\right) 
+2h|\varphi_1|^2|\varphi_2|^2 - \frac{g_{1}}{2}(\varphi_1\varphi_2\partial_\mu\varphi_1^*\partial^\mu\varphi_2^*+{\rm c.c.}) 
- \frac{g_{2}}{2}(\varphi_1\varphi_2^*\partial_\mu\varphi_1^*\partial^\mu\varphi_2 +{\rm c.c.}) \, ,
\ee
where c.c.\ denotes complex conjugate, $m_i>0$ are the mass parameters of the bosons, and $\lambda_i>0$ the self-coupling constants. This Lagrangian is invariant under 
transformations $\varphi_i\to e^{i\alpha_i}\varphi_i$, $\alpha_i\in \mathbb{R}$, 
resulting in a global $U(1)\times U(1)$ symmetry group, and we have introduced three different inter-species couplings: 
one non-derivative coupling whose strength is given by the dimensionless coupling constant $h$, and two derivative couplings with coupling constants $g_{1}$, $g_{2}$, 
which have mass dimension $-2$. 
The model is non-renormalizable and thus has to be considered as an effective theory. This will play no role in our discussion because we are only interested in the 
tree-level potential and the quasiparticle excitations, calculated from the tree-level propagator. We thus never have to introduce a momentum cutoff, which would be necessary if
we were to compute loops, for instance in an extension to nonzero temperatures, which has been done within the two-particle irreducible formalism for a single superfluid \cite{Alford:2013koa}. 
Our Lagrangian is very similar to the Ginzburg-Landau model usually considered 
for superfluid nuclear matter in the core of a neutron star, where the two fields correspond to the Cooper pair condensates of neutrons and protons 
\cite{Alpar:1984zz,Alford:2007np,2015arXiv150400570K}, for approaches on a more microscopic level, see for instance Refs.\ \cite{sauls1989superfluidity,Comer:2002dm,Chamel:2006rc}. 
The only differences are that we consider a relativistic setup and two neutral fields, not taking into account the electric charge of the proton. 

Via Noether's theorem, the symmetry gives rise to two conserved currents,
\begin{subequations}\label{j1j2a}
\bea 
j_1^\mu &=& i(\varphi_1\partial^\mu\varphi_1^*-\varphi_1^*\partial^\mu\varphi_1) + g|\varphi_1|^2i(\varphi_2\partial^\mu\varphi_2^*-\varphi_2^*\partial^\mu\varphi_2) \, , \\[2ex] 
j_2^\mu &=& i(\varphi_2\partial^\mu\varphi_2^*-\varphi_2^*\partial^\mu\varphi_2) + g|\varphi_2|^2i(\varphi_1\partial^\mu\varphi_1^*-\varphi_1^*\partial^\mu\varphi_1) \, ,
\eea  
\end{subequations}
where we have abbreviated the difference of the two entrainment couplings,
\be
g \equiv \frac{g_{1}- g_{2}}{2} \, . 
\ee
Later we shall also need the sum of them,
\be
G \equiv \frac{g_{1}+ g_{2}}{2} \, .
\ee

\subsection{Tree-level potential and phase diagram}
\label{sec:tree}

We write the complex fields as 
\be\label{shift}
\varphi_i = \langle\varphi_i\rangle + {\rm fluctuations} \, . 
\ee
In this section, we neglect the fluctuations, i.e., the fields are solely given by their expectation values $\langle\varphi_i\rangle$ (the ``condensates''), which we parameterize 
by their modulus $\rho_i$ and their phase $\psi_i$,
\be \label{rhopsi}
\langle\varphi_i\rangle = \frac{\rho_i}{\sqrt{2}}e^{i\psi_i} \, .
\ee
In this approximation, the Lagrangian (\ref{L}) can simply be read as (the negative of) a Ginzburg-Landau free energy, 
where the fields correspond to the order parameters. In this paper, we shall not include the fluctuations in our calculation of the phase diagram. For low temperatures and small coupling strengths
we expect our mean-field approach to be a good approximation. We shall always work at zero temperature, and the numerical values of the coupling strengths we consider in this paper are always well below 1, the self-couplings $\lambda_1,\lambda_2 \le 0.3$, the non-entrainment coupling mostly zero, only 
in Fig.\ \ref{fignn} we set $h=0.05$, and the entrainment coupling $g$, in units of $m_1m_2$ or, where we work in the ultrarelativistic limit, in units of $\mu_1\mu_2$, smaller than or equal to 0.2. 
In any case, our approach should be seen as a first approximation to the possible real-world applications we have in mind. Consequently, any refinement of the approximation -- before improving the model itself -- 
that is not expected to change the results qualitatively is questionable from a phenomenological point of view.

In the parametrization (\ref{rhopsi}) the conserved currents (\ref{j1j2a}) become
\begin{subequations}
\bea 
j_1^\mu &=& \rho_1^2\partial^\mu\psi_1 +\frac{g}{2} \rho_1^2\rho_2^2\partial^\mu\psi_2 \, , \\[2ex]
j_2^\mu &=& \frac{g}{2} \rho_1^2\rho_2^2\partial^\mu\psi_1 + \rho_2^2\partial^\mu\psi_2 \, .
\eea  
\end{subequations}
This form has to be compared to the general form of the currents (\ref{j1j2}). We see that a nonzero entrainment coefficient ${\cal A}$ arises due to the derivative coupling. 
(Notice that there is no entrainment for $g_1=g_2$.)
We shall therefore use the terms derivative coupling and entrainment coupling synonymously.
 
We restrict ourselves to uniform condensates and fluid velocities, $\partial^\mu\rho_i=0$, $\partial^\mu\psi_i = {\rm const}$. 
The phases $\psi_i$ depend linearly on time and space, with 
$\partial_0\psi_i = \mu_i$ and $\nabla\psi_i =-\mu_i \vec{v}_i$, see Eq.\ (\ref{super}). Actually, the chemical potentials $\mu_i$ 
should be introduced via 
${\cal H} - \mu_1{\cal N}_1 - \mu_2{\cal N}_2$, where ${\cal H}$ is the Hamiltonian and ${\cal N}_i = j^0_i$ the charge densities. It is well known that this is 
equivalent to introducing the chemical potential just like a background temporal gauge field, $\partial_0\varphi_i\to (\partial_0-i\mu_i)\varphi_i$. 
In appendix \ref{app0} we prove that this replacement has do be done in all derivative terms, i.e., also for the entrainment coupling, see also Ref.\ \cite{Stetina:2015exa}.
Consequently, we can indeed introduce the chemical potentials directly in the phase of the condensates. The same arguments can be applied to the superfluid velocity, or more precisely 
to the spatial components of the momentum conjugate to the current, such that we can write more generally 
${\cal H}-p_{1\mu}j_1^\mu-p_{2\mu}j_2^\mu$, and the vectors $\vec{p}_i$ can be viewed as spatial components of a background gauge field. 

The tree-level potential becomes 
\be \label{U}
U =  -\frac{p_1^2-m_1^2}{2}\rho_1^2+\frac{\lambda_1}{4}\rho_1^4 -\frac{p_2^2-m_2^2}{2}\rho_2^2+\frac{\lambda_2}{4}\rho_2^4
-\frac{h+gp_{12}^2}{2}\rho_1^2\rho_2^2 \, , 
\ee 
where
\be
p_i^2 \equiv \partial_\mu\psi_i\partial^\mu\psi_i = \mu_i^2(1-v_i^2) \, , \qquad p_{12}^2 \equiv \partial_\mu\psi_1\partial^\mu\psi_2 = \mu_1\mu_2(1-\vec{v}_1\cdot\vec{v}_2)
\, .
\ee
As explained in Sec.\ \ref{sec:hydro}, $p_1$ and $p_2$ are the chemical potentials in the rest frames of the fluids. 
Here we keep the general notation $p$ introduced in the previous section, deviating slightly from the notation in Refs.\ \cite{2013PhRvD..87f5001A,Schmitt:2013nva}, where 
$\partial_\mu\psi\partial^\mu\psi$ was instead denoted by $\sigma^2$.

The potential $U$ needs to be bounded from below. This requires $\lambda_1,\lambda_2>0$ (which we shall always assume)
and $h+gp_{12}^2<\sqrt{\lambda_1\lambda_2}$. In particular, the potential is bounded for 
arbitrary negative values of $h+gp_{12}^2$. Notice that the boundedness of the potential depends on the chemical potentials, which enter $p_{12}$ (and also on the 
fluid velocities). This is not a problem as long as we identify the unbounded region and always work with externally fixed chemical potentials in the bounded region. 
 
\begin{figure} [t]
\begin{center}
\hbox{\includegraphics[width=0.5\textwidth]{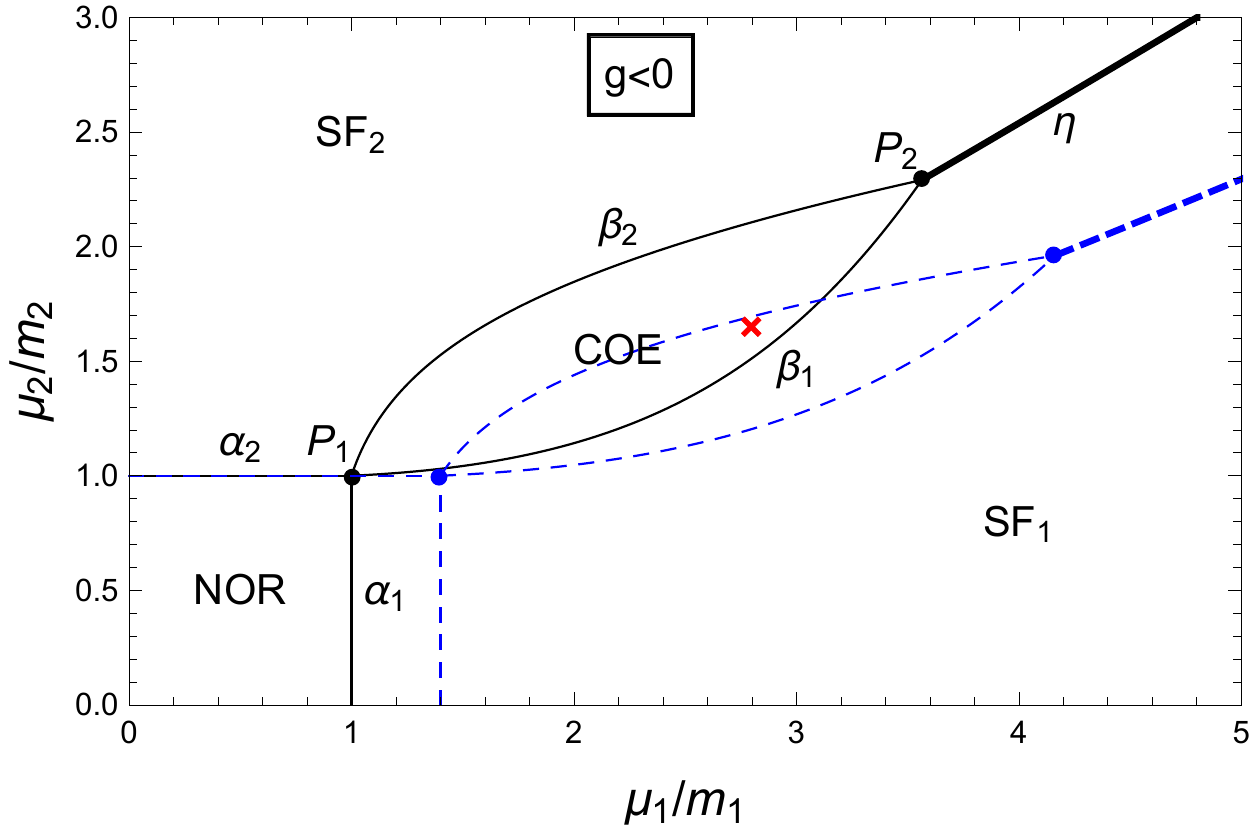}\includegraphics[width=0.5\textwidth]{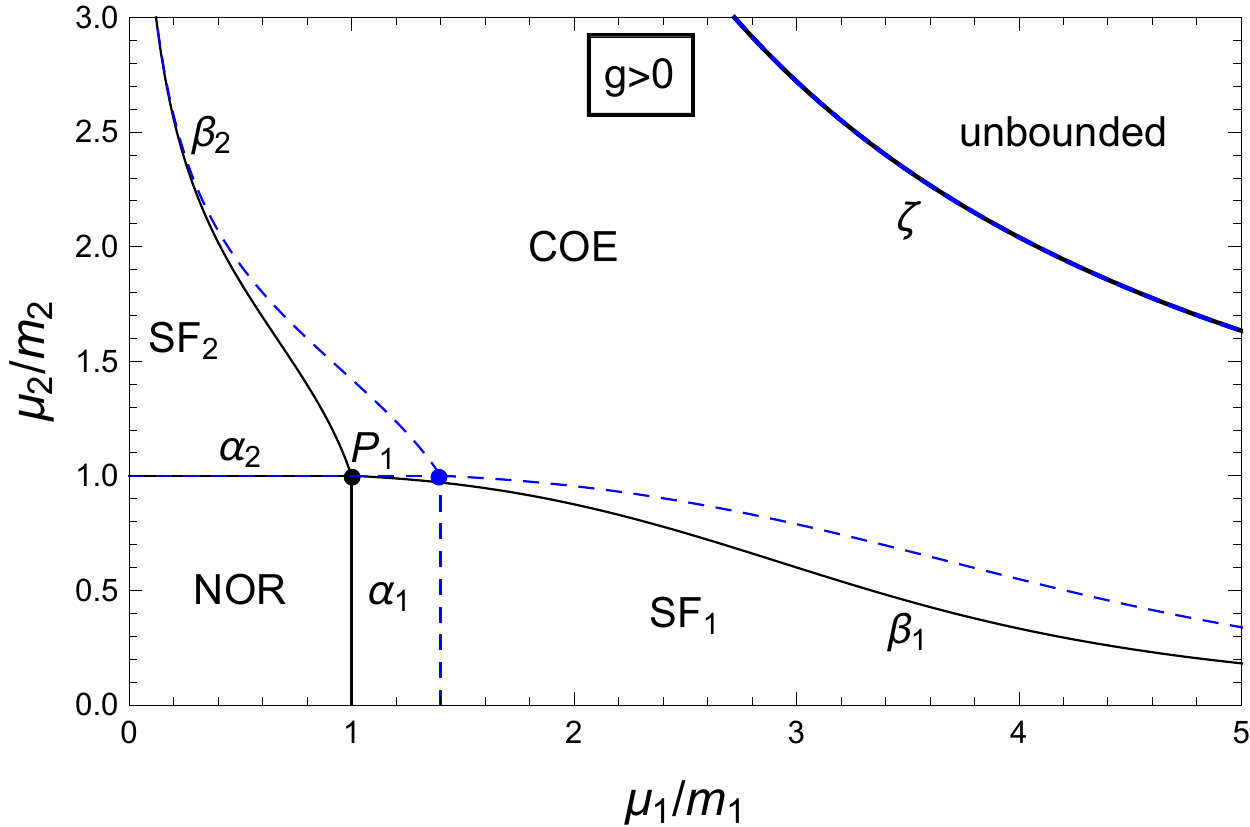}}
\caption{Phase diagrams in the plane of the two chemical potentials $\mu_1$, $\mu_2$ for two different signs of the entrainment coupling $g$ 
with the same magnitude $|g|=0.03/(m_1 m_2)$ and vanishing non-entrainment coupling $h=0$. We have chosen a mass ratio $m_2/m_1 = 1.5$ and the self-coupling 
constants are chosen to be $\lambda_1=0.3$, $\lambda_2=0.2$. Solid (black) lines correspond to vanishing fluid velocities, $v_1=v_2=0$, while dashed (blue) lines 
correspond to $v_1=0.7$, $v_2=0$ (in units of the speed of light). 
Thin lines are second-order phase transitions, while the thick lines in the left panel are phase transitions of first order. In 
the upper right corner of the right panel the potential is unbounded from below. The lines and points are labelled, and their expressions are given in Table \ref{table0}. NOR denotes 
the phase of no condensate, ${\rm SF}_1$ and ${\rm SF}_2$ the superfluid phases where only one field condenses, and COE the superfluid phase where both condensates coexist. This 
phase is most relevant for our discussion in Sec.\ \ref{sec:instabilities}, and for Figs.\ \ref{figdisp} -- \ref{figUpDown} we choose a point within this phase,
marked here with a (red) cross. }
\label{figphases}
\end{center}
\end{figure}

Our first goal is to find the phase structure of the model within the given uniform ansatz. To this end, we need to minimize the potential $U$
with respect to the condensates $\rho_1$ and $\rho_2$,
\be
0 = \frac{\partial U}{\partial\rho_1} = \frac{\partial U}{\partial\rho_2} \, . 
\ee
We first identify the four different phases that are solutions of these equations and that are distinguished by their residual symmetry group.
One solution is the normal phase [``NOR'', residual group $U(1)\times U(1)$], where there is no condensate, 
\be
\rho_1=\rho_2=0 \, , \qquad U_{\rm NOR}=0 \, . 
\ee
The other solutions are determined by the equations 
\begin{subequations}
\bea
p_1^2-m_1^2-\lambda_1\rho_1^2 + (h+gp_{12}^2)\rho_2^2 &=& 0 \, , \\[2ex]
p_2^2-m_2^2-\lambda_2\rho_2^2 + (h+gp_{12}^2)\rho_1^2 &=& 0 \, . 
\eea
\end{subequations}
If only field 1 condenses and forms a superfluid [``${\rm SF}_1$'', residual group $U(1)$], we have 
\be
\rho_1^2 = \rho_{01}^2 \equiv \frac{p_{1}^2-m_{1}^2}{\lambda_1} \, , \qquad \rho_{2} = 0 \, , \qquad U_{\rm SF_{1}} = -\frac{\lambda_{1}}{4}\rho_{01}^4 \, ,
\ee
and analogously for field 2 (``${\rm SF}_2$'').
Finally, both condensates may coexist (``COE'', residual group ${\bf 1}$), 
\bea \label{COE}
\rho_1^2 = \frac{\lambda_2(p_1^2-m_1^2)+(h+gp_{12}^2)(p_2^2-m_2^2)}{\lambda_1\lambda_2-(h+gp_{12}^2)^2} \, , 
\qquad \rho_2^2 = \frac{\lambda_1(p_2^2-m_2^2)+(h+gp_{12}^2)(p_1^2-m_1^2)}{\lambda_1\lambda_2-(h+gp_{12}^2)^2} \, , \non[2ex]\qquad 
U_{\rm COE} = - \frac{\lambda_1(p_2^2-m_2^2)^2+\lambda_2(p_1^2-m_1^2)^2+2(h+gp_{12}^2)(p_1^2-m_1^2)(p_2^2-m_2^2)}{4[\lambda_1\lambda_2-(h+gp_{12}^2)^2]}
\, .
\eea
These results can be used to determine the global minimum of the free energy for given $\mu_1$, $\mu_2$, $\vec{v}_1$, and $\vec{v}_2$. We shall restrict ourselves to $h=0$, i.e., 
the coupling between the two fields is only given by the derivative terms. We have checked that the presence of two different inter-fluid couplings, although giving rise to a 
more complicated phase structure, does not give a  
qualitative change to our main results which concern the instabilities discussed in Sec.\ \ref{sec:instabilities}.

Let us first briefly discuss the trivial situation without coupling, $g=0$, and without any velocities, $\vec{v}_1=\vec{v}_2=0$. Bose-Einstein condensation occurs for chemical 
potentials larger than the mass of the bosons. Therefore, in an uncoupled system, there is no condensate for $\mu_1<m_1$, $\mu_2<m_2$, there is exactly one condensate if exactly one 
of the chemical potentials becomes larger than the corresponding mass, and there are 
two condensates if both chemical potentials are larger than the corresponding masses, $\mu_1>m_1$, $\mu_2>m_2$. This can easily be checked 
from the above general expressions. A coupling between the two condensates can disfavor or favor coexistence of the two condensates, depending on the sign of the 
coupling constant. In our convention, $g<0$ disfavors and $g>0$ favors the COE phase. We therefore present two different phase diagrams in Fig.\ \ref{figphases} for these two 
qualitatively different cases\footnote{In a system of neutrons and protons inside a neutron star, the results of Ref.\ \cite{Chamel:2006rc} suggest that the entrainment 
coupling $g$ is negative. This can be seen 
by rewriting the free energy (\ref{U}) as
\be
U = U(\vec{v}_1=\vec{v}_2=0) + \frac{\mu_1^2\rho_1^2}{2}v_1^2+\frac{\mu_2^2\rho_2^2}{2}v_2^2 +\frac{g\mu_1\mu_2\rho_1^2\rho_2^2}{2}\vec{v}_1\cdot\vec{v}_2 \, .
\ee
By comparing this expression with the non-relativistic version in Ref.\ \cite{Chamel:2006rc} and using the results from the fermionic microscopic theory therein, 
we conclude $g<0$ (with $|g|$ depending on the baryon density). In view of our results it is thus an interesting question whether  
the entrainment coupling between neutrons and protons may forbid the coexistence of both condensates, in particular under circumstances (i.e., at a given temperature and 
baryon density) in which each of the condensates would be allowed to exist on its own.}.
Without loss of generality we can restrict ourselves to $\mu_1,\mu_2>0$ because the chemical potentials enter 
the free energy $U$ only quadratically and through the combination $g\mu_1\mu_2$, and $g$ only enters in this combination. All phase transition lines and critical points are 
given by simple analytical expressions, see Table\ \ref{table0}.

\begin{table*}[t]
\begin{tabular}{|c||c|} 
\hline
\rule[-1.5ex]{0em}{4ex} 
 $\alpha_i$ & $\displaystyle{\mu_{i} = \gamma_i m_{i}}$ \\[1ex] \hline
\rule[-1.5ex]{0em}{8ex} 
 $\;\;\beta_{1/2}\;\;$ & $\displaystyle{\mu_{2/1} = \gamma_{2/1}^2\left[\sqrt{\frac{g^2\mu_{1/2}^2}{4}\rho_{01/02}^4(1-\vec{v}_1\cdot\vec{v}_2)^2+\frac{m_{2/1}^2}{\gamma_{2/1}^2}}
-\frac{g\mu_{1/2}}{2}\rho_{01/02}^2(1-\vec{v}_1\cdot\vec{v}_2)\right]}$ \\[4ex] \hline
\rule[-1.5ex]{0em}{6ex}
 $\eta$ & $\displaystyle{\mu_1 = \gamma_1\sqrt{m_1^2+\sqrt{\lambda_1\lambda_2}\rho_{02}^2}}$ \\[2ex] \hline
\rule[-1.5ex]{0em}{6ex}
 $\zeta$ & $\displaystyle{\mu_2 = \frac{\sqrt{\lambda_1\lambda_2}}{g\mu_1(1-\vec{v}_1\cdot\vec{v}_2)}}$ \\[3ex] \hline\hline
\rule[-1.5ex]{0em}{4ex}
 $P_1$ & $\displaystyle{(\gamma_1m_1,\gamma_2 m_2)}$ \\[1ex] \hline
\rule[-1.5ex]{0em}{7ex}
 $P_2$ & $\;\;(\mu_1,\mu_2)$ with $\displaystyle{\mu_{1/2}^2 = \frac{\gamma_{1/2}^2}{2\sqrt{\lambda_{2/1}}}\left[\sqrt{\left(\sqrt{\lambda_2}m_1^2-\sqrt{\lambda_1}m_2^2\right)^2
+\frac{4(\lambda_1\lambda_2)^{3/2}}{g^2\gamma_1^2\gamma_2^2(1-\vec{v}_1\cdot\vec{v}_2)^2}}\pm(\sqrt{\lambda_2}m_1^2-\sqrt{\lambda_1}m_2^2)\right]}\;\;$ \\[3ex] \hline
\end{tabular}
\caption{Expressions for the phase transition lines $\alpha_1$, $\alpha_2$, $\beta_1$, $\beta_2$ (second order), $\eta$ (first order), the critical line $\zeta$ for the unboundedness 
of the potential,
and the critical points $P_1$ and $P_2$ in the phase diagrams of Fig.\ \ref{figphases} (as in the phase diagrams, we have set the non-entrainment coupling $h=0$ for simplicity). 
We have abbreviated the Lorentz factors $\gamma_i = 1/\sqrt{1-v_i^2}$ and the condensates in the absence of a second field
$\rho_{0i}^2=(p_i^2-m_i^2)/\lambda_i$, $i=1,2$. 
}
\label{table0}
\end{table*}

We see in the left panel that the region for the COE phase gets squeezed by the entrainment coupling. Let us explain the various phase transitions by 
following a horizontal line in the phase diagram: for $1<\mu_2/m_2\lesssim 2.3$ we start in the ${\rm SF}_2$ phase and, upon increasing $\mu_1$, reach the line $\beta_2$. At this line
both condensates change continuously, in particular the condensate of field 1 becomes nonzero. 
This second-order phase transition line is found from $U_{{\rm SF}_2} = U_{\rm COE}$. It is easy to check that it is identical to the line where $\rho_1$ from 
Eq.\ (\ref{COE}) is zero. By further increasing $\mu_1$ we leave the COE phase through a second-order phase transition line $\beta_1$ and reach the phase ${\rm SF}_1$. 
For $\mu_2/m_2\gtrsim 2.3$, we do not reach the COE phase by increasing $\mu_1$, although there is a region where the COE phase is allowed, i.e., where both $\rho_1$ and $\rho_2$ from 
Eq.\ (\ref{COE}) assume real nonzero values. However, the free energy of this state turns out to be larger than the free energies of the phases ${\rm SF}_1$ and ${\rm SF_2}$. 
Therefore, there is a direct first-order phase transition line $\eta$ between ${\rm SF}_1$ and ${\rm SF}_2$. For larger values of $|g|$ the region of the COE phase becomes smaller, 
and beyond a critical value the COE phase completely disappears from the phase diagram. 
This critical value is reached when the points $P_1$ and $P_2$ coincide, and it is given by
\be
g = -\frac{\sqrt{\lambda_1\lambda_2}}{m_1m_2}\frac{\sqrt{(1-v_1^2)(1-v_2^2)}}{1-\vec{v}_1\cdot\vec{v}_2} \, . 
\ee
The right panel shows the scenario where the coupling is in favor of the COE phase. In this case, there is a region for large chemical potentials 
where the tree-level potential becomes unbounded. 

In each panel we show the phase structure for the case of zero velocities and for the case where the velocity of superfluid 1 is nonzero. The effect of the nonzero velocity 
on the ${\rm SF}_1$ phase is 
very simple: condensation occurs if $p_1>m_1$, i.e., if the chemical potential measured in the rest frame of the fluid
is sufficiently large. We work, however, with fixed $\mu_1$, i.e., we fix the chemical potential measured in the frame where the fluid moves with velocity $\vec{v}_1$. 
Therefore, the chemical potential relevant for condensation, $p_1 = \mu_1\sqrt{1-v_1^2}$, is reduced by a nonzero $\vec{v}_1$ through a standard Lorentz factor, and thus a nonzero 
velocity effectively disfavors condensation of the given field. The effect of the velocity on the COE phase is a bit more complicated, in this case there is no frame in which 
the velocity dependence can be eliminated. For either sign of the coupling $g$, a nonzero velocity reduces the region of the COE phase, i.e., there is a parameter region in which, 
for zero velocity, the COE phase is preferred, but which is taken over by a single-condensate phase or the NOR phase at nonzero velocity. For negative values of the entrainment 
coupling $g$, 
see left panel of Fig.\ \ref{figphases}, a nonzero velocity can also work in favor of the COE phase:
there are points in the single-superfluid phase ${\rm SF}_1$ at zero $\vec{v}_1$ which undergo a phase transition to the COE phase at nonzero $\vec{v}_1$.

We emphasize that the discussion in this section has been on a purely thermodynamic level, in the sense that the velocities have been treated as external 
parameters in the same way as the chemical potentials. This is possible when the velocity fields are constant in space and time. It is very natural from the point of view
of the covariant formalism since chemical potential and superfluid velocity are different components of the same four-vector, the conjugate momentum $\partial^\mu\psi$.
This ``generalized'' thermodynamics can be carried further to compute Landau's critical velocity from ``generalized'' susceptibilities 
\cite{2003JETPL..78..574A,2008JLTP..150..612K}. We shall discuss this connection after we have computed the quasiparticle excitations, from which Landau's critical velocity as well 
as dynamical instabilities can be 
computed. 

\subsection{Quasiparticle propagator and Goldstone modes}
\label{sec:goldstone}

In this section we are mainly interested in the quasiparticle excitations, which are obtained from the poles of the quasiparticle propagator. Using the simplest 
possible approximation, we work with the tree-level propagator, which is straightforwardly read off from the Lagrangian. Here, in the main text, we simply present the propagator in 
momentum space and then directly proceed with evaluating the excitation energies. In appendix \ref{app1}, we present a more detailed derivation, embedded in the formalism of thermal field theory.
This illustrates the meaning of 
the excitation energies in a system in thermal equilibrium and prepares the extension of the present calculation to 
nonzero temperatures in future work.

The calculation of the excitation energies requires to include the fluctuations that we had dropped in Eq.\ (\ref{shift}),
\be \label{fluc}
\varphi_i = \frac{e^{i\psi_i}}{\sqrt{2}}(\rho_i+\phi_i'+i\phi_i'') \, .
\ee
Here, the fluctuations are parameterized in terms of real part $\phi_i'$ and imaginary part $\phi_i''$, and we have  separated 
the space-time dependent 
phase factor $e^{i\psi_i}$ for convenience, i.e., we work with fluctuation fields transformed by this factor. The reason is that only in the basis of the transformed 
fields the propagator is diagonal in momentum space. The inverse propagator is obtained from the quadratic terms in the fluctuations and can be written as 
\be
S^{-1} = \left(\begin{array}{cc} S^{-1}_{11} & S^{-1}_{12} \\ S^{-1}_{21} & S^{-1}_{22} \end{array}\right) \, , 
\ee
where
\begin{subequations} \label{Sinv}
\bea
S^{-1}_{11/22} &=& \left(\begin{array}{cc} -K^2+\lambda_{1/2}(3\rho_{1/2}^2-\rho_{01/02}^2)  & 
2iK\cdot\partial\psi_{1/2} 
\\ -2iK\cdot\partial\psi_{1/2} & 
-K^2+\lambda_{1/2}(\rho_{1/2}^2-\rho_{01/02}^2)   \end{array}\right) \non[2ex]
&&-\left(\begin{array}{cc}  (h+gp_{12}^2)\rho_{2/1}^2  & 
-ig\rho_{2/1}^2 K\cdot\partial\psi_{2/1} 
\\ ig\rho_{2/1}^2 K\cdot\partial\psi_{2/1} & 
 (h+gp_{12}^2)\rho_{2/1}^2  \end{array}\right) \, , \allowdisplaybreaks\\[2ex]
S^{-1}_{12/21} &=& \frac{\rho_1\rho_2}{2}\left(\begin{array}{cc} GK^2-4(h+gp_{12}^2)\;\;  & 
2igK\cdot\partial\psi_{1/2} \\ -2igK\cdot\partial\psi_{2/1} & 
-gK^2  \end{array}\right) \, ,
\eea
\end{subequations}
with the four-momentum $K=(k_0,\vec{k})$, and $K\cdot\partial\psi_{1/2} = K_\mu \partial^\mu\psi_{1/2}$. Note that here also the sum of the two derivative coupling constants $G$ appears, while in the tree-level potential of the previous section only their difference $g$ 
had entered.
This form of the propagator is general and holds for all possible phases. We are only interested in the excitation energies of the COE phase, where both condensates are nonzero. 
In this case, \mbox{$\lambda_{1/2}(\rho_{1/2}^2-\rho_{01/02}^2) -(h+gp_{12}^2)\rho_{2/1}^2 =0$}, and we can simplify
\be
S^{-1}_{11/22} = \left(\begin{array}{cc} -K^2+2\lambda_{1/2}\rho_{1/2}^2\;\;  & 
2iK\cdot\partial\psi_{1/2} 
\\ -2iK\cdot\partial\psi_{1/2} & 
-K^2  \end{array}\right)
+
ig\rho_{2/1}^2K\cdot\partial\psi_{2/1}\left(\begin{array}{cc} 0  & 
1 \\ -1 & 0  \end{array}\right) \, .
\ee
The excitation energies are given by ${\rm det}\,S^{-1}=0$. Due to the symmetry of the determinant under $K\to -K$, the excitations come in 4 pairs: if $k_0=\epsilon_{r,\vec{k}}$ is a zero, then also 
$k_0=-\epsilon_{r,-\vec{k}}$, ($r=1,\ldots ,4$). In appendix \ref{app1} we show that only one energy of each pair has to be kept in order to compute the thermodynamic properties of the system. In the COE phase, the solutions of ${\rm det}\, S^{-1}=0$ correspond to the two Goldstone modes and two massive modes.   
For the simplest scenario, let us set the superfluid velocities, the mass parameters, and the non-entrainment coupling to zero, $\nabla\psi_1=\nabla\psi_2=m_1=m_2=h=0$.
Then, the energy of the massive modes has the form $\epsilon_{\vec{k}} = M + {\cal O}(k^2)$ with the two masses 
\be
M = \sqrt{6}\left[\mu_{1/2} + \frac{\mu_{2/1}^3}{2\lambda_{2/1}}\,g + {\cal O}(g^2)\right] \, .
\ee
These modes are of no further interest to us since we shall focus on the low-energy properties of the system. Also, we should recall that most of the real-world 
superfluids we have in mind 
are of fermionic nature. Therefore, our bosonic approach can at best be a low-energy effective description. In a fermionic superfluid there is a massless mode too because of the 
Goldstone theorem, but typically there is no stable massive mode. There rather is a continuum of states for energies 
larger than twice the fermionic pairing gap \cite{Fukushima:2005gt,2015JPSJ...84d4003Y}, and thus at these energies our effective bosonic description breaks down. 

The energies of the Goldstone modes have the form $\epsilon_{\vec{k}} = u k + {\cal O}(k^3)$, where
\be
u  = \frac{1}{\sqrt{3}}\left[1\pm\frac{\mu_1\mu_2}{2\sqrt{\lambda_1\lambda_2}}\,g + {\cal O}(g^2)\right] \, .
\ee
We see that the effect of a small entrainment coupling is to split the two Goldstone modes, with one mode becoming faster and one mode becoming slower. One can 
check that the behavior of a non-entrainment coupling $h$ is different: if we set the entrainment couplings to zero, $G=g=0$, but keep a nonzero $h$, we find that one mode remains unperturbed by 
the coupling and the other acquires a larger speed. 

After these preparations we can now consider nonzero fluid velocities and discuss the resulting instabilities. 

\section{Dynamical and energetic instabilities}
\label{sec:instabilities}
 
\subsection{Results for sound modes and identification of instabilities}

Having identified the regions in parameter space where both species 1 and 2 become superfluid and having derived the quasiparticle propagator $S$ for this
phase, we now compute the excitation energies numerically. We focus on the two Goldstone modes and do not discuss the massive modes any further. 
For small momenta, the dispersion relations of the Goldstone modes are linear, $\epsilon_{\vec{k}} = u k$, 
and their slopes $u$ can also be computed from the linearized hydrodynamic equations, i.e., if we are only interested in the low-momentum behavior we may alternatively employ 
Eq.\ (\ref{detu}) instead of ${\rm det}\,S^{-1}=0$. In general, this is not true. For instance, in a single superfluid at nonzero temperature, there is 
only one Goldstone mode, but there are two different sound modes, usually called first and second sound. Only for small temperatures, the Goldstone mode is well 
approximated by first sound, in general neither first nor second sound corresponds to the Goldstone mode. In our zero-temperature approximation of two coupled superfluids, 
we have ``two first sounds'' which coincide with the Goldstone modes at low momentum.

\begin{figure}[t]
\begin{center}
\hbox{\includegraphics[width=0.28\textwidth]{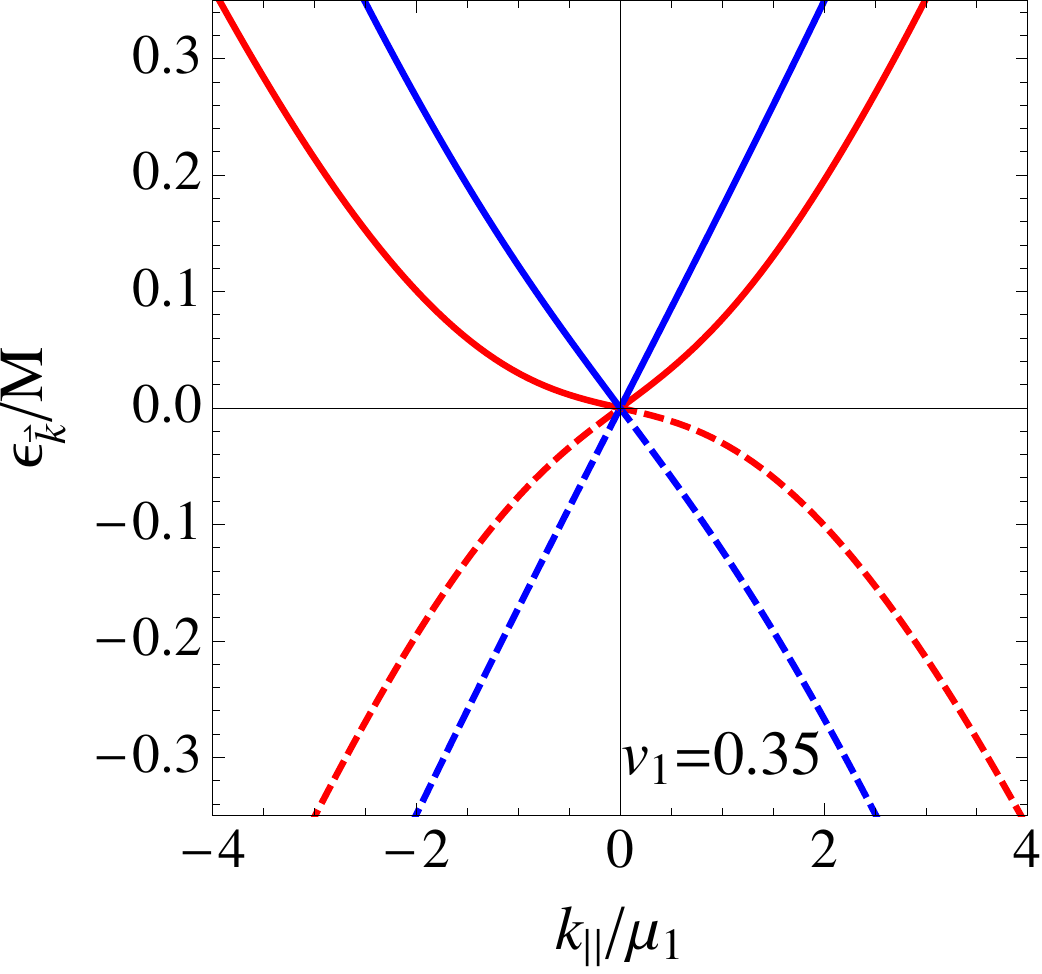}\includegraphics[width=0.235\textwidth]{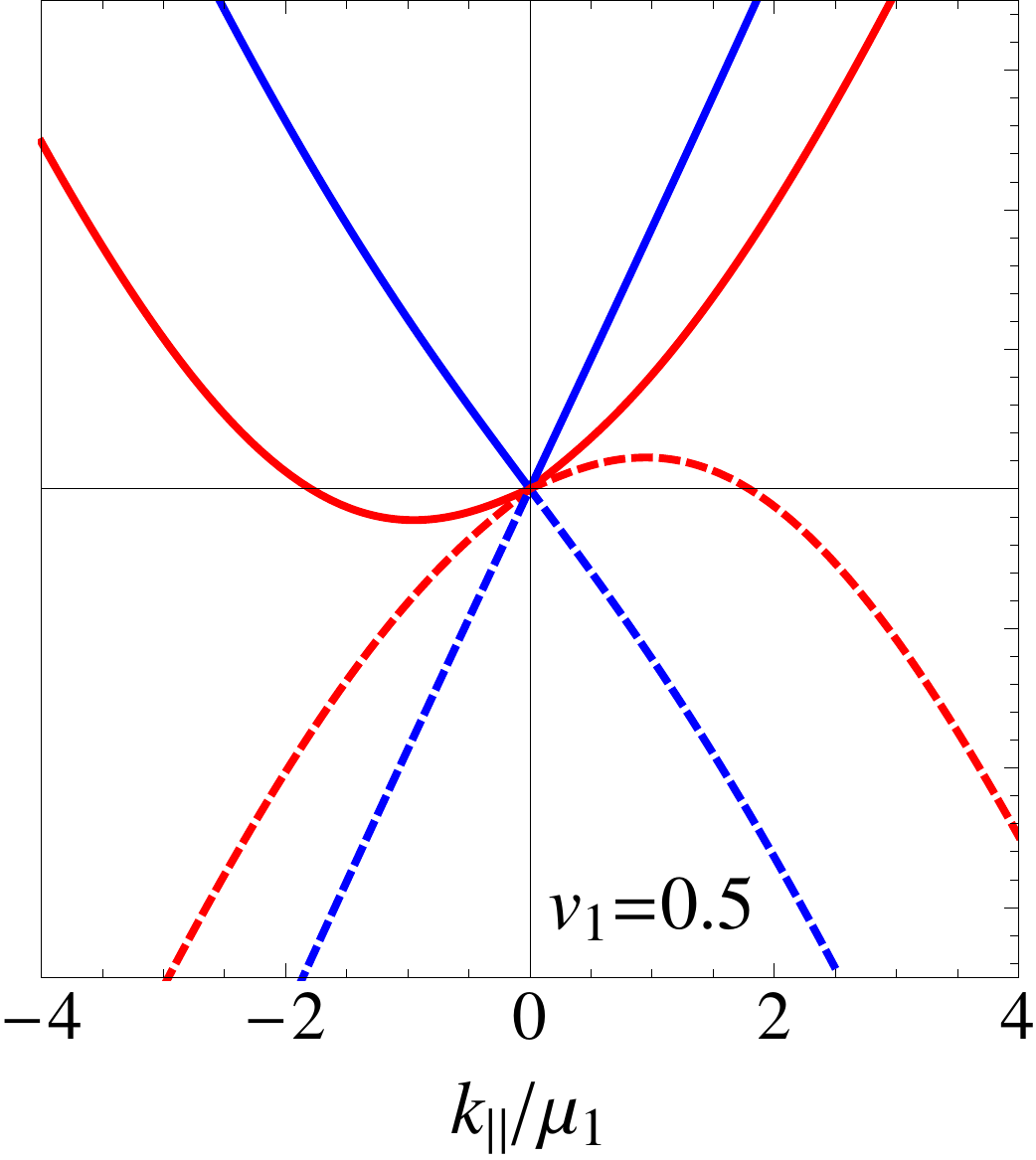}
\includegraphics[width=0.235\textwidth]{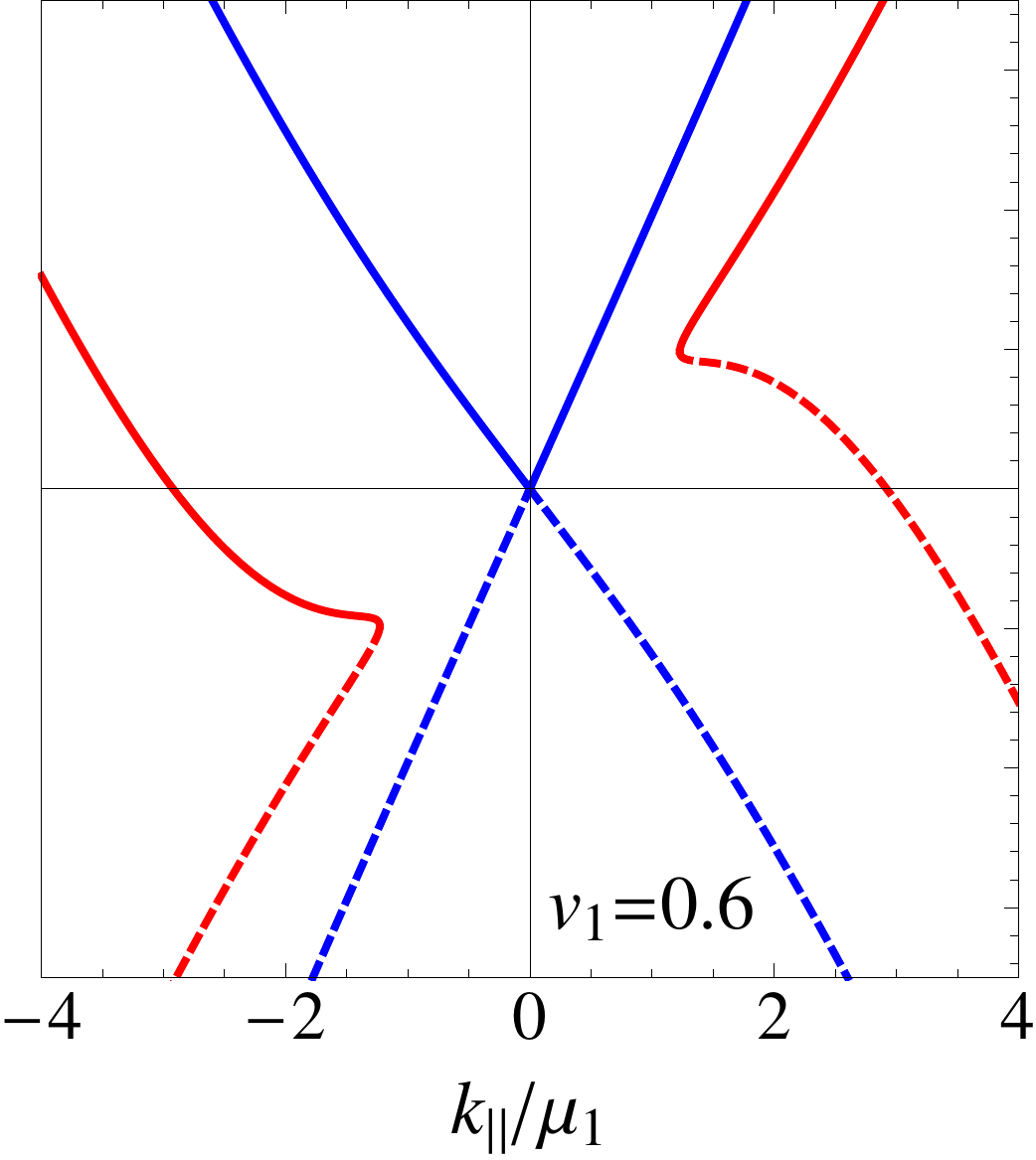}\includegraphics[width=0.235\textwidth]{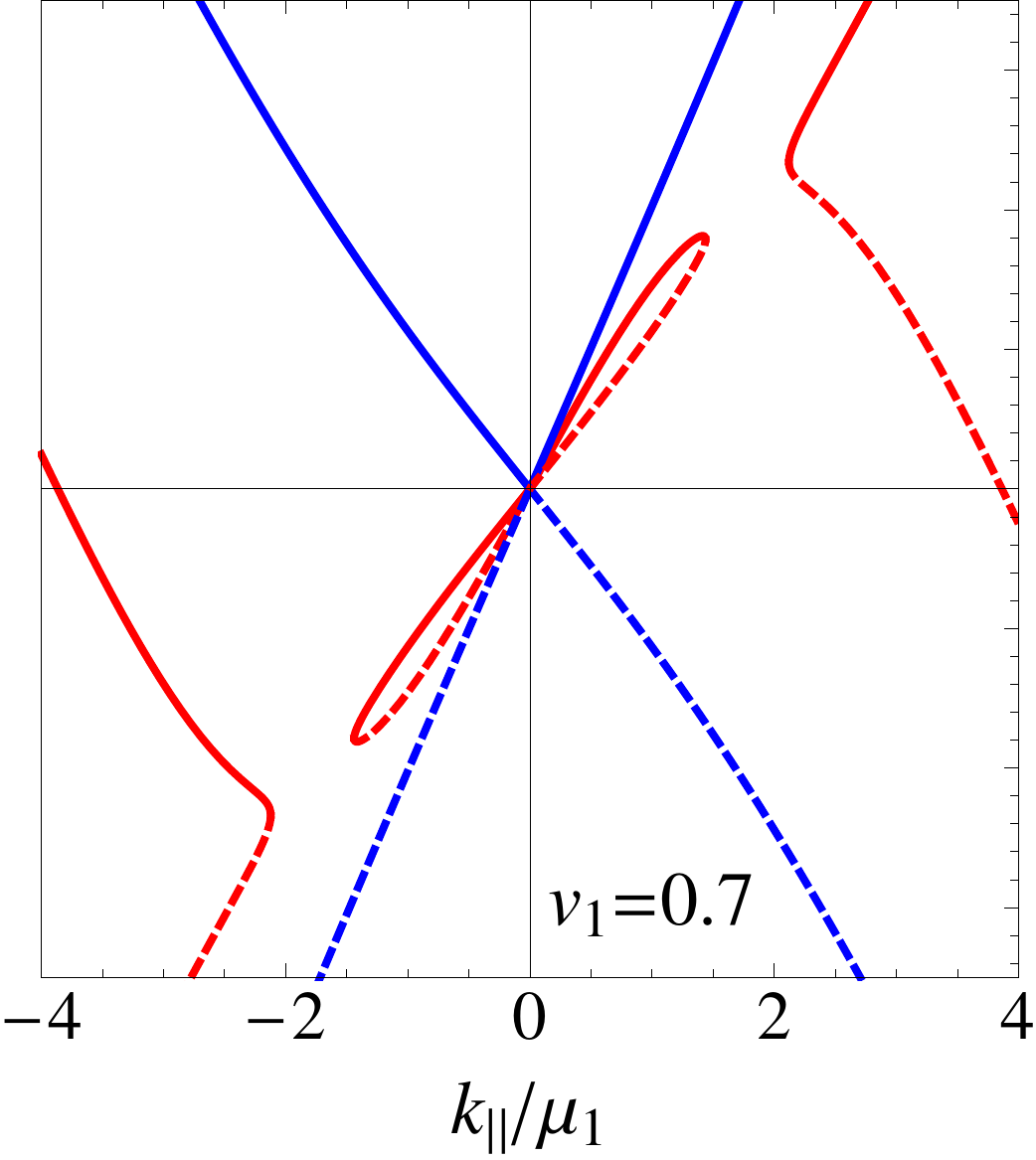}}
\caption{Goldstone dispersion relations with the parameters of the phase diagram in the left panel of Fig.\ \ref{figphases} with $\mu_1/m_1=2.8$, $\mu_2/m_2=1.67$ (marked point in that 
phase diagram), and four different velocities $v_1$, parallel ($k_{||}>0$) and anti-parallel ($k_{||}<0$) to 
$\vec{v}_1$. 
The excitation energy $\epsilon_{\vec{k}}$ is normalized to the mass $M$ of the lighter of the two massive modes. For small momenta the dispersion relations are linear, their 
slope is shown in Fig.\ \ref{figpolar} for all angles. For a given $k_{||}$ 
we show all four massless solutions of ${\rm det}\,S^{-1}=0$, including the negative ``mirror branches'' (dashed lines). 
From the second panel on, the branches that were positive for vanishing superflow (solid lines) acquire negative energies for certain momenta. 
In the third and fourth panels there are gaps in the curves for certain momenta where $\epsilon_{\vec{k}}$ is complex, 
indicating a dynamical instability. }
\label{figdisp}
\end{center}
\end{figure}
\begin{figure}[t]
\begin{center}
\hbox{\includegraphics[width=0.27\textwidth]{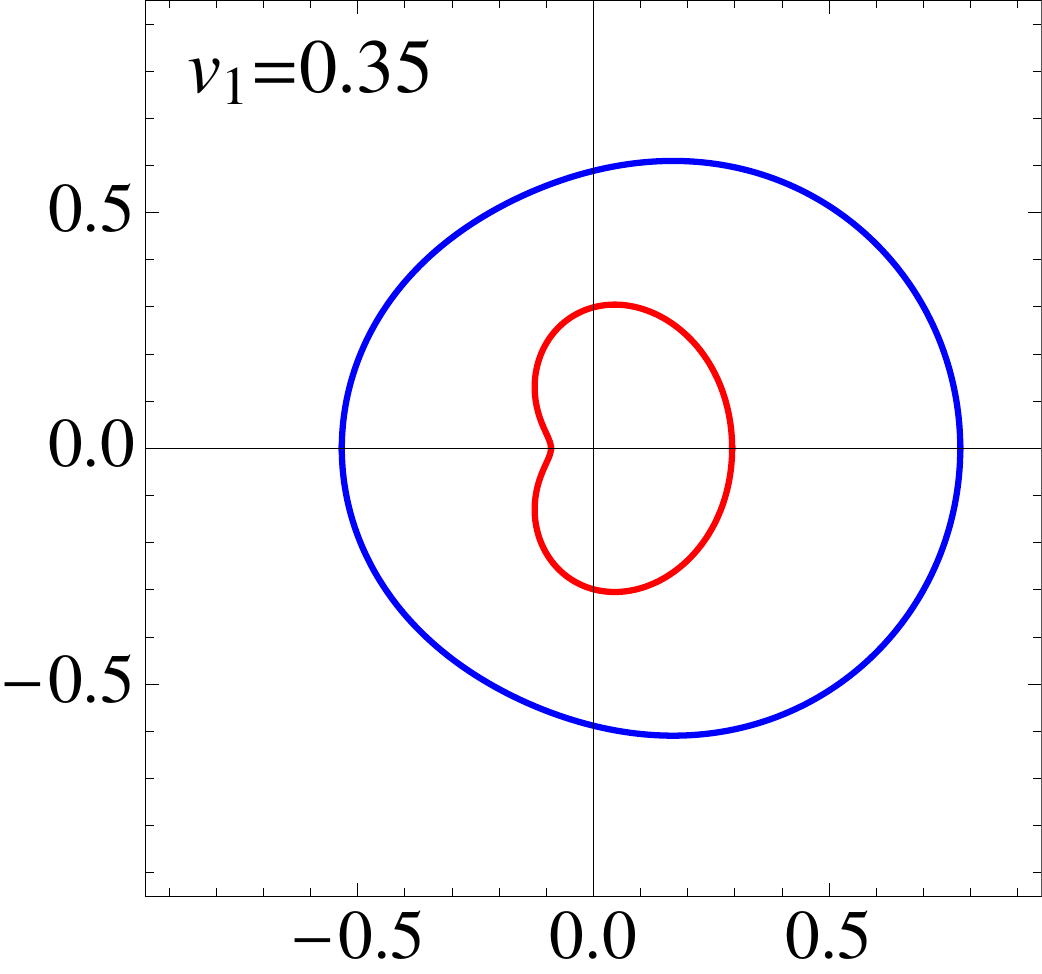}\,\includegraphics[width=0.235\textwidth]{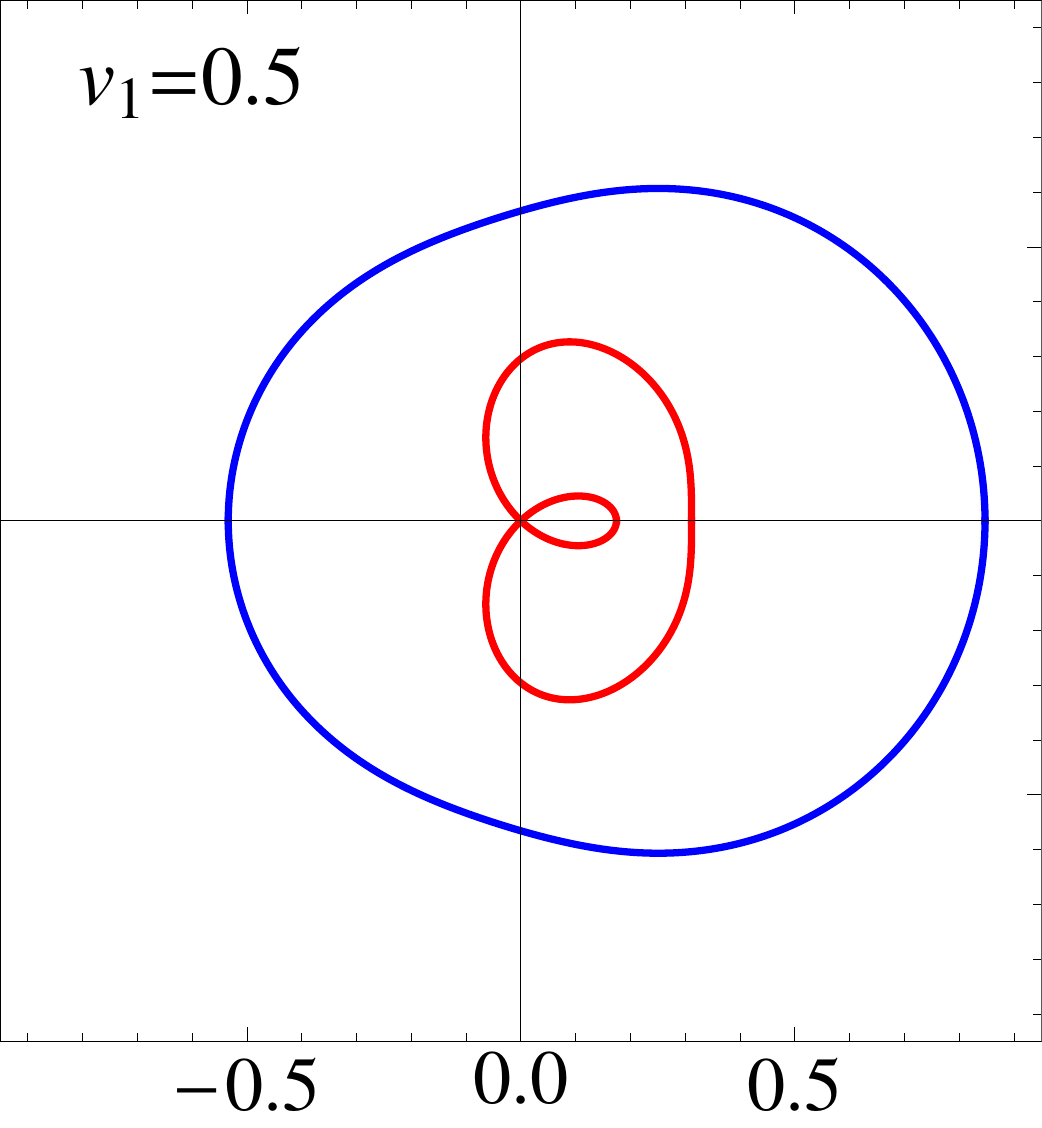}
\includegraphics[width=0.235\textwidth]{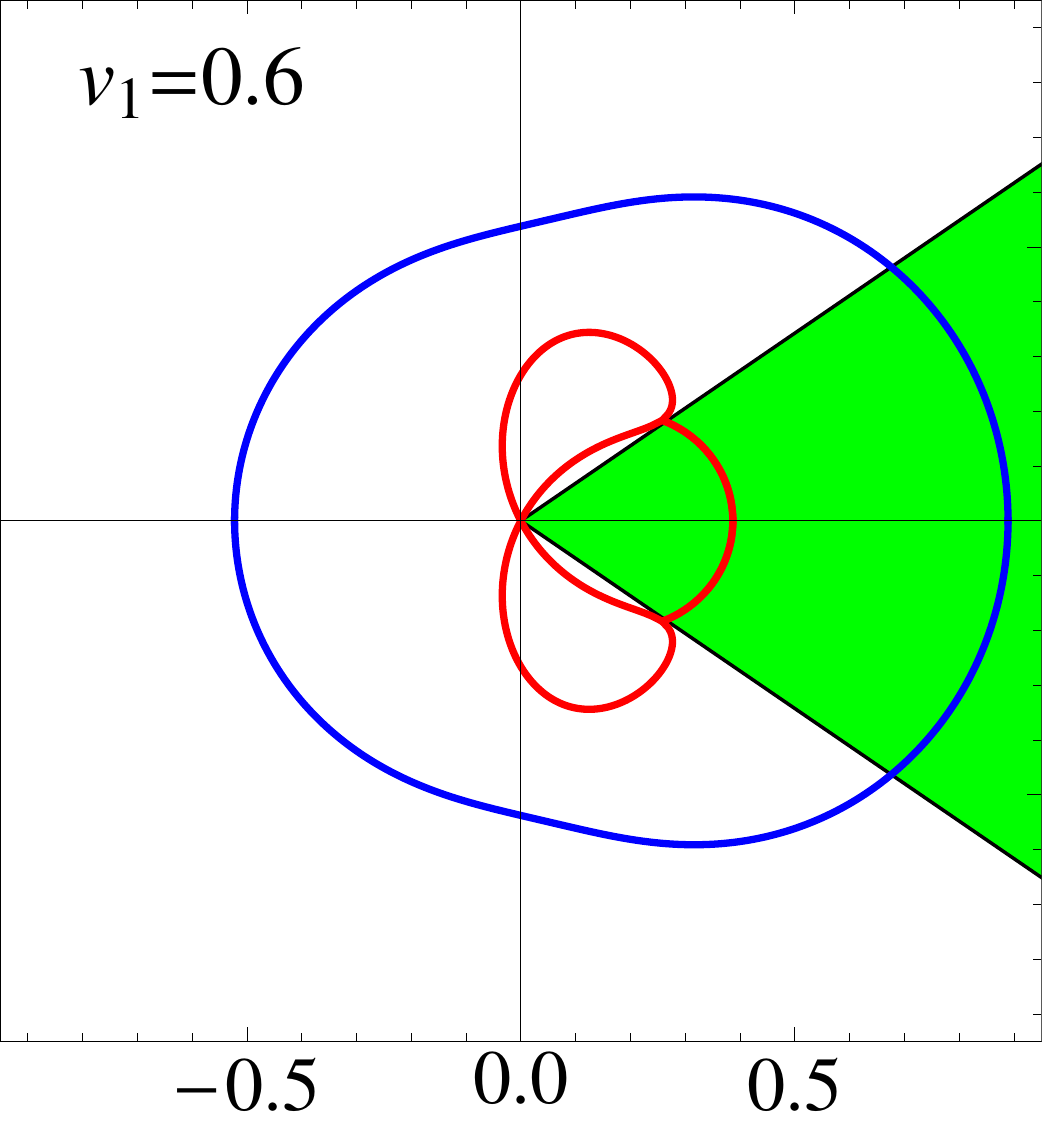}\,\includegraphics[width=0.235\textwidth]{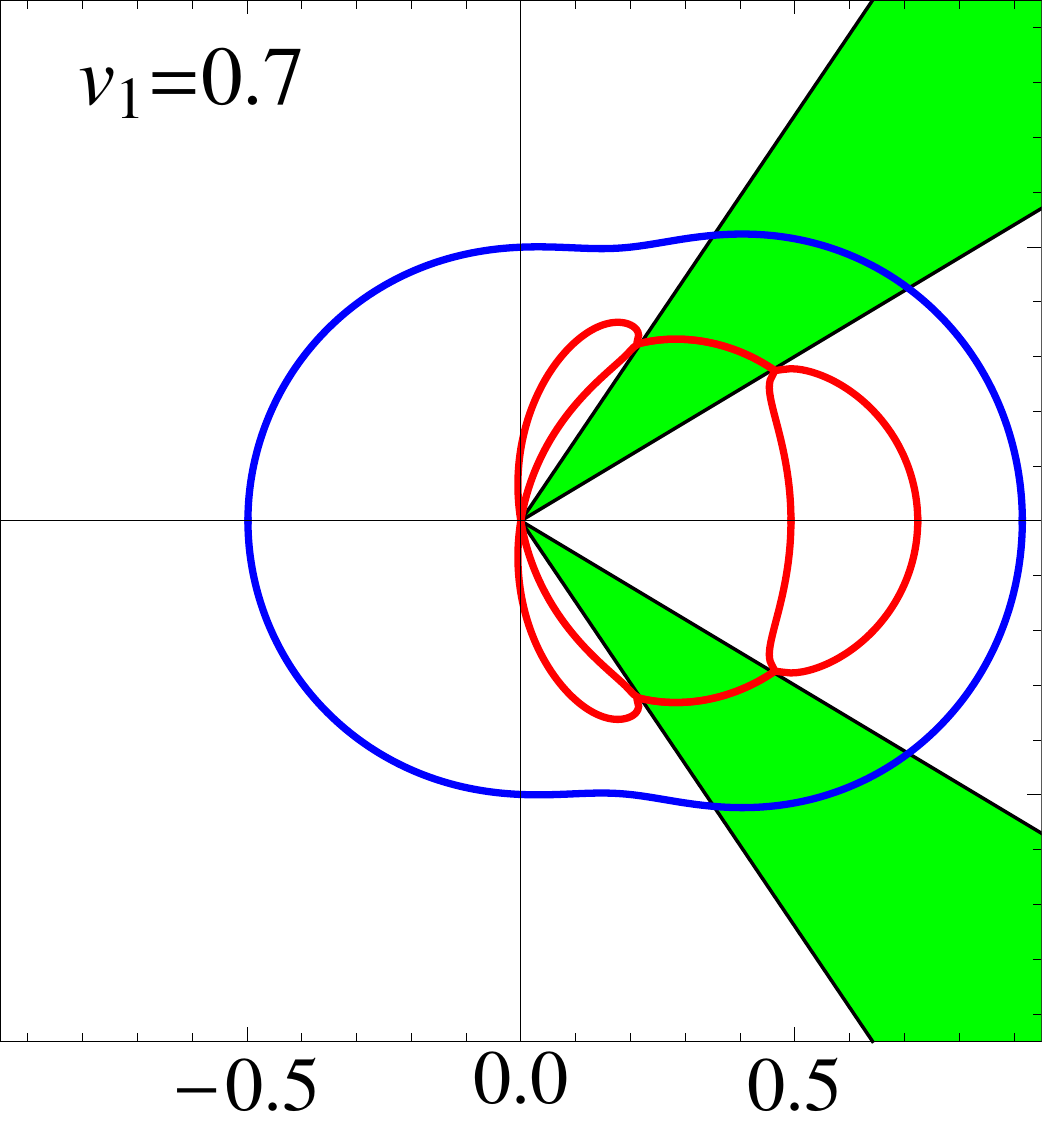}}
\caption{Real part of the sound speeds with the parameters of Fig.\ \ref{figdisp} for all angles between the wave vector $\vec{k}$ and the velocity $\vec{v}_1$, 
and four different magnitudes $v_1$ (the distance from the origin to the curve is the speed of 
sound for a given angle). The velocity $\vec{v}_1$ points to the right. In the two last panels there are unstable directions for which the sound speed becomes complex and the 
real parts of two branches coincide, indicated by the shaded (green) areas.}
\label{figpolar}
\end{center}
\end{figure}

Our microscopic model contains several parameters, and a complete survey of the parameter space is very tedious and not necessary for our purpose. Of course, many details depend on the 
specific choice of the parameters. For instance, had we allowed for a non-entrainment coupling $h$ in addition to the entrainment coupling $g$, the topology of the phase diagrams
in Fig.\ \ref{figphases} would have been different for certain choices of $h$ and $g$. 
However, the phase structure is not our main concern, we are rather interested in the instabilities at nonzero fluid velocities. 
To this end, we consider a particular point of the phase diagram in the left panel of Fig.\ \ref{figphases}, 
marked with a (red) cross, i.e., we again restrict ourselves to the case of a pure entrainment coupling, setting $h=0$. The behavior of this point 
in terms of dynamical and energetic instabilities is generic in a sense that we will discuss in Sec.\ \ref{sec:analysis}. 
In fact, for the two-component superfluid, the following discussion about dynamical and energetic instabilities would be quantitatively the same for a non-entrainment coupling, 
entrainment is {\it not} crucial. Therefore, (the non-relativistic limit of) our results will also be relevant for dilute atomic gases, where entrainment is believed to be negligible.
For a system of two normal fluids, however, entrainment does make a qualitative 
difference for the dynamical instability, see Fig.\ \ref{fignn} and discussion at the end of Sec.\ \ref{sec:analysis}.
The entrainment terms themselves contain two different coupling constants and in principle we can choose both of them independently. 
We have seen that for the tree-level potential only the combination $g = (g_1-g_2)/2$ 
matters, while in the propagator both constants enter separately. One can check, however, that for the linear, low-energy part of the dispersion, again only $g$
matters, such that we can keep working with the single entrainment parameter $g$. Only in Fig.\ \ref{figdisp} we show dispersion relations that go beyond linear order in momentum,
and in this case we have made the choice $g_2=0$, such that $g=G$ (again with quantitative, but for our conclusions irrelevant, changes if $g_1$ and $g_2$ are chosen 
differently). 

In Fig.\ \ref{figdisp} we show the dispersion relations of the two Goldstone modes for four different values of $v_1$, with $v_2=0$, i.e., 
the calculation is done in the rest frame of superfluid 2. We show 
all four massless solutions of ${\rm det}\,S^{-1}=0$. As explained in the previous section, for each solution $\epsilon_{r,\vec{k}}$ (solid lines),  
$-\epsilon_{r,-\vec{k}}$ is also a solution
(dashed lines). Several observations are obvious from the four panels. First of all we see the trivial effect that 
a nonzero superflow (here of superfluid 1) leads to anisotropic dispersion relations, in particular to different sound speeds parallel and anti-parallel to the superflow
(first panel). Beyond a certain value of the superflow, negative excitation energies appear for small momenta (second panel), before the energies become 
complex at small momenta (third panel), and this complex region moves to larger momenta (fourth panel). Complementary information for the same parameter set is shown 
in Fig.\ \ref{figpolar}. The four panels in this figure show less in the sense that only the sound speeds are plotted (i.e., the slope of the Goldstone dispersion at small 
momenta), but they show more in the sense that these speeds are shown for {\it all} angles between the direction of the sound wave and the superflow. Also, 
in Fig.\ \ref{figpolar} we have restricted ourselves to positive excitations, i.e., only the branches of the upper half of Fig.\ \ref{figdisp} are shown in Fig.\ \ref{figpolar}.
For instance, for $v_1=0.5$, there is a branch with negative energy in the upstream direction (anti-parallel to $\vec{v}_1$), see the lower (red) solid line in the second panel of 
Fig.\ \ref{figdisp}. At the same time, the ``mirror branch'' in the downstream direction has acquired positive energy, the upper (red) dashed curve in the same panel. The latter is shown 
as a solid curve in the second panel of Fig.\ \ref{figpolar}, which also shows that this branch exists for all angles in the half-space $\vec{k}\cdot\vec{v}_1>0$.

\begin{figure} [t]
\begin{center}
\hbox{\includegraphics[width=0.5\textwidth]{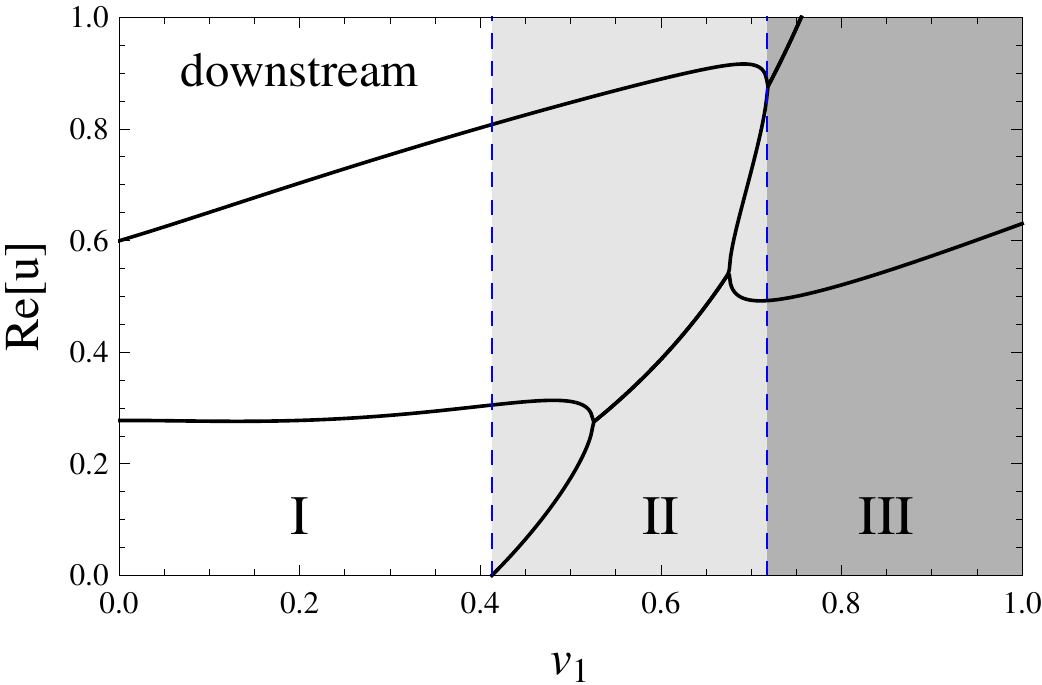}\includegraphics[width=0.5\textwidth]{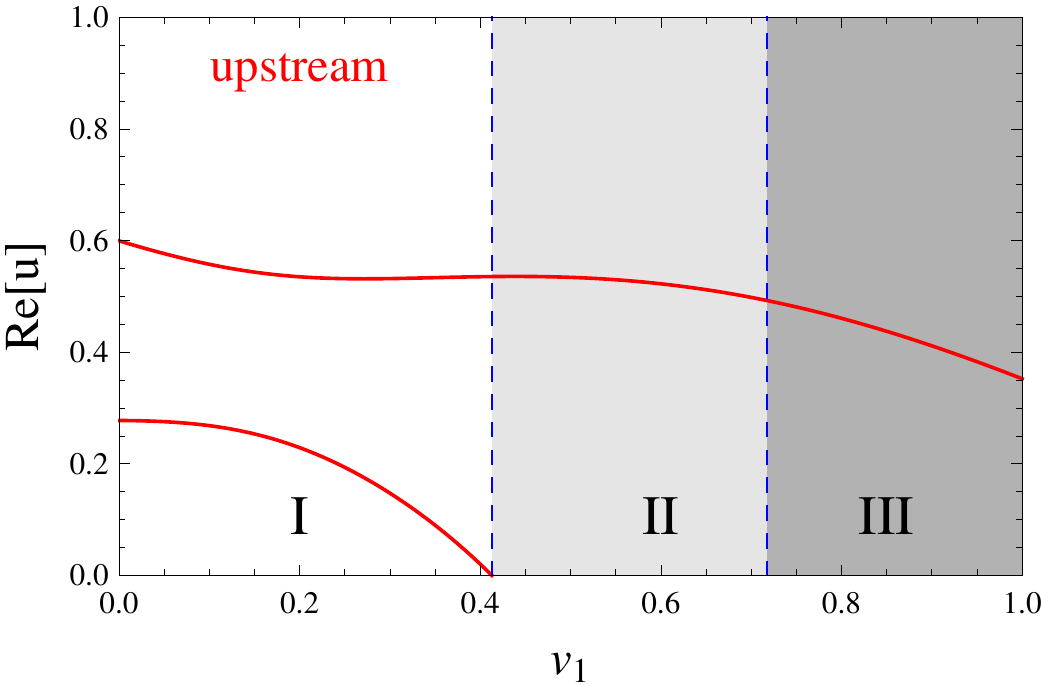}}
\caption{Real part of the sound speeds parallel (``downstream'') and anti-parallel (``upstream'') to the velocity of superfluid 1, for the parameters of Figs.\ \ref{figdisp} and
\ref{figpolar}. Region I: stable; region II: energetically unstable and containing a dynamically unstable region; region III: single-superfluid phase ${\rm SF}_2$ preferred.}
\label{figUpDown}
\end{center}
\end{figure}

The most relevant information of Figs.\ \ref{figdisp} and \ref{figpolar} is extracted in Fig.\ \ref{figUpDown}, which we now use to discuss the various instabilities. 
\begin{itemize}
\item {\it energetic instability}: the transition from region I to region II is defined by the point where the excitation energy of one of the 
Goldstone modes becomes negative. Such a point is well known in the theory of superfluidity. It exists even for a single superfluid and the corresponding critical velocity
is nothing but Landau's critical velocity. Compared to Landau's original argument, our calculation is a generalization to a two-component superfluid system and to the relativistic 
case. Since in our system the onset of negative excitation energies is equivalent to the sound speed becoming zero, we can also use Eq.\ (\ref{detu})
to compute the critical velocity. From this equation we easily read off that $u=0$ occurs for ${\rm det}\, \chi_0=0$  
which, in turn, occurs at the point where one of the eigenvalues of $\chi_0$ changes its sign; 
both eigenvalues are positive in region I, one eigenvalue is negative in region II. 
Consequently, in our approximation, where the quasiparticle modes and the sound modes coincide, 
Landau's critical velocity is a manifestation of the negativity of the ``current susceptibility'', 
the second derivative of the pressure with respect to the spatial components of the conjugate 
four-momentum $p^\mu=\partial^\mu\psi$. A more common susceptibility is the ``number susceptibility'', the 
second derivative with respect to the chemical potential, which is the {\it temporal} component of the 
conjugate four-momentum. In this case, the negativity implies that the density decreases
by increasing the corresponding chemical potential, which indicates an instability. This is completely analogous to the spatial components: here, a negative susceptibility 
implies that the three-current decreases by increasing the corresponding velocity, 
again indicating an unstable situation. In general, one has to check both stability criteria 
separately; Landau's critical velocity does not necessarily coincide with the onset of a negative current susceptibility. An example is a single superfluid at nonzero temperature, 
where neither of the two sound modes coincides with the quasiparticle mode and thus the connection between the quasiparticle energy and the 
susceptibility cannot be made.

In Fig.\ \ref{figVc} we show Landau's critical velocity $v_L$ for all values of the chemical potentials for which the COE phase is preferred in the absence of any fluid 
velocity. We see that 
for negative values of the entrainment coupling {\it all} states that where stable at $\vec{v}_1=0$ become energetically unstable at some critical velocity. This is different 
for positive values of the entrainment coupling where there is a region in the phase diagram with no instability, indicated by $v_L=1$. This case is also interesting because we find a region 
with vanishing critical velocity, i.e., an unstable COE state which appeared to be stable in the calculation based on the tree-level potential. 
This means that the COE phase may well be the global 
minimum of the potential $U$ within our ansatz of a uniform condensate, but it may be a saddle point if an anisotropic or inhomogeneous condensate is allowed for. 

\begin{figure}[t]
\begin{center}
\begin{minipage}{9.3cm}
\vspace{0.15cm}
\includegraphics[width=9.3cm]{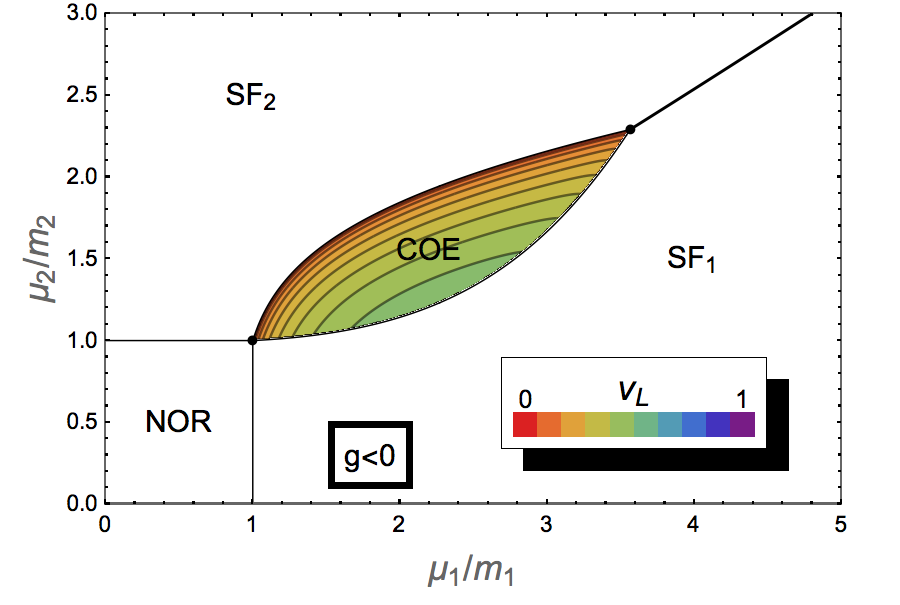}
\end{minipage}
\begin{minipage}{8.5cm}
\hspace{-0.8cm}\includegraphics[width=8.4cm]{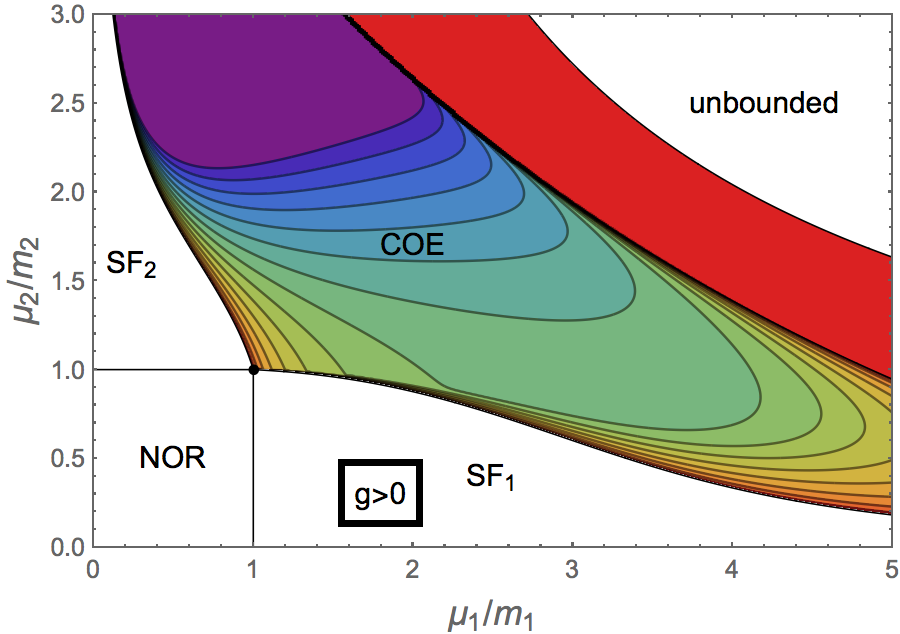}
\end{minipage}
\caption{Landau's critical velocity $v_L$ (``energetic instability'') in the COE phase of the two phase diagrams from Fig.\ \ref{figphases}, i.e., for negative (left panel) 
and positive (right panel) values of the entrainment coupling $g$. 
The asymmetry between the two superfluids arises because we work, without loss of generality, in the rest frame of superfluid 2, i.e., the critical velocity
refers to $\vec{v}_1$. 
In the right panel, the COE region is separated from the region where the tree-level potential is unbounded from below
by a band where $v_L=0$, i.e., states in this band are energetically unstable already in the static case although they appear stable within our uniform ansatz.
The critical velocity for the two-stream instability is, where it exists, larger than Landau's critical velocity throughout the phase diagram.  
The phases where only one field condenses, ${\rm SF}_1$ and ${\rm SF}_2$, also become unstable beyond certain critical velocities, but these are not shown here.}
\label{figVc}
\end{center}
\end{figure}

Besides the negativity of the susceptibility, we can describe the energetic instability also in the following intuitive way, using the picture of the sound waves: 
if a sound mode propagates ``upstream'', it is natural to expect that it is slowed down compared to the situation without superflow. 
Landau's critical velocity
is the point at which the mode is slowed down so much that it comes to rest, and for larger velocities it appears to propagate in the ``wrong'' direction. 
It then shows up as an additional 
downstream mode, which is manifest in all Figs.\ \ref{figdisp} -- \ref{figUpDown}, for instance in 
the left panel of Fig.\ \ref{figUpDown} where
at the transition between regions I and II a third mode appears in the downstream direction.  
Rephrased in this way, the energetic instability encountered here is analogous to the 
Chandrasekhar-Friedman-Schutz (CFS) instability \cite{PhysRevLett.24.611,1978ApJ...221..937F} (including the $r$-mode instability \cite{Andersson:1997xt}) known from 
astrophysics\footnote{In the astrophysical literature such an instability is called {\it secular} instability. 
Here we use the term {\it energetic} instability synonymously, following the condensed matter literature, see for instance Ref.\ \cite{PhysRevA.87.063610}.}: 
in that case, certain oscillatory modes of a neutron star can be ``dragged'' by the rotation of the star, such that from a distant observer they appear to propagate, say, 
counter-clockwise, 
while in the co-rotating frame they propagate clockwise. This is exactly the same kind of behavior as we observe here, we just have to replace the oscillatory mode of the 
neutron star by the sound mode and the angular velocity of the star by the velocity of the superfluid. It is very instructive to develop this analogy a bit further: in a rotating neutron 
star, the instability is realized due to the emission of gravitational waves, which are able to transfer angular momentum from the system. In the same way, as already pointed out
by Landau in his original argument, the negative excitation energies lead to an instability if there is a mechanism that can exchange momentum with the superfluid, 
for example the presence of the walls of a capillary in which the superfluid flows. The exchange of angular momentum or momentum is crucial for the instability to set in,
and knowledge about how this exchange works is required to determine the 
time scale on which the instability operates. This is in contrast to the {\it dynamical} instability, where a time scale is inherent and which we discuss next.

\item {\it dynamical instability}: in a sub-region of region II, two of the sound speeds acquire an imaginary part of opposite sign and equal magnitude, while their real parts 
coincide (since the polynomial from which the sound speeds are computed has real coefficients, for any given solution also the complex conjugate is a solution). This means that the 
amplitude of one of the modes is damped, while it increases exponentially for the other one, with a time scale given by the magnitude of the imaginary part. 
This kind of instability is well-known in two-fluid or multi-fluid plasmas \cite{Buneman:1959zz,1963PhRvL..10..279F,2001AmJPh..69.1262A}, 
and is termed two-stream instability or counterflow instability. 
In plasma physics, usually the fluids are electrically charged, and the calculation of the sound modes is somewhat different because the hydrodynamic equations
become coupled to Maxwell's equations. This provides an ``indirect'' coupling between the two fluids, while we have coupled the two fluids ``directly'' through a coupling 
term in the Lagrangian. In the context of two-component 
superfluids the two-stream instability has been discussed in a non-relativistic context \cite{2004MNRAS.354..101A,PhysRevA.87.063610,2015EPJD...69..126A} and, without reference to superfluidity, 
in a relativistic context \cite{Samuelsson:2009up}. It has also been discussed in a single, relativistic superfluid at nonzero temperature \cite{Schmitt:2013nva}. 
For a truly dynamical discussion of the two-stream instability one has to go beyond linearized hydrodynamics \cite{Hawke:2013haa} and possibly take into account the formation of
vortex rings and turbulence \cite{2011PhRvA..83f3602I}.  
Our present calculation only yields the time scale of the exponential growth at the onset of the instability.

\item {\it phase transition to single-superfluid phase}:  
as the phase diagram in the left panel of Fig.\ \ref{figphases} shows, a point within the COE phase with fixed chemical potentials 
will simply leave the COE region beyond a critical velocity. Therefore, in region III a different phase, even within our simple uniform ansatz,
is preferred, in this case the phase ${\rm SF}_2$, where only field 2 forms a condensate. This instability is also seen in the sound modes because the COE 
phase ceases to be a local minimum at that point. In other words, the phase transition is of second order.  

\end{itemize}

\subsection{Further analysis of instabilities and comparison to normal fluids} 
\label{sec:analysis}

In the results shown so far, the dynamical instability occurs in an energetically unstable region, i.e., a complex sound speed only occurs if there is already a negative excitation 
energy. In other words, the two branches that merge in the left panel of Fig.\ \ref{figUpDown} are not the two ``original'' downstream modes, but one downstream mode and the original 
upstream mode that has changed its direction. Two questions arise immediately: 

\begin{enumerate}

\item[$(i)$] Does the occurrence of complex sound speeds have any physical meaning if they occur in an 
energetically unstable region? 

\item[$(ii)$] Is this behavior generic or can a dynamical instability arise even in the absence of an energetic instability? 

\end{enumerate}

As for point $(i)$, we will give a brief qualitative discussion, while for point $(ii)$ we will give a definite answer within our approach.

$(i)$ The problem that seems to arise is that an energetically unstable system will choose a different configuration, either another equilibrium state with lower free energy than the
one that exhibits the negative excitation energies, or it will refuse to be in equilibrium altogether. In either case, the two-stream instability we have observed may well be absent 
in the new configuration, simply because the calculation in the energetically unstable state is not valid since this state is not realized. 
As mentioned above, the energetic instability
is, in our approximation, identical to a negative current susceptibility. 
Negative number susceptibilities are well-known indicators of an instability, for instance for Cooper-paired systems  
with mismatched Fermi surfaces \cite{Gubankova:2006gj,Deng:2006ed,Huang:2006kr}, where the resolution may be phase separation, i.e., a spatially inhomogeneous 
state with paired regions separated from unpaired regions. In our present calculation, the negative susceptibility may as well be cured by an inhomogeneous state which we 
have not included into our ansatz, for example in the form of stratification of the two superfluid components \cite{2004LaPhL...1...50Y,2009JPhCS.150c2057M} or  
a crystalline structure of the condensates \cite{Landea:2014naa}. While these solutions to the problem concern equilibrium states, the fate of an energetically unstable state 
in a real physical system, be it in a neutron star or in ultra-cold atoms, may be more complicated. As mentioned above, the realization of the energetic (or secular, in the 
astrophysical terminology) instability depends on a mechanism that is able to transfer momentum to and from the system. If such a mechanism is absent or operates on a large
time scale it is thus conceivable (depending on the actual physical situation) 
that the two-component superfluid becomes unstable only at the larger critical velocity where the 
two-stream instability sets in. 

\begin{figure} [t]
\begin{center}
\hbox{\includegraphics[width=0.355\textwidth]{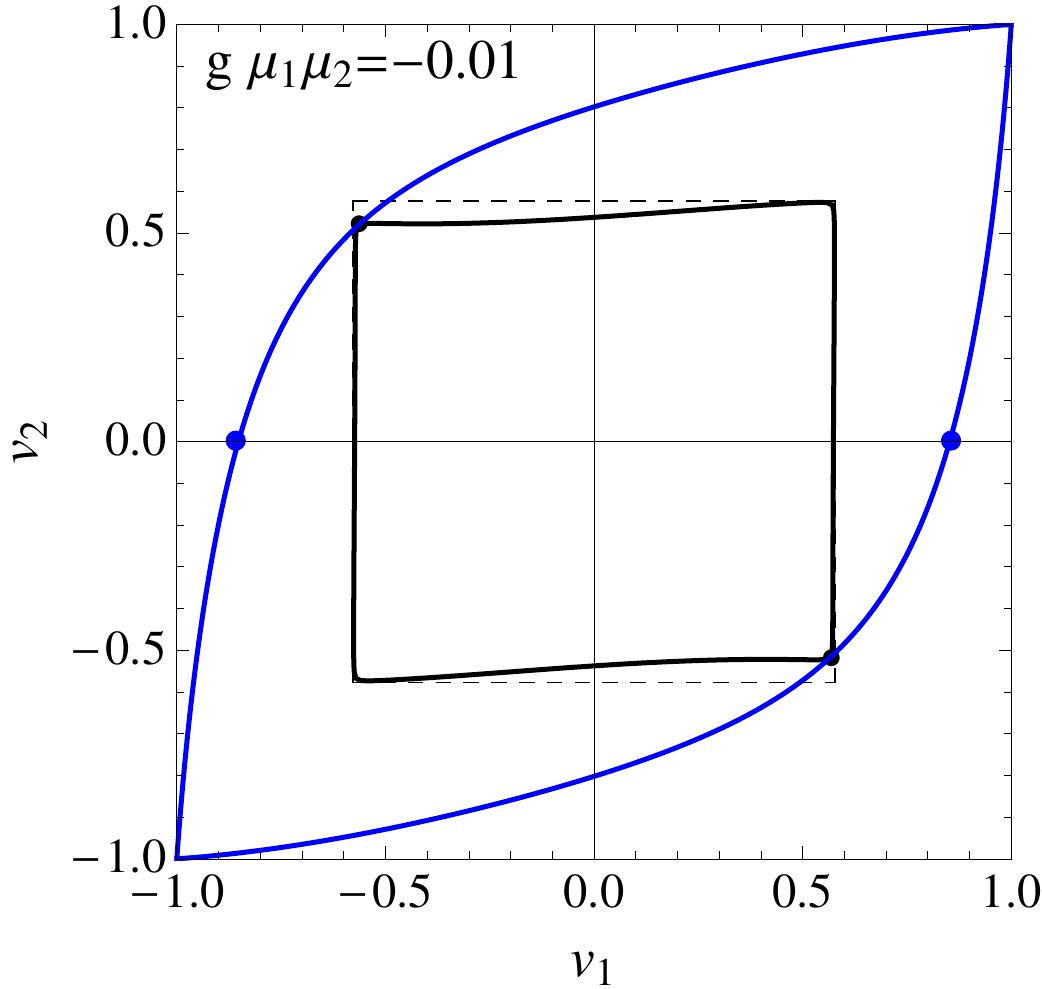}\includegraphics[width=0.31\textwidth]{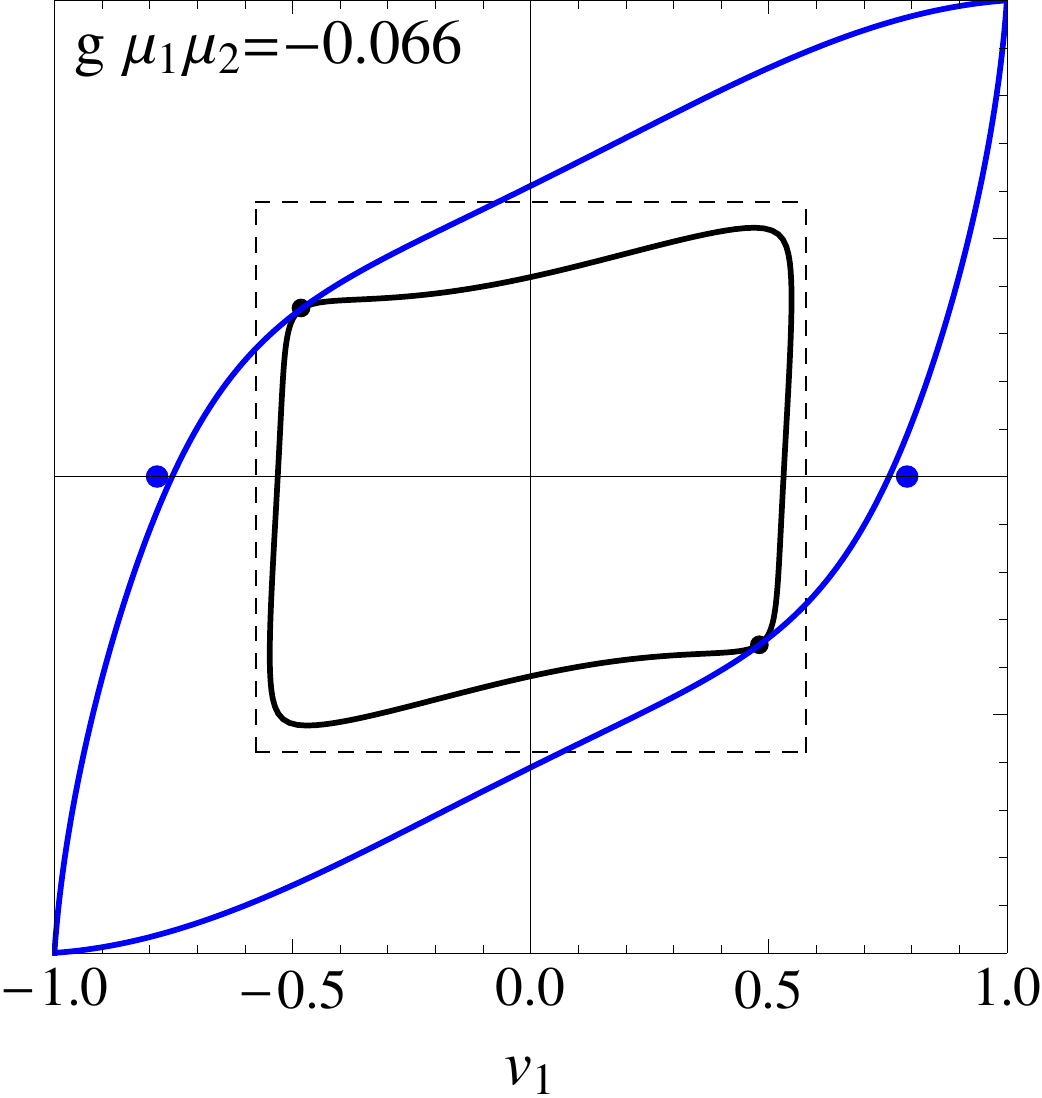}
\includegraphics[width=0.31\textwidth]{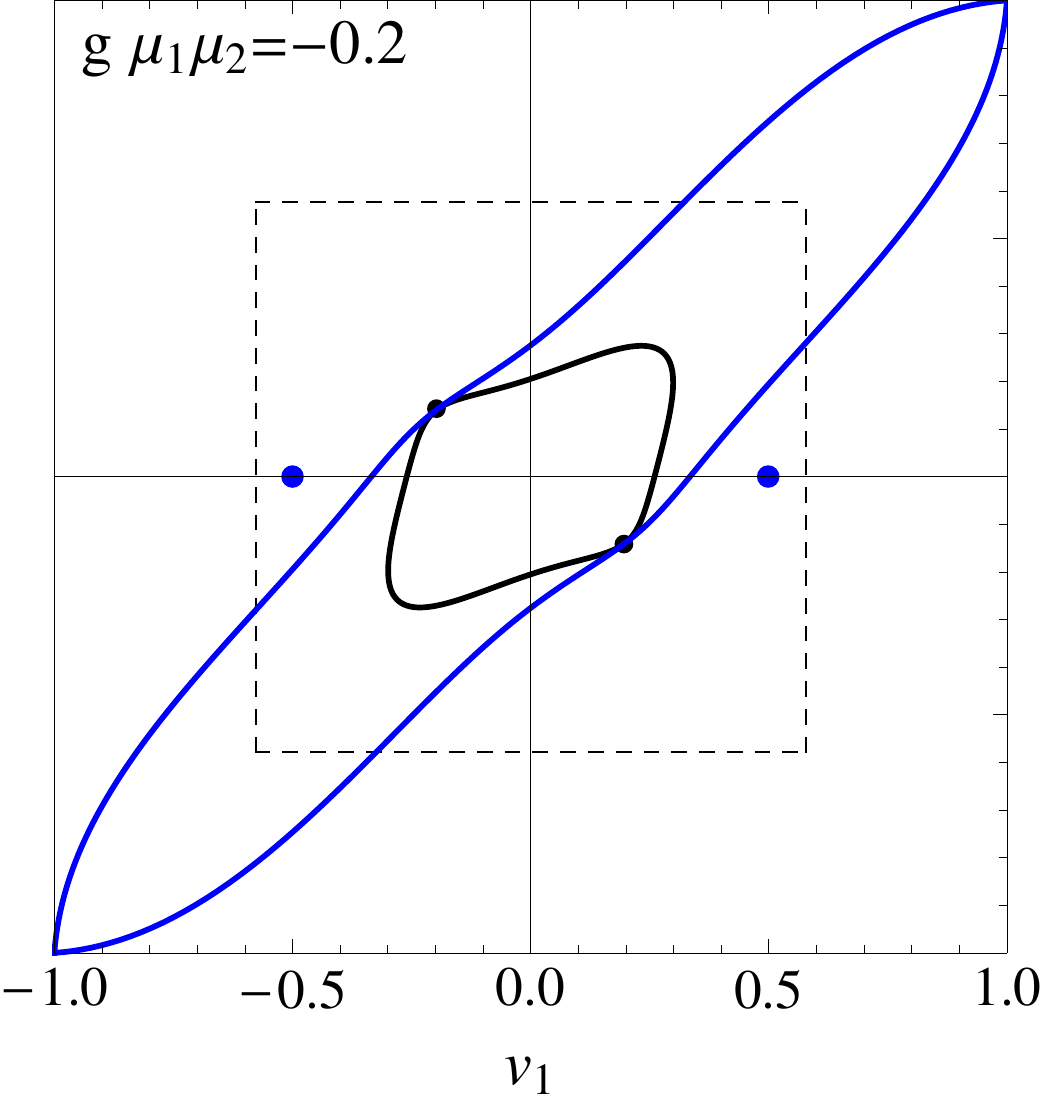}}
\caption{Landau's critical velocity $v_L$ [inner (black) solid curve] and critical velocity for the two-stream instability $v_{\rm two-stream}$ 
[outer (blue) solid curve] for two superfluids moving with velocities $v_1$ and $v_2$
parallel [${\rm sgn}(v_1v_2)>0$] or anti-parallel [${\rm sgn}(v_1v_2)<0$] to each other for negative entrainment coupling $g$ of three different strengths, and $m_1=m_2=0$, 
$\lambda_1=0.3$, $\lambda_2=0.2$. (Angles in this plot correspond to different ratios $v_1/v_2$, not to angles between $\vec{v}_1$ and $\vec{v}_2$, which are always aligned or
anti-aligned.)  The dashed square shows Landau's critical velocity $1/\sqrt{3}$ in the absence of a coupling between the fluids. The large (blue) dots on the horizontal axis mark the 
analytical result for small coupling strength for the onset of the two-stream instability (\ref{v2stream}), while the small (black) dots mark the points where both 
critical velocities coincide.} 
\label{figv1v2}
\end{center}
\end{figure}

$(ii)$
First of all we recall that the negative excitation energies and Landau's critical velocity are frame dependent. So far we have worked in the rest frame of one of the 
two superfluids. But one can well imagine that there is a third meaningful reference frame, for instance the rest frame of the entropy fluid if we consider 
nonzero temperatures, which is the most convenient rest frame in a field-theoretical calculation \cite{2013PhRvD..87f5001A}. 
Therefore, for a general analysis of Landau's critical velocity, we should allow both fluids to move with nonzero velocities $\vec{v}_1$ and $\vec{v}_2$. 
(In Landau's original argument for a single superfluid a second reference frame is given by the capillary.) Of course, the 
calculation becomes very tedious if we allow for three different vectors with arbitrary directions: $\vec{v}_1$, $\vec{v}_2$, and the wave vector $\vec{k}$. Therefore,
we restrict ourselves to the case where all these vectors are aligned with each other. As we have seen in Fig.\ \ref{figpolar}, it is the downstream direction where energetic and
dynamical instabilities set in ``first'' (i.e., for the smallest velocities). Therefore, it {\it is} a restriction to consider only aligned $\vec{v}_1$, $\vec{v}_2$, 
but once they are aligned it is no further restriction to align $\vec{k}$ too if we are interested in the critical velocities.
 
In Fig.\ \ref{figv1v2} we plot Landau's critical velocity $v_L$ and the critical velocity for the two-stream instability $v_{\rm two-stream}$
for arbitrary values of $v_1$ and $v_2$, and three different values of the entrainment coupling. If the fluids were uncoupled, Landau's critical velocity for each of the fluids
is $1/\sqrt{3}$ in the ultra-relativistic limit (for simplicity, we have set the mass parameters to zero in this plot), 
irrespective of the velocity of the other fluid. This is indicated by the dashed square. A nonzero 
entrainment coupling reduces Landau's critical velocity, and it tends to do more so if the fluids move in opposite directions, where ${\rm sgn}\,(v_1v_2)<0$. (This is 
different if we choose the opposite sign for the coupling constant, $g>0$: in this case Landau's critical velocity is {\it enhanced} for anti-aligned flow.) This is a similar effect
as in the phase diagrams of Fig.\ \ref{figphases}: an entrainment coupling $g<0$ disfavors the COE phase, it is stable only in a smaller region compared to the uncoupled case
in the $\mu_1$-$\mu_2$ plane 
(Fig.\ \ref{figphases}) or in the $v_1$-$v_2$ plane (Fig.\ \ref{figv1v2}). The behavior of $v_{\rm two-stream}$,  the outer instability curve, 
is easy to understand in the following way: one might expect the two-stream instability to depend only on the relative velocity of the two superfluids. So, if relativistic effects were 
neglected, one might expect two straight lines given by $v_2-v_1=\pm {\rm const}$. The actual curves are different for two reasons: first, relativistic effects bend the 
curves in the $v_1$-$v_2$ plane according to the relativistic addition of velocities and, second, the velocities also have a nontrivial effect on the condensates of the superfluids,
as already discussed in the context of the phase diagram in the $\mu_1$-$\mu_2$ plane. 

For $v_2=0$ and small values of the entrainment coupling, one can find a simple analytical expression for the critical velocity,
\be \label{v2stream}
v_{\rm two-stream} = \frac{\sqrt{3}}{2}\left[1+\frac{\lambda_2\mu_1^4+16\lambda_1\mu_2^4}{32\lambda_1\lambda_2\mu_1\mu_2}\,g + {\cal O}(g^2)\right] \, .
\ee
This result is obtained by setting the discriminant for the polynomial in the sound velocity $u$ of degree 4 to zero \cite{1922}. It shows that for $g\to 0$ the critical 
velocity is  the relativistic sum of the two Landau critical velocities of each fluid $1/\sqrt{3}$. The entrainment coupling reduces or enhances this result, depending on the sign of the coupling constant. 
We have indicated the value (\ref{v2stream}) in Fig.\ \ref{figv1v2} as (blue) dots on the horizontal axis and see that the approximation becomes worse for larger couplings.  
In interpreting Eq.\ (\ref{v2stream}) we have to keep in mind that we have determined the critical velocity at fixed 
$\mu_1$, $\mu_2$, i.e., we have fixed the chemical potentials in the "lab frame", not in the respective rest frames of the fluids. Therefore, increasing $v_1$ while keeping $v_2=0$ affects the condensate 1 and not the 
condensate 2 (in addition to making the two fluids move with respect to each other). This asymmetry is manifest in the term proportional to $g$ in Eq.\ (\ref{v2stream}), and it becomes manifest even in the limit 
$g\to 0$ if we include nonzero masses $m_1$, $m_2$. We have checked that if we work with fixed chemical potentials in the rest frames of the superfluids $p_1$, $p_2$, the critical velocity $v_{\rm two-stream}$ does become
symmetric in the two fluids, and the $g\to 0$ limit is the relativistic sum of the two Landau critical velocities for all masses $m_1$, $m_2$. This includes the non-relativistic limit, in agreement with Ref.\ \cite{2015EPJD...69..126A}. 

The main conclusion from Fig.\ \ref{figv1v2} regarding the above question $(ii)$ is obvious: $v_L\le v_{\rm two-stream}$ for all ratios $v_2/v_1$,  
and there exists one ratio $v_2/v_1$ where $v_L = v_{\rm two-stream}$. 
In other words, the scenario shown in Fig.\ \ref{figUpDown} {\it is} generic, we have not found any region 
in the parameter space where the two-stream instability sets in at smaller velocities than the energetic instability. A general, rigorous proof of this statement is difficult
because the sound modes are solutions of quartic equations. Thus, strictly speaking, we have not rigorously proven this
statement, but, besides the results shown in Fig.\ \ref{figv1v2} we have checked many other parameter sets, including a different sign of the entrainment coupling and including
a non-entrainment coupling. The situation we are asking for is a merger of the two original downstream modes, i.e., the two curves in region I in the left panel of 
Fig.\ \ref{figUpDown}. We have in particular looked at parameter sets where the curves of these modes cross in the absence of any coupling. But, if a coupling is switched on, 
no imaginary part
but rather an ``avoided crossing'' develops at this point. Therefore, in all cases we have considered, the qualitative conclusion about the order of the two critical velocities remains. 

At this point, let us come back to the modes that we have discussed in Sec.\ \ref{sec:hydro}. We pointed out that a two-fluid system allows for a richer spectrum of massless modes if 
one or both of the fluids are normal fluids rather than superfluids. In a superfluid, the oscillations are constrained because chemical potential and superfluid velocity are both 
related to the phase of the condensate. As a consequence, only longitudinal oscillations are allowed. 
If exactly one of the fluids is a normal fluid, one additional transverse mode appears, given in Eq.\ (\ref{vnk1}). This mode has a fixed form, for all possible couplings between the 
fluids, and thus does not couple to the other, longitudinal modes. Therefore, the discussion of energetic and dynamical instabilities reduces to exactly the same modes as discussed
in this section for the two-component superfluid. Of course, for a specific discussion
we need the generalized pressure 
as a function of the Lorentz scalars $p_1^2$, $p_1^2$, $p_{12}^2$, of which our microscopic theory only provides one specific example. An example for a system of one normal fluid 
and one superfluid is a single superfluid at nonzero temperature, for which the generalized pressure has been derived from a microscopic model  
in the low-temperature approximation, keeping terms of order $T^4$ \cite{Carter:1995if,2013PhRvD..87f5001A,Schmitt:2014eka}, 
see for instance Eq.\ (4.45) in Ref.\ \cite{Schmitt:2014eka}. In its regime of validity, i.e., low temperatures compared to the 
chemical potential and also small superfluid velocities, there is no dynamical instability, at least not in an energetically stable 
regime \cite{2013PhRvD..87f5001A}. At arbitrary temperatures below the 
critical temperature a covariant form of the generalized pressure, based on a microscopic model, is unknown to the best of our knowledge, 
and one has to apply more complicated methods. Within the self-consistent 2-particle-irreducible 
formalism, a two-stream instability was indeed found at nonzero temperatures, 
remarkably in an energetically {\it stable} regime, for velocities slightly below Landau's critical velocity \cite{Schmitt:2013nva}.

\begin{figure} [t]
\begin{center}
\hbox{\includegraphics[width=0.5\textwidth]{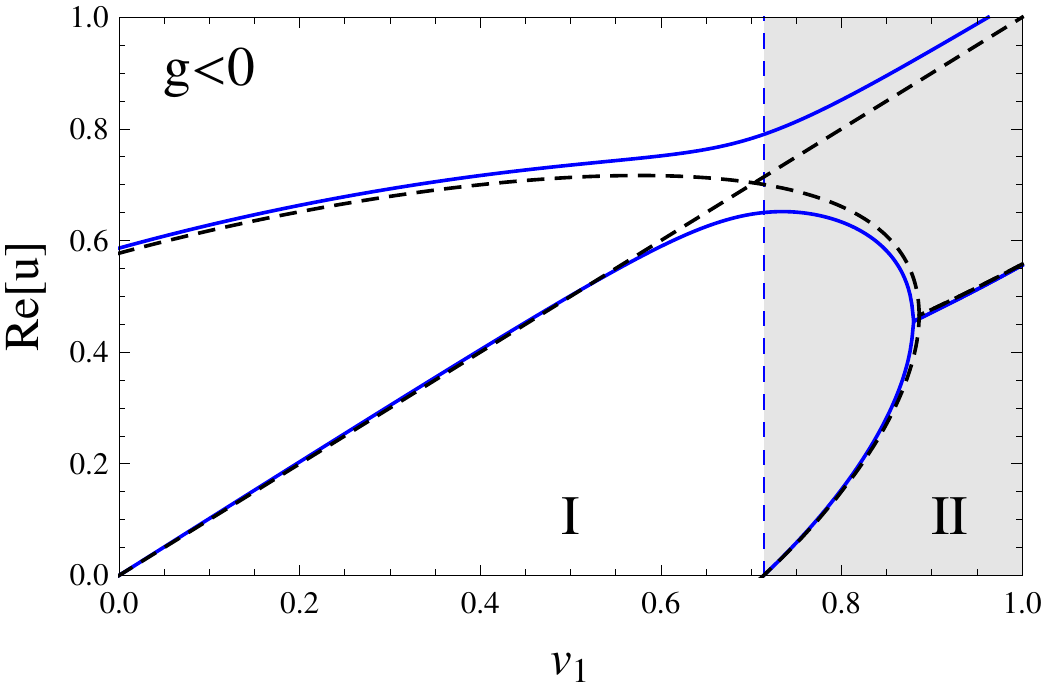}\includegraphics[width=0.5\textwidth]{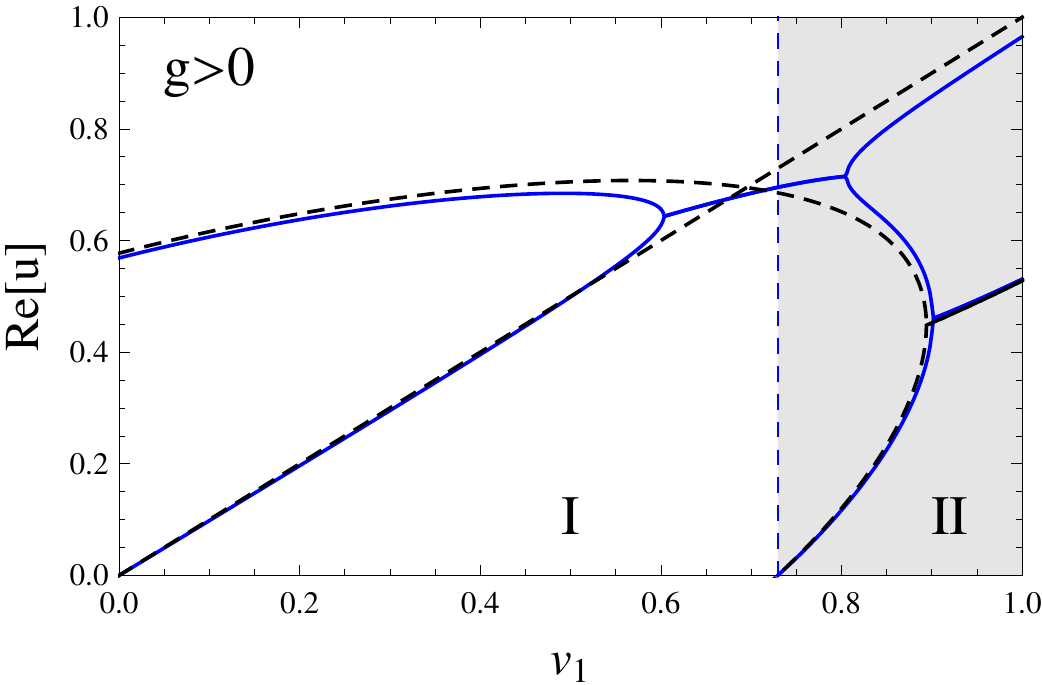}}
\caption{Real part of the sound speeds for transverse modes in the downstream direction, in a system of two {\it normal} fluids with entrainment [solid (blue) lines]. 
Regions I and II are energetically stable and unstable regions, respectively. For a positive 
entrainment coupling, there is a two-stream instability in an energetically stable regime,  
which does not occur in the system of two coupled superfluids, where the two-stream
instability is always in region II. Here
we have used the equation of state of the superfluid system and chose the ultrarelativistic limit, $m_1=m_2=0$, the ratio of chemical potentials $\mu_2/\mu_1=1.5$, 
the self-couplings $\lambda_1=0.3$, $\lambda_2=0.2$, the non-entrainment coupling $h=0.05$, and the entrainment coupling $g\mu_1\mu_2=-0.003$ (left panel) and 
$g\mu_1\mu_2=+0.003$ (right panel). The dashed (black) lines are the results with the same parameters, but in the absence of entrainment, $g=0$.}
\label{fignn}
\end{center}
\end{figure}

For the case of two normal fluids, we discuss two intriguing 
observations. Firstly, in the absence of entrainment, but
in the presence of a non-entrainment coupling, we have found sound modes whose speed has a very simple analytical form, see Eqs.\ (\ref{dmu20}) and (\ref{dmu10}). Just like
the two-component superfluid, they show an energetic instability and a dynamical instability, with corresponding critical velocities given 
by Eqs.\ (\ref{dmu20vcr}) and (\ref{dmu10vcr}). As these equations make obvious, the critical velocity for the energetic instability is always smaller than 
that of the dynamical instability. We have thus found an example for a set of (transverse) 
modes where there is a very simple proof that for all values of the coupling constant and for 
all equations of state for each single fluid the dynamical instability can only occur if the system is already energetically unstable. 
Secondly, let us discuss the modes (\ref{dmu20}) and (\ref{dmu10}) in the 
presence of entrainment. In this case, their analytical form is extremely 
complicated, so we again resort to a numerical evaluation. To this end, we need to specify an equation of state, and we simply use the pressure from our 
microscopic model of the complex fields, i.e., $P=-U_{\rm COE}$ with $U_{\rm COE}$ from Eq.\ (\ref{COE}). This is one particular example where both 
fluids enter symmetrically, and we simply reinterpret the pressure in the sense that the conjugate four-momentum is not related to any phase, thus allowing for 
non-longitudinal oscillations. The results for a certain choice of parameters
is shown in Fig.\ \ref{fignn}. The dashed (black) lines show the result in the absence of entrainment: these are the modes from Eq.\ (\ref{dmu20}) just 
discussed, with one mode becoming negative (in the upstream direction, not shown in the figure) and, at a larger velocity, two modes that merge and acquire complex 
values.
This is very similar to the superfluid case in Fig.\ \ref{figUpDown}.
But, additionally, there is the transverse mode $u=v_1\cos\theta$ which, in the absence of entrainment, is unaffected by the other modes. In the presence of 
entrainment, the modes couple. Mathematically speaking, the polynomial does not factorize in this case.  
In particular, for a certain sign of the entrainment coupling constant, in our convention $g>0$, there is a two-stream instability 
due to the presence of this mode\footnote{We may also consider the generalization of the modes (\ref{dmu10}) in the presence of entrainment, i.e., the modes that are continuously 
connected to the uncoupled oscillations of the fluid at rest. In this case, it is the zero mode $u=0$ instead of the mode $u=v_1\cos\theta$ which, in the presence of entrainment, 
couples to the modes (\ref{dmu10}), also resulting in a dynamical instability.}. 
This is remarkable because of two reasons: it is an example where entrainment, as opposed to a non-entrainment coupling, 
 leads to a qualitative difference regarding the dynamical
instability. (In the case of the two-superfluid system, although we have only shown results with entrainment coupling, the qualitative conclusions would not have changed had we 
worked with a non-entrainment coupling.) And, it is an example for a two-stream instability occurring in an energetically stable regime: none of the modes has negative energy at the 
point where two of the modes become complex.

\section{Summary and Outlook}
\label{sec:summary}

We have investigated quasiparticle excitations in a relativistic two-component superfluid, starting from a bosonic field-theoretical model for two complex
scalar fields and  including an inter-species derivative coupling that gives rise to entrainment between the two fluids. We have focused on the simplest hydrodynamic
situation where the fluid velocities are uniform in space and time and have restricted ourselves to zero temperature. 
In this scenario, we have calculated the phase structure of the system in an ``extended'' 
grand canonical ensemble, where besides the two chemical potentials also the two fluid velocities serve as externally given parameters. This is very natural because 
both quantities are different components of the same four-momentum conjugate to the conserved current. By including fluctuations about the condensates, 
we have computed the dispersion relations
of the two Goldstone modes in the phase where both complex fields condense. Besides this microscopic approach, we have also presented a general derivation of the sound modes 
from two-fluid relativistic hydrodynamics in the linear approximation. While this approach reproduces the Goldstone dispersion relations of the two-component superfluid at low energy, 
it also allows for a study of the sound modes in two-component {\it normal} fluids. 

Our main focus has been on the instabilities that arise at nonzero fluid velocities. We have systematically analyzed energetic and dynamical instabilities.
The energetic instability manifests itself in negative excitation energies beyond a certain critical velocity. This critical velocity - or rather a critical surface in the space of both chemical potentials and fluid velocities
-- is a generalization of Landau's critical velocity for a single superfluid. With the help of the hydrodynamic equations, we have pointed out that this 
instability is, within our zero-temperature approximation, 
equivalent to a negative eigenvalue of the ``current susceptibility'' matrix, the matrix of second derivatives of the pressure with 
respect to the fluid velocities (analogous to the number susceptibilities which are second derivatives with respect to the chemical potentials). In a certain parameter 
range this instability even occurs in the limit of vanishing fluid velocities. Besides the energetic instability we have also 
computed the critical velocity for the two-stream instability. This dynamical instability manifests itself in a complex sound velocity, whose imaginary part determines the time scale 
on which the unstable mode grows. 

As a result of this analysis, we have found that, in the case of a two-component superfluid  
at zero temperature, the dynamical instability can only occur in the presence of an energetic
instability. In other words, one mode must acquire negative energy before it can couple to another mode to develop an exponentially growing amplitude. 
Our numerical evaluation suggests that this is a general result, holding throughout the parameter space and for both kinds of couplings we have considered, 
non-entrainment and entrainment couplings. We have complemented this
result by an analysis of two normal fluids, which allow for non-longitudinal oscillations. This is in contrast to a superfluid, where a constraint on the oscillations is given
by the relation of both chemical potential and superfluid velocity to the phase of the condensate. Most importantly, we
have found a set of transverse modes in a two-component normal fluid that couple to each other if entrainment is present and that show a two-stream instability in the {\it absence} of 
an energetic instability.

Our study raises several questions and can be extended in various directions. A direct extension would be to include temperature effects. This seems straightforward but 
will complicate the calculation significantly since it introduces a third fluid given by the entropy current. It would nevertheless be promising to study 
dynamical instabilities in this case since for instance in 
superfluid neutron star matter temperature effects can be expected to be relevant. Another obvious extension is to include electromagnetism, i.e., to promote one or both 
of the global symmetry groups of our Lagrangian to a local one. This would firstly make the connection to ordinary plasma physics where the fluid components are usually electrically 
charged. It would secondly also be relevant for the interior of neutron stars, where a system of neutrons and protons can form a two-component system of one superfluid
and one superconductor. It would be interesting to see whether this changes any of the conclusions regarding the dynamical instability, especially in view of its possible relevance
for pulsar glitches. Furthermore, one should ask whether the energetic instability can be resolved by the formation of an inhomogeneous superfluid, possibly 
through a phase separation of the two components. And, subsequently, it would be interesting to find out whether such an inhomogeneous state, if energetically stable, 
will still suffer a dynamical instability. This can be relevant not only for superfluids in the astrophysical context, but also for superfluids in the laboratory 
such as Bose-Fermi mixtures in cold atomic gases.

\begin{acknowledgments}
We would like to thank Nils Andersson for valuable comments and discussions. We acknowledge support from the Austrian Science 
Fund (FWF) under project no.\ W1252, and from the {\mbox NewCompStar} network, COST Action MP1304. A.S.\ is also supported by the FWF under project no.\ P26328
and the Science \& Technology Facilities Council (STFC) in the form of an Ernest Rutherford Fellowship, and S.S.\ is supported by the Schr\"{o}dinger Fellowship of the FWF, 
project no.\ J3639. 
\end{acknowledgments}

\appendix

\section{Chemical potentials in the presence of an entrainment coupling}
\label{app0}

In this appendix we prove that the chemical potentials $\mu_1$, $\mu_2$ are introduced in the Lagrangian by replacing all derivatives -- in the kinetic term {\it and} the
derivative coupling --  by covariant derivatives which contain the chemical potentials in the form of a temporal background gauge field. The main idea of the proof is to introduce 
a compact matrix notation in which the derivation then proceeds very similarly to the standard scenario without derivative coupling. As an aside, we shall point out a complication 
arising from the functional integration over the canonical momenta, which produces a nontrivial field-dependent factor in the presence of a derivative coupling.

We first introduce real and imaginary parts of the complex fields,
\be \label{ReIm}
\varphi_i = \frac{1}{\sqrt{2}}(\varphi'_i+i\varphi''_i) \, .
\ee
[Here we do not have to separate the condensates and the fluctuations; note that the real and imaginary parts of the fluctuations introduced in Eq.\ (\ref{fluc}) are
transformed by the phase of the condensate, and thus denoted by a different symbol.] 
In this basis, the Lagrangian becomes
\bea
{\cal L} &=& \sum_{i=1,2}\left[\frac{1}{2}\partial_\mu\varphi'_i\partial^\mu\varphi'_i+\frac{1}{2}\partial_\mu\varphi''_i\partial^\mu\varphi''_i
-\frac{m_i^2}{2}(\varphi_i'^2+\varphi_i''^2)-\frac{\lambda_i}{4}(\varphi_i'^2+\varphi_i''^2)^2\right] + \frac{h}{2}(\varphi_1'^2+\varphi_1''^2)(\varphi_2'^2+\varphi_2''^2) \non[2ex]
&&-\frac{g_1}{4}\left[(\varphi_1'\varphi_2'-\varphi_1''\varphi_2'')(\partial_\mu\varphi_1'\partial^\mu\varphi_2'-
\partial_\mu\varphi_1''\partial^\mu\varphi_2'')+(\varphi_1'\varphi_2''+\varphi_1''\varphi_2')(\partial_\mu\varphi_1'\partial^\mu\varphi_2''+
\partial_\mu\varphi_1''\partial^\mu\varphi_2')\right] \non[2ex]
&&-\frac{g_2}{4}\left[(\varphi_1'\varphi_2'+\varphi_1''\varphi_2'')(\partial_\mu\varphi_1'\partial^\mu\varphi_2'+
\partial_\mu\varphi_1''\partial^\mu\varphi_2'')+(\varphi_1'\varphi_2''-\varphi_1''\varphi_2')(\partial_\mu\varphi_1'\partial^\mu\varphi_2''-
\partial_\mu\varphi_1''\partial^\mu\varphi_2')\right] \, .
\eea
By introducing the vector $\vec{\varphi}$ through 
\be
\vec{\varphi} = \left(\begin{array}{c} \vec{\varphi}_1 \\[1ex] \vec{\varphi}_2 \end{array}\right) \, , \qquad 
\vec{\varphi}_i = \left(\begin{array}{c} \varphi_i' \\[1ex] \varphi_i'' \end{array}\right) \, , 
\ee
we can write the Lagrangian in the compact form
\be \label{L1}
{\cal L} = \frac{1}{2}\Big[\partial_\mu\vec{\varphi}^T Q^{-1} \partial^\mu\vec{\varphi} - \vec{\varphi}^T(m^2+Y)\vec{\varphi}\Big]
\, ,
\ee
with $m^2 = {\rm diag}(m_1^2,m_2^2)$. Here we have abbreviated 
\be 
Q^{-1} \equiv \left(\begin{array}{cc} 1 & A \\[1ex] A^T & 1 \end{array}\right) \, , \qquad 
Y \equiv \frac{1}{2} \left(\begin{array}{cc} \lambda_1 Y_{11} & -hY_{12} \\[1ex] -hY_{21} & \lambda_2 Y_{22} \end{array}\right) \, ,
\ee
where 
\be
A \equiv -\frac{1}{2}(GY_{12} +g\tau_2Y_{12}\tau_2) \, , \qquad \tau_2=\left(\begin{array}{cc} 0 & -1 \\ 1 & 0 \end{array}\right) \,  , \qquad 
Y_{ij} \equiv \left(\begin{array}{cc} \varphi_i'\varphi_j' & \varphi_i'\varphi_j'' \\[1ex] \varphi_i''\varphi_j' & \varphi_i''\varphi_j'' \end{array}\right) \, .
\ee
The compact form (\ref{L1}) seems to suggest that the Lagrangian is quadratic in the fields. This is of course not true, one has to keep in mind 
that the matrices $Q^{-1}$ and $Y$ contain terms quadratic in the fields too. The point was to write the Lagrangian in a form that shows the derivative terms explicitly, and absorb 
all remaining structure in the most compact way. This facilitates the introduction of the conjugate momenta and the chemical potentials, and also makes the structure of the 
Lagrangian very transparent: $Y$ contains the non-derivative 
self-couplings (diagonal terms, proportional to $\lambda_i$), and the non-derivative cross-coupling (off-diagonal terms, proportional to 
$h$), while $Q^{-1}$ contains the derivative cross-couplings (off-diagonal terms, proportional to $g_i$) and the kinetic terms (diagonal). We do not include derivative self-couplings, which 
would occur diagonally in $Q^{-1}$. 

The canonical momenta conjugate to the fields are defined as 
\bea
\pi_i'&=&\frac{\partial {\cal L}}{\partial(\partial_0\varphi_i')}  \, , \qquad  \pi_i''=\frac{\partial {\cal L}}{\partial(\partial_0\varphi_i'')}  \, , 
\eea
which can be compactly written as 
\be\label{pidphi}
\vec{\pi} = Q^{-1} \,\partial^0\vec{\varphi} \, , 
\ee
with the vector $\vec{\pi}$ defined analogously to $\vec{\varphi}$. Below we shall need the inverse relation $\partial^0\vec{\varphi} = Q\vec{\pi}$ with 
\be \label{Qinv}
Q = \left(\begin{array}{cc} (1-AA^T)^{-1} & -A(1-A^TA)^{-1} \\[1ex] -A^T(1-AA^T)^{-1} & (1-A^TA)^{-1} \end{array}\right) \, .
\ee
The Hamiltonian is given by the Legendre transform of ${\cal L}$ with respect to the pair of variables $(\partial^0\vec{\varphi}, \vec{\pi}$),
\bea
{\cal H} &=& \vec{\pi}^T\partial^0\vec{\varphi}-{\cal L} = \frac{1}{2}\Big[\vec{\pi}^T Q\vec{\pi}+\nabla\vec{\varphi}^T\cdot Q^{-1}\nabla\vec{\varphi}
+\vec{\varphi}^T(m^2+Y)\vec{\varphi}\Big]  \, . 
\eea
Introducing two chemical potentials for the two conserved charges amounts to a shifted Hamiltonian ${\cal H}-\mu_1{\cal N}_1-\mu_2{\cal N}_2$ with the two charge densities 
given by the temporal components of the currents, ${\cal N}_i = j_i^0$. The currents (\ref{j1j2a}) can be written as 
\be
j_1^\mu = -\vec{\varphi}_1^T\tau_2\left(\partial^\mu\vec{\varphi}_1-\frac{g}{2}\tau_2Y_{12}\tau_2\partial^\mu\vec{\varphi}_2\right) \, , \qquad
j_2^\mu = -\vec{\varphi}_2^T\tau_2\left(\partial^\mu\vec{\varphi}_2-\frac{g}{2}\tau_2Y_{21}\tau_2\partial^\mu\vec{\varphi}_1\right) \, . 
\ee
Now, using the temporal component of these expressions and inserting $\partial^0\vec{\varphi} = Q\vec{\pi}$ with $Q$ from Eq.\ (\ref{Qinv}), one computes
the remarkably simple relations
\begin{subequations}
\bea
j_1^0 &=& -\vec{\varphi}_1^T\tau_2\vec{\pi}_1 = \varphi_1'\pi_1''-\varphi_1''\pi_1'  \, ,\\[2ex]
j_2^0 &=& -\vec{\varphi}_2^T\tau_2\vec{\pi}_2 = \varphi_2'\pi_2''-\varphi_2''\pi_2'  \, ,
\eea
\end{subequations}
and thus $\mu_1j_1^0 + \mu_2j_2^0 = - \vec{\varphi}^T\mu\tau_2\vec{\pi}$ with $\mu={\rm diag}(\mu_1,\mu_2)$. 
(Note that $\tau_2$ is a matrix in the space of real and imaginary parts, while $\mu$ is a matrix in the space of fields 1 and 2.) 
The partition function is
\be
Z = \int {\cal D}\varphi_1{\cal D}\varphi_2{\cal D}\pi_1{\cal D}\pi_2\,\exp\left[-\int_X\left({\cal H}-\mu_1{\cal N}_1 - \mu_2{\cal N}_1
-\vec{\pi}^T\partial_0\vec{\varphi}\right)\right]  \, ,
\ee
with the abbreviation
\be
\int_X\equiv \int_0^{1/T} d\tau \int d^3x \, ,
\ee
where $\tau$ is the imaginary time. 
Therefore, we compute 
\bea \label{Lmu}
-\left({\cal H}-\mu_1j_1^0 - \mu_2j_2^0-\vec{\pi}^T\partial_0\vec{\varphi}\right) &=& -\frac{1}{2}\vec{\pi}^TQ\vec{\pi} + (\partial_0\vec{\varphi}+\mu\tau_2\vec{\varphi})^T\vec{\pi}
-\frac{1}{2}\left[\nabla\vec{\varphi}^T\cdot Q^{-1}\nabla\vec{\varphi}+\vec{\varphi}^T(m^2+Y)\vec{\varphi}\right] \non[2ex]
&=&-\frac{1}{2}\vec{\Pi}^TQ\vec{\Pi} +\frac{1}{2}\Big[(D_\mu\vec{\varphi})^T Q^{-1} D^\mu\vec{\varphi} - \vec{\varphi}^T(m^2+Y)\vec{\varphi}\Big] \, ,
\eea
with the shifted momenta $\vec{\Pi} = \vec{\pi}-Q^{-1}(\partial_0\vec{\varphi}+\mu\tau_2\vec{\varphi})$. 
The second term in the second line is the ``new'' Lagrangian; it is identical to the original Lagrangian (\ref{L1}), but with the derivatives replaced by the 
covariant derivatives $D^\mu = \partial^\mu+\delta^\mu_0 \mu\tau_2$. Consequently, the chemical potentials are effectively introduced by replacing all derivatives
in the Lagrangian by covariant derivatives. They can thus equivalently be introduced in the phase of the 
condensates, as we do in the main text.  

The integration over the shifted momenta can now easily be 
performed. In the presence of a derivative coupling, this integration induces a nontrivial, i.e., field-dependent, factor in the integrand of the 
partition function, 
\be
Z =  \int {\cal D}\varphi_1{\cal D}\varphi_2\,({\rm det}\,Q)^{-1/2}\exp\left\{\frac{1}{2}\int_X
\Big[(D_\mu\vec{\varphi})^T Q^{-1} D^\mu\vec{\varphi} - \vec{\varphi}^T(m^2+Y)\vec{\varphi}\Big]\right\} \, , 
\ee
with 
\be
{\rm det}\,Q = \frac{1}{\left(1-\frac{G^2}{4}|\varphi_1|^2|\varphi_2|^2\right)\left(1-\frac{g^2}{4}|\varphi_1|^2|\varphi_2|^2\right)} \, .
\ee
If we expand ${\rm det}\,Q$ around the condensates $\rho_1$, $\rho_2$ 
and only keep the lowest order contribution we can write   
\bea \label{ZQ0}
Z \simeq   [{\rm det}\,Q^{(0)}]^{-1/2} \int {\cal D}\varphi_1{\cal D}\varphi_2\,\exp\left[-\frac{V}{T}U+ \frac{1}{2}\int_X {\cal L}^{(2)} +\ldots \right] \, ,  
\eea
where
\be \label{Q0}
{\rm det}\,Q^{(0)}  = \frac{1}{\left(1-\frac{G^2}{4}\rho_1^2\rho_2^2\right)\left(1-\frac{g^2}{4}\rho_1^2\rho_2^2\right)} \, ,
\ee
and where we have also employed the expansion in the exponent, keeping terms up to second order in the fluctuations.

\section{Grand canonical potential and excitation energies}
\label{app1}

The thermodynamics of the system is obtained from the grand canonical potential density (``free energy density'' or short ``free energy'')  
\be \label{free}
\Omega = -\frac{T}{V} \ln Z \, , 
\ee
with the three-volume $V$ and the partition function $Z$ from Eq.\ (\ref{ZQ0}). 
Inserting the fluctuations from Eq.\ (\ref{fluc}) into the Lagrangian (\ref{L}), we find terms up to fourth order in the fluctuations. The first-order terms cancel if one applies the equations of motion, and  
we neglect all terms of third and fourth order. The remaining second-order terms read
\bea
{\cal L}^{(2)} &=& \frac{1}{2}\sum_{i=1,2}\Big[(\partial\phi_i')^2+(\partial\phi_i'')^2 +|\phi_i|^2(p_i^2-m_i^2)
+2\partial\psi_i\cdot(\phi_i'\partial\phi_i''-\phi_i''\partial\phi_i')-\lambda_i\rho_i^2(3\phi_i'^2+\phi_i''^2)\Big] \non[2ex]
&&+\frac{h+gp_{12}^2}{2}\Big(\rho_1^2|\phi_2|^2+\rho_2^2|\phi_1|^2+4\rho_1\rho_2\phi_1'\phi_2'\Big)
-\frac{\rho_1\rho_2}{2}\Big(G\partial\phi_1'\cdot\partial\phi_2'-g\partial\phi_1''\cdot\partial\phi_2''\Big) \non[2ex]
&&+\frac{g}{2}\Big[\rho_1^2\partial\psi_1\cdot(\phi_2'\partial\phi_2''-\phi_2''\partial\phi_2')
+\rho_2^2\partial\psi_2\cdot(\phi_1'\partial\phi_1''-\phi_1''\partial\phi_1')+2\rho_1\rho_2(\phi_1'\partial\psi_1\cdot\partial\phi_2''
+2\phi_2'\partial\psi_2\cdot\partial\phi_1'')\Big] \, ,
\eea
where we have assumed $\rho_1$ and $\rho_2$ to be constant, and where $|\phi_i|^2 \equiv \phi_i'^2+\phi_i''^2$.
We introduce the Fourier transformed fields via 
\be
\phi_i'(X) = \frac{1}{\sqrt{TV}}\sum_K e^{-iK\cdot X}\phi_i'(K) \, , \qquad \phi_i''(X) = \frac{1}{\sqrt{TV}}\sum_K e^{-iK\cdot X}\phi_i''(K) \, , 
\ee
with the space-time four-vector $X=(-i\tau,\vec{x}) $ and four-momentum $K=(k_0,\vec{k})$, where $k_0=-i\omega_n$ with the bosonic Matsubara frequencies 
$\omega_n = 2\pi n T$, $n\in \mathbb{Z}$. Then, the second-order terms in the fluctuations can be written as  
\be
\int_X {\cal L}^{(2)}= -\frac{1}{2}\sum_K \phi(-K)^T\frac{S^{-1}(K)}{T^2}\phi(K) \, , 
\ee
where we have abbreviated 
\be
\phi(K) = \left(\begin{array}{c} \phi_1'(K) \\ \phi_1''(K) \\ \phi_2'(K) \\ \phi_2''(K) \end{array}\right) \, .
\ee
In this basis, the inverse tree-level propagator in momentum space $S^{-1}(K)$ is a $4\times 4$ matrix, given in Eqs.\ (\ref{Sinv}), and the free energy (\ref{free}) becomes 
\be \label{Om1}
\Omega = U + \frac{1}{2}\frac{T}{V}\sum_K\ln{\rm det}\frac{Q^{(0)}S^{-1}(K)}{T^2} \, , 
\ee
where the determinant is taken over $4\times 4$ space. The determinant of the inverse propagator is a polynomial in $k_0$ of degree 8, which we can write in terms of its zeros, $k_0 = \epsilon_{r,\vec{k}}$. 
As mentioned in the main text, these zeros can be grouped in 4 pairs, $\{\epsilon_{r,\vec{k}},-\epsilon_{r,-\vec{k}}\}$, and we can thus write 
\be
{\rm det}\,S^{-1} = \left(1-\frac{G^2}{4}\rho_1^2\rho_2^2\right)\left(1-\frac{g^2}{4}\rho_1^2\rho_2^2\right)\prod_{r=1}^4(k_0-\epsilon_{r,\vec{k}})(k_0+\epsilon_{r,-\vec{k}}) \, .
\ee
The prefactor is exactly cancelled by ${\rm det}\,Q^{(0)}$, see Eq.\ (\ref{Q0}). 
This is important since otherwise it would yield an unphysical, divergent contribution to the free energy. 
 
As an illustrative example for a thermodynamic quantity, let us write down the charge densities $n_i=-\partial \Omega/\partial\mu_i$ at nonzero temperatures. 
With the help of Eq.\ (\ref{Om1}) and $\ln\,{\rm det} S^{-1} = \Tr\ln S^{-1}$ we find  
\bea \label{ni}
n_i&=& -\frac{\partial U}{\partial \mu_i}- \frac{1}{2}\frac{T}{V}\sum_K\Tr\left[S\frac{\partial S^{-1}}{\partial\mu_i}\right] \non[2ex] 
&=& -\frac{\partial U}{\partial \mu_i}- \frac{1}{2}\int\frac{d^3k}{(2\pi)^3}\;T\hspace{-0.2cm}\sum_{n=-\infty}^\infty \frac{F(k_0,\vec{k})}{\prod_{r=1}^4(k_0-\epsilon_{r,\vec{k}})
(k_0+\epsilon_{r,-\vec{k}})} \, ,
\eea
where, in the second step, we have rewritten the sum over four-momentum as a discrete sum over Matsubara frequencies $n$ and an integral over three-momentum $\vec{k}$, 
and $F(k_0,\vec{k})$ is a complicated function without poles in $k_0$ which obeys the symmetry $F(-k_0,-\vec{k})=F(k_0,\vec{k})$. 
The sum over Matsubara frequencies creates 8 terms, each for one of the poles. 
Due to the symmetries of $\epsilon_{r,\vec{k}}$ and $F(k_0,\vec{k})$ these terms give the same result pairwise under the $\vec{k}$-integral (this is easily seen with the new integration 
variable $\vec{k}\to -\vec{k}$ in one term of each pair). Therefore, we can restrict ourselves to 4 terms, 
\be
n_i =  -\frac{\partial U}{\partial \mu_i} + \frac{1}{2}\sum_{r=1}^4\int\frac{d^3k}{(2\pi)^3} \frac{F(\epsilon_{r,\vec{k}},\vec{k})}{(\epsilon_{r,\vec{k}}+\epsilon_{r,-\vec{k}})
\prod_s (\epsilon_{r,\vec{k}}-\epsilon_{s,\vec{k}})(\epsilon_{r,\vec{k}}+\epsilon_{s,-\vec{k}})}\coth\frac{\epsilon_{r,\vec{k}}}{2T} \, ,
\ee
where the product over $s$ runs over the three integers different from $r$. There are two
contributions due to $\coth\frac{\epsilon}{2T} = 1+2f(\epsilon)$ with the Bose distribution function $f(\epsilon)=1/(e^{\epsilon/T}-1)$. The first one is temperature independent 
(more precisely, there is no {\it explicit} dependence on $T$, in general a temperature dependence enters through the condensates). This contribution is infinite and has to be 
regularized. The second contribution depends on temperature explicitly. It is the ``usual'' integral over the Bose distribution, here with a complicated, momentum-dependent prefactor.
This prefactor is trivial in the NOR phase, where there are no condensates, and particles and antiparticles carry positive and negative unit charges for each of the species 
separately. In the COE phase, where both fields condense, the 2 massive and 2 massless modes each contribute to both charge densities in a nontrivial way. Note that in the thermal 
contribution alone we do {\it not} have 8 terms that are pairwise equal. Instead, now it is crucial to work with positive excitation energies, otherwise one obtains 
unphysical, negative occupation numbers. We do not further evaluate the charge densities or any other thermodynamic quantity since our main focus in this work is on 
the excitation energies $\epsilon_{r,\vec{k}}$ themselves. 

\bibliography{refs}

\begin{thebibliography}{57}
\expandafter\ifx\csname natexlab\endcsname\relax\def\natexlab#1{#1}\fi
\expandafter\ifx\csname bibnamefont\endcsname\relax
  \def\bibnamefont#1{#1}\fi
\expandafter\ifx\csname bibfnamefont\endcsname\relax
  \def\bibfnamefont#1{#1}\fi
\expandafter\ifx\csname citenamefont\endcsname\relax
  \def\citenamefont#1{#1}\fi
\expandafter\ifx\csname url\endcsname\relax
  \def\url#1{\texttt{#1}}\fi
\expandafter\ifx\csname urlprefix\endcsname\relax\def\urlprefix{URL }\fi
\providecommand{\bibinfo}[2]{#2}
\providecommand{\eprint}[2][]{\url{#2}}

\bibitem[{\citenamefont{Tisza}(1938)}]{tisza38}
\bibinfo{author}{\bibfnamefont{L.}~\bibnamefont{Tisza}},
  \bibinfo{journal}{Nature} \textbf{\bibinfo{volume}{141}},
  \bibinfo{pages}{913} (\bibinfo{year}{1938}).

\bibitem[{\citenamefont{Landau}(1941)}]{landau41}
\bibinfo{author}{\bibfnamefont{L.}~\bibnamefont{Landau}},
  \bibinfo{journal}{Phys. Rev.} \textbf{\bibinfo{volume}{60}},
  \bibinfo{pages}{356} (\bibinfo{year}{1941}).

\bibitem[{\citenamefont{{Tuoriniemi} et~al.}(2002)\citenamefont{{Tuoriniemi},
  {Martikainen}, {Pentti}, {Sebedash}, {Boldarev}, and
  {Pickett}}}]{2002JLTP..129..531T}
\bibinfo{author}{\bibfnamefont{J.}~\bibnamefont{{Tuoriniemi}}},
  \bibinfo{author}{\bibfnamefont{J.}~\bibnamefont{{Martikainen}}},
  \bibinfo{author}{\bibfnamefont{E.}~\bibnamefont{{Pentti}}},
  \bibinfo{author}{\bibfnamefont{A.}~\bibnamefont{{Sebedash}}},
  \bibinfo{author}{\bibfnamefont{S.}~\bibnamefont{{Boldarev}}},
  \bibnamefont{and}
  \bibinfo{author}{\bibfnamefont{G.}~\bibnamefont{{Pickett}}},
  \bibinfo{journal}{Journal of Low Temperature Physics}
  \textbf{\bibinfo{volume}{129}}, \bibinfo{pages}{531} (\bibinfo{year}{2002}).

\bibitem[{\citenamefont{Rysti et~al.}(2012)\citenamefont{Rysti, Tuoriniemi, and
  Salmela}}]{PhysRevB.85.134529}
\bibinfo{author}{\bibfnamefont{J.}~\bibnamefont{Rysti}},
  \bibinfo{author}{\bibfnamefont{J.}~\bibnamefont{Tuoriniemi}},
  \bibnamefont{and} \bibinfo{author}{\bibfnamefont{A.}~\bibnamefont{Salmela}},
  \bibinfo{journal}{Phys. Rev. B} \textbf{\bibinfo{volume}{85}},
  \bibinfo{pages}{134529} (\bibinfo{year}{2012}).

\bibitem[{\citenamefont{{Ferrier-Barbut}
  et~al.}(2014)\citenamefont{{Ferrier-Barbut}, {Delehaye}, {Laurent}, {Grier},
  {Pierce}, {Rem}, {Chevy}, and {Salomon}}}]{2014Sci...345.1035F}
\bibinfo{author}{\bibfnamefont{I.}~\bibnamefont{{Ferrier-Barbut}}},
  \bibinfo{author}{\bibfnamefont{M.}~\bibnamefont{{Delehaye}}},
  \bibinfo{author}{\bibfnamefont{S.}~\bibnamefont{{Laurent}}},
  \bibinfo{author}{\bibfnamefont{A.~T.} \bibnamefont{{Grier}}},
  \bibinfo{author}{\bibfnamefont{M.}~\bibnamefont{{Pierce}}},
  \bibinfo{author}{\bibfnamefont{B.~S.} \bibnamefont{{Rem}}},
  \bibinfo{author}{\bibfnamefont{F.}~\bibnamefont{{Chevy}}}, \bibnamefont{and}
  \bibinfo{author}{\bibfnamefont{C.}~\bibnamefont{{Salomon}}},
  \bibinfo{journal}{Science} \textbf{\bibinfo{volume}{345}},
  \bibinfo{pages}{1035} (\bibinfo{year}{2014}), \eprint{1404.2548}.

\bibitem[{\citenamefont{{Delehaye} et~al.}(2015)\citenamefont{{Delehaye},
  {Laurent}, {Ferrier-Barbut}, {Jin}, {Chevy}, and
  {Salomon}}}]{2015arXiv151006709D}
\bibinfo{author}{\bibfnamefont{M.}~\bibnamefont{{Delehaye}}},
  \bibinfo{author}{\bibfnamefont{S.}~\bibnamefont{{Laurent}}},
  \bibinfo{author}{\bibfnamefont{I.}~\bibnamefont{{Ferrier-Barbut}}},
  \bibinfo{author}{\bibfnamefont{S.}~\bibnamefont{{Jin}}},
  \bibinfo{author}{\bibfnamefont{F.}~\bibnamefont{{Chevy}}}, \bibnamefont{and}
  \bibinfo{author}{\bibfnamefont{C.}~\bibnamefont{{Salomon}}}
  (\bibinfo{year}{2015}), \eprint{1510.06709}.

\bibitem[{\citenamefont{Gusakov et~al.}(2009)\citenamefont{Gusakov, Kantor, and
  Haensel}}]{Gusakov:2009kc}
\bibinfo{author}{\bibfnamefont{M.~E.} \bibnamefont{Gusakov}},
  \bibinfo{author}{\bibfnamefont{E.~M.} \bibnamefont{Kantor}},
  \bibnamefont{and} \bibinfo{author}{\bibfnamefont{P.}~\bibnamefont{Haensel}},
  \bibinfo{journal}{Phys. Rev.} \textbf{\bibinfo{volume}{C79}},
  \bibinfo{pages}{055806} (\bibinfo{year}{2009}), \eprint{0904.3467}.

\bibitem[{\citenamefont{Alford et~al.}(1999)\citenamefont{Alford, Rajagopal,
  and Wilczek}}]{Alford:1998mk}
\bibinfo{author}{\bibfnamefont{M.~G.} \bibnamefont{Alford}},
  \bibinfo{author}{\bibfnamefont{K.}~\bibnamefont{Rajagopal}},
  \bibnamefont{and} \bibinfo{author}{\bibfnamefont{F.}~\bibnamefont{Wilczek}},
  \bibinfo{journal}{Nucl. Phys.} \textbf{\bibinfo{volume}{B537}},
  \bibinfo{pages}{443} (\bibinfo{year}{1999}), \eprint{hep-ph/9804403}.

\bibitem[{\citenamefont{Alford et~al.}(2008)\citenamefont{Alford, Schmitt,
  Rajagopal, and Sch{\"a}fer}}]{Alford:2007xm}
\bibinfo{author}{\bibfnamefont{M.~G.} \bibnamefont{Alford}},
  \bibinfo{author}{\bibfnamefont{A.}~\bibnamefont{Schmitt}},
  \bibinfo{author}{\bibfnamefont{K.}~\bibnamefont{Rajagopal}},
  \bibnamefont{and}
  \bibinfo{author}{\bibfnamefont{T.}~\bibnamefont{Sch{\"a}fer}},
  \bibinfo{journal}{Rev.Mod.Phys.} \textbf{\bibinfo{volume}{80}},
  \bibinfo{pages}{1455} (\bibinfo{year}{2008}), \eprint{0709.4635}.

\bibitem[{\citenamefont{Sch{\"a}fer}(2000)}]{Schafer:2000tw}
\bibinfo{author}{\bibfnamefont{T.}~\bibnamefont{Sch{\"a}fer}},
  \bibinfo{journal}{Phys. Rev.} \textbf{\bibinfo{volume}{D62}},
  \bibinfo{pages}{094007} (\bibinfo{year}{2000}), \eprint{hep-ph/0006034}.

\bibitem[{\citenamefont{Schmitt}(2005)}]{Schmitt:2004et}
\bibinfo{author}{\bibfnamefont{A.}~\bibnamefont{Schmitt}},
  \bibinfo{journal}{Phys. Rev.} \textbf{\bibinfo{volume}{D71}},
  \bibinfo{pages}{054016} (\bibinfo{year}{2005}), \eprint{nucl-th/0412033}.

\bibitem[{\citenamefont{{Andreev} and {Bashkin}}(1975)}]{1976JETP...42..164A}
\bibinfo{author}{\bibfnamefont{A.~F.} \bibnamefont{{Andreev}}}
  \bibnamefont{and} \bibinfo{author}{\bibfnamefont{E.~P.}
  \bibnamefont{{Bashkin}}}, \bibinfo{journal}{Zh. Eksp. Teor. Fiz.}
  \textbf{\bibinfo{volume}{69}}, \bibinfo{pages}{319} (\bibinfo{year}{1975}),
  \bibinfo{note}{[Sov.\ Phys.\ JETP {\bf 42}, 164 (1976)]}.

\bibitem[{\citenamefont{{Shevchenko} and {Fil}}(2007)}]{2007JETP..105..135S}
\bibinfo{author}{\bibfnamefont{S.~I.} \bibnamefont{{Shevchenko}}}
  \bibnamefont{and} \bibinfo{author}{\bibfnamefont{D.~V.} \bibnamefont{{Fil}}},
  \bibinfo{journal}{Soviet Journal of Experimental and Theoretical Physics}
  \textbf{\bibinfo{volume}{105}}, \bibinfo{pages}{135} (\bibinfo{year}{2007}).

\bibitem[{\citenamefont{Schmitt}(2014)}]{Schmitt:2013nva}
\bibinfo{author}{\bibfnamefont{A.}~\bibnamefont{Schmitt}},
  \bibinfo{journal}{Phys. Rev.} \textbf{\bibinfo{volume}{D89}},
  \bibinfo{pages}{065024} (\bibinfo{year}{2014}), \eprint{1312.5993}.

\bibitem[{\citenamefont{Law et~al.}(2001)\citenamefont{Law, Chan, Leung, and
  Chu}}]{PhysRevA.63.063612}
\bibinfo{author}{\bibfnamefont{C.~K.} \bibnamefont{Law}},
  \bibinfo{author}{\bibfnamefont{C.~M.} \bibnamefont{Chan}},
  \bibinfo{author}{\bibfnamefont{P.~T.} \bibnamefont{Leung}}, \bibnamefont{and}
  \bibinfo{author}{\bibfnamefont{M.-C.} \bibnamefont{Chu}},
  \bibinfo{journal}{Phys. Rev. A} \textbf{\bibinfo{volume}{63}},
  \bibinfo{pages}{063612} (\bibinfo{year}{2001}).

\bibitem[{\citenamefont{{Andersson} et~al.}(2004)\citenamefont{{Andersson},
  {Comer}, and {Prix}}}]{2004MNRAS.354..101A}
\bibinfo{author}{\bibfnamefont{N.}~\bibnamefont{{Andersson}}},
  \bibinfo{author}{\bibfnamefont{G.~L.} \bibnamefont{{Comer}}},
  \bibnamefont{and} \bibinfo{author}{\bibfnamefont{R.}~\bibnamefont{{Prix}}},
  \bibinfo{journal}{Mon.Not.Roy.Astron.Soc.} \textbf{\bibinfo{volume}{354}},
  \bibinfo{pages}{101} (\bibinfo{year}{2004}).

\bibitem[{\citenamefont{{Ishino} et~al.}(2011)\citenamefont{{Ishino},
  {Tsubota}, and {Takeuchi}}}]{2011PhRvA..83f3602I}
\bibinfo{author}{\bibfnamefont{S.}~\bibnamefont{{Ishino}}},
  \bibinfo{author}{\bibfnamefont{M.}~\bibnamefont{{Tsubota}}},
  \bibnamefont{and}
  \bibinfo{author}{\bibfnamefont{H.}~\bibnamefont{{Takeuchi}}},
  \bibinfo{journal}{\pra} \textbf{\bibinfo{volume}{83}}, \bibinfo{eid}{063602}
  (\bibinfo{year}{2011}), \eprint{1106.0884}.

\bibitem[{\citenamefont{{Abad} et~al.}(2015)\citenamefont{{Abad}, {Recati},
  {Stringari}, and {Chevy}}}]{2015EPJD...69..126A}
\bibinfo{author}{\bibfnamefont{M.}~\bibnamefont{{Abad}}},
  \bibinfo{author}{\bibfnamefont{A.}~\bibnamefont{{Recati}}},
  \bibinfo{author}{\bibfnamefont{S.}~\bibnamefont{{Stringari}}},
  \bibnamefont{and} \bibinfo{author}{\bibfnamefont{F.}~\bibnamefont{{Chevy}}},
  \bibinfo{journal}{European Physical Journal D} \textbf{\bibinfo{volume}{69}},
  \bibinfo{eid}{126} (\bibinfo{year}{2015}), \eprint{1411.7560}.

\bibitem[{\citenamefont{Wu and Niu}(2001)}]{PhysRevA.64.061603}
\bibinfo{author}{\bibfnamefont{B.}~\bibnamefont{Wu}} \bibnamefont{and}
  \bibinfo{author}{\bibfnamefont{Q.}~\bibnamefont{Niu}},
  \bibinfo{journal}{Phys. Rev. A} \textbf{\bibinfo{volume}{64}},
  \bibinfo{pages}{061603} (\bibinfo{year}{2001}).

\bibitem[{\citenamefont{{Andreev} and
  {Melnikovsky}}(2003)}]{2003JETPL..78..574A}
\bibinfo{author}{\bibfnamefont{A.~F.} \bibnamefont{{Andreev}}}
  \bibnamefont{and} \bibinfo{author}{\bibfnamefont{L.~A.}
  \bibnamefont{{Melnikovsky}}}, \bibinfo{journal}{Soviet Journal of
  Experimental and Theoretical Physics Letters} \textbf{\bibinfo{volume}{78}},
  \bibinfo{pages}{574} (\bibinfo{year}{2003}), \eprint{cond-mat/0304019}.

\bibitem[{\citenamefont{{Andreev} and
  {Melnikovsky}}(2006)}]{2006JETP..103..944A}
\bibinfo{author}{\bibfnamefont{A.~F.} \bibnamefont{{Andreev}}}
  \bibnamefont{and} \bibinfo{author}{\bibfnamefont{L.~A.}
  \bibnamefont{{Melnikovsky}}}, \bibinfo{journal}{Soviet Journal of
  Experimental and Theoretical Physics} \textbf{\bibinfo{volume}{103}},
  \bibinfo{pages}{944} (\bibinfo{year}{2006}).

\bibitem[{\citenamefont{{Kravchenko} and {Fil}}(2008)}]{2008JLTP..150..612K}
\bibinfo{author}{\bibfnamefont{L.~Y.} \bibnamefont{{Kravchenko}}}
  \bibnamefont{and} \bibinfo{author}{\bibfnamefont{D.~V.} \bibnamefont{{Fil}}},
  \bibinfo{journal}{Journal of Low Temperature Physics}
  \textbf{\bibinfo{volume}{150}}, \bibinfo{pages}{612} (\bibinfo{year}{2008}),
  \eprint{0807.0726}.

\bibitem[{\citenamefont{Peralta et~al.}(2006)\citenamefont{Peralta, Melatos,
  Giacobello, and Ooi}}]{Peralta:2006um}
\bibinfo{author}{\bibfnamefont{C.}~\bibnamefont{Peralta}},
  \bibinfo{author}{\bibfnamefont{A.}~\bibnamefont{Melatos}},
  \bibinfo{author}{\bibfnamefont{M.}~\bibnamefont{Giacobello}},
  \bibnamefont{and} \bibinfo{author}{\bibfnamefont{A.}~\bibnamefont{Ooi}},
  \bibinfo{journal}{Astrophys. J.} \textbf{\bibinfo{volume}{651}},
  \bibinfo{pages}{1079} (\bibinfo{year}{2006}), \eprint{astro-ph/0607161}.

\bibitem[{\citenamefont{Haskell and Melatos}(2015)}]{Haskell:2015jra}
\bibinfo{author}{\bibfnamefont{B.}~\bibnamefont{Haskell}} \bibnamefont{and}
  \bibinfo{author}{\bibfnamefont{A.}~\bibnamefont{Melatos}},
  \bibinfo{journal}{Int. J. Mod. Phys.} \textbf{\bibinfo{volume}{D24}},
  \bibinfo{pages}{1530008} (\bibinfo{year}{2015}), \eprint{1502.07062}.

\bibitem[{\citenamefont{Pethick et~al.}(2010)\citenamefont{Pethick, Chamel, and
  Reddy}}]{pethick2010superfluid}
\bibinfo{author}{\bibfnamefont{C.}~\bibnamefont{Pethick}},
  \bibinfo{author}{\bibfnamefont{N.}~\bibnamefont{Chamel}}, \bibnamefont{and}
  \bibinfo{author}{\bibfnamefont{S.}~\bibnamefont{Reddy}},
  \bibinfo{journal}{Progress of Theoretical Physics Supplement}
  \textbf{\bibinfo{volume}{186}}, \bibinfo{pages}{9} (\bibinfo{year}{2010}).

\bibitem[{\citenamefont{Chamel et~al.}(2012)\citenamefont{Chamel, Pearson, and
  Goriely}}]{Chamel:2012pk}
\bibinfo{author}{\bibfnamefont{N.}~\bibnamefont{Chamel}},
  \bibinfo{author}{\bibfnamefont{J.~M.} \bibnamefont{Pearson}},
  \bibnamefont{and} \bibinfo{author}{\bibfnamefont{S.}~\bibnamefont{Goriely}},
  \bibinfo{journal}{ASP Conf. Ser.} \textbf{\bibinfo{volume}{466}},
  \bibinfo{pages}{203} (\bibinfo{year}{2012}), \eprint{1206.6926}.

\bibitem[{\citenamefont{Andersson et~al.}(2012)\citenamefont{Andersson,
  Glampedakis, Ho, and Espinoza}}]{Andersson:2012iu}
\bibinfo{author}{\bibfnamefont{N.}~\bibnamefont{Andersson}},
  \bibinfo{author}{\bibfnamefont{K.}~\bibnamefont{Glampedakis}},
  \bibinfo{author}{\bibfnamefont{W.~C.~G.} \bibnamefont{Ho}}, \bibnamefont{and}
  \bibinfo{author}{\bibfnamefont{C.~M.} \bibnamefont{Espinoza}},
  \bibinfo{journal}{Phys. Rev. Lett.} \textbf{\bibinfo{volume}{109}},
  \bibinfo{pages}{241103} (\bibinfo{year}{2012}), \eprint{1207.0633}.

\bibitem[{\citenamefont{Carter and Langlois}(1995)}]{Carter:1995if}
\bibinfo{author}{\bibfnamefont{B.}~\bibnamefont{Carter}} \bibnamefont{and}
  \bibinfo{author}{\bibfnamefont{D.}~\bibnamefont{Langlois}},
  \bibinfo{journal}{Phys.Rev.} \textbf{\bibinfo{volume}{D51}},
  \bibinfo{pages}{5855} (\bibinfo{year}{1995}), \eprint{hep-th/9507058}.

\bibitem[{\citenamefont{{Alford} et~al.}(2013)\citenamefont{{Alford},
  {Mallavarapu}, {Schmitt}, and {Stetina}}}]{2013PhRvD..87f5001A}
\bibinfo{author}{\bibfnamefont{M.~G.} \bibnamefont{{Alford}}},
  \bibinfo{author}{\bibfnamefont{S.~K.} \bibnamefont{{Mallavarapu}}},
  \bibinfo{author}{\bibfnamefont{A.}~\bibnamefont{{Schmitt}}},
  \bibnamefont{and}
  \bibinfo{author}{\bibfnamefont{S.}~\bibnamefont{{Stetina}}},
  \bibinfo{journal}{\prd} \textbf{\bibinfo{volume}{87}}, \bibinfo{eid}{065001}
  (\bibinfo{year}{2013}), \eprint{1212.0670}.

\bibitem[{\citenamefont{Schmitt}(2015)}]{Schmitt:2014eka}
\bibinfo{author}{\bibfnamefont{A.}~\bibnamefont{Schmitt}},
  \bibinfo{journal}{Lect. Notes Phys.} \textbf{\bibinfo{volume}{888}}
  (\bibinfo{year}{2015}), \eprint{1404.1284}.

\bibitem[{\citenamefont{Adamenko et~al.}(2008)\citenamefont{Adamenko,
  Nemchenko, Slipko, and Wyatt}}]{PhysRevB.77.144515}
\bibinfo{author}{\bibfnamefont{I.~N.} \bibnamefont{Adamenko}},
  \bibinfo{author}{\bibfnamefont{K.~E.} \bibnamefont{Nemchenko}},
  \bibinfo{author}{\bibfnamefont{V.~A.} \bibnamefont{Slipko}},
  \bibnamefont{and} \bibinfo{author}{\bibfnamefont{A.~F.~G.}
  \bibnamefont{Wyatt}}, \bibinfo{journal}{Phys. Rev. B}
  \textbf{\bibinfo{volume}{77}}, \bibinfo{pages}{144515}
  (\bibinfo{year}{2008}).

\bibitem[{\citenamefont{Alford et~al.}(2014)\citenamefont{Alford, Mallavarapu,
  Schmitt, and Stetina}}]{Alford:2013koa}
\bibinfo{author}{\bibfnamefont{M.~G.} \bibnamefont{Alford}},
  \bibinfo{author}{\bibfnamefont{S.~K.} \bibnamefont{Mallavarapu}},
  \bibinfo{author}{\bibfnamefont{A.}~\bibnamefont{Schmitt}}, \bibnamefont{and}
  \bibinfo{author}{\bibfnamefont{S.}~\bibnamefont{Stetina}},
  \bibinfo{journal}{Phys. Rev.} \textbf{\bibinfo{volume}{D89}},
  \bibinfo{pages}{085005} (\bibinfo{year}{2014}), \eprint{1310.5953}.

\bibitem[{\citenamefont{Alpar et~al.}(1984)\citenamefont{Alpar, Langer, and
  Sauls}}]{Alpar:1984zz}
\bibinfo{author}{\bibfnamefont{M.~A.} \bibnamefont{Alpar}},
  \bibinfo{author}{\bibfnamefont{S.~A.} \bibnamefont{Langer}},
  \bibnamefont{and} \bibinfo{author}{\bibfnamefont{J.~A.} \bibnamefont{Sauls}},
  \bibinfo{journal}{Astrophys. J.} \textbf{\bibinfo{volume}{282}},
  \bibinfo{pages}{533} (\bibinfo{year}{1984}).

\bibitem[{\citenamefont{Alford and Good}(2008)}]{Alford:2007np}
\bibinfo{author}{\bibfnamefont{M.~G.} \bibnamefont{Alford}} \bibnamefont{and}
  \bibinfo{author}{\bibfnamefont{G.}~\bibnamefont{Good}},
  \bibinfo{journal}{Phys. Rev.} \textbf{\bibinfo{volume}{B78}},
  \bibinfo{pages}{024510} (\bibinfo{year}{2008}), \eprint{0712.1810}.

\bibitem[{\citenamefont{{Kobyakov} et~al.}(2015)\citenamefont{{Kobyakov},
  {Samuelsson}, {Marklund}, {Lundh}, {Bychkov}, and
  {Brandenburg}}}]{2015arXiv150400570K}
\bibinfo{author}{\bibfnamefont{D.}~\bibnamefont{{Kobyakov}}},
  \bibinfo{author}{\bibfnamefont{L.}~\bibnamefont{{Samuelsson}}},
  \bibinfo{author}{\bibfnamefont{M.}~\bibnamefont{{Marklund}}},
  \bibinfo{author}{\bibfnamefont{E.}~\bibnamefont{{Lundh}}},
  \bibinfo{author}{\bibfnamefont{V.}~\bibnamefont{{Bychkov}}},
  \bibnamefont{and}
  \bibinfo{author}{\bibfnamefont{A.}~\bibnamefont{{Brandenburg}}}
  (\bibinfo{year}{2015}), \eprint{1504.00570}.

\bibitem[{\citenamefont{Sauls}(1989)}]{sauls1989superfluidity}
\bibinfo{author}{\bibfnamefont{J.}~\bibnamefont{Sauls}}, in
  \emph{\bibinfo{booktitle}{Timing Neutron Stars}}, edited by
  \bibinfo{editor}{\bibfnamefont{H.}~\bibnamefont{{\"O}gelman}}
  \bibnamefont{and} \bibinfo{editor}{\bibfnamefont{E.}~\bibnamefont{van~den
  Heuvel}} (\bibinfo{publisher}{Springer Netherlands}, \bibinfo{year}{1989}),
  vol. \bibinfo{volume}{262} of \emph{\bibinfo{series}{NATO ASI Series}}, pp.
  \bibinfo{pages}{457--490}.

\bibitem[{\citenamefont{Comer and Joynt}(2003)}]{Comer:2002dm}
\bibinfo{author}{\bibfnamefont{G.}~\bibnamefont{Comer}} \bibnamefont{and}
  \bibinfo{author}{\bibfnamefont{R.}~\bibnamefont{Joynt}},
  \bibinfo{journal}{Phys.Rev.} \textbf{\bibinfo{volume}{D68}},
  \bibinfo{pages}{023002} (\bibinfo{year}{2003}), \eprint{gr-qc/0212083}.

\bibitem[{\citenamefont{Chamel and Haensel}(2006)}]{Chamel:2006rc}
\bibinfo{author}{\bibfnamefont{N.}~\bibnamefont{Chamel}} \bibnamefont{and}
  \bibinfo{author}{\bibfnamefont{P.}~\bibnamefont{Haensel}},
  \bibinfo{journal}{Phys. Rev.} \textbf{\bibinfo{volume}{C73}},
  \bibinfo{pages}{045802} (\bibinfo{year}{2006}), \eprint{nucl-th/0603018}.

\bibitem[{\citenamefont{Stetina}(2015)}]{Stetina:2015exa}
\bibinfo{author}{\bibfnamefont{S.}~\bibnamefont{Stetina}}, Ph.D. thesis,
  \bibinfo{school}{Technische Universit{\"a}t Wien} (\bibinfo{year}{2015}),
  \eprint{1502.00122}.

\bibitem[{\citenamefont{Fukushima and Iida}(2005)}]{Fukushima:2005gt}
\bibinfo{author}{\bibfnamefont{K.}~\bibnamefont{Fukushima}} \bibnamefont{and}
  \bibinfo{author}{\bibfnamefont{K.}~\bibnamefont{Iida}},
  \bibinfo{journal}{Phys. Rev.} \textbf{\bibinfo{volume}{D71}},
  \bibinfo{pages}{074011} (\bibinfo{year}{2005}), \eprint{hep-ph/0501276}.

\bibitem[{\citenamefont{{Yamamura} and {Yamamoto}}(2015)}]{2015JPSJ...84d4003Y}
\bibinfo{author}{\bibfnamefont{H.}~\bibnamefont{{Yamamura}}} \bibnamefont{and}
  \bibinfo{author}{\bibfnamefont{D.}~\bibnamefont{{Yamamoto}}},
  \bibinfo{journal}{Journal of the Physical Society of Japan}
  \textbf{\bibinfo{volume}{84}}, \bibinfo{eid}{044003} (\bibinfo{year}{2015}),
  \eprint{1403.5621}.

\bibitem[{\citenamefont{Chandrasekhar}(1970)}]{PhysRevLett.24.611}
\bibinfo{author}{\bibfnamefont{S.}~\bibnamefont{Chandrasekhar}},
  \bibinfo{journal}{Phys. Rev. Lett.} \textbf{\bibinfo{volume}{24}},
  \bibinfo{pages}{611} (\bibinfo{year}{1970}).

\bibitem[{\citenamefont{{Friedman} and {Schutz}}(1978)}]{1978ApJ...221..937F}
\bibinfo{author}{\bibfnamefont{J.~L.} \bibnamefont{{Friedman}}}
  \bibnamefont{and} \bibinfo{author}{\bibfnamefont{B.~F.}
  \bibnamefont{{Schutz}}}, \bibinfo{journal}{\apj}
  \textbf{\bibinfo{volume}{221}}, \bibinfo{pages}{937} (\bibinfo{year}{1978}).

\bibitem[{\citenamefont{Andersson}(1998)}]{Andersson:1997xt}
\bibinfo{author}{\bibfnamefont{N.}~\bibnamefont{Andersson}},
  \bibinfo{journal}{Astrophys. J.} \textbf{\bibinfo{volume}{502}},
  \bibinfo{pages}{708} (\bibinfo{year}{1998}), \eprint{gr-qc/9706075}.

\bibitem[{\citenamefont{Ozawa et~al.}(2013)\citenamefont{Ozawa, Pitaevskii, and
  Stringari}}]{PhysRevA.87.063610}
\bibinfo{author}{\bibfnamefont{T.}~\bibnamefont{Ozawa}},
  \bibinfo{author}{\bibfnamefont{L.~P.} \bibnamefont{Pitaevskii}},
  \bibnamefont{and}
  \bibinfo{author}{\bibfnamefont{S.}~\bibnamefont{Stringari}},
  \bibinfo{journal}{Phys. Rev. A} \textbf{\bibinfo{volume}{87}},
  \bibinfo{pages}{063610} (\bibinfo{year}{2013}).

\bibitem[{\citenamefont{Buneman}(1959)}]{Buneman:1959zz}
\bibinfo{author}{\bibfnamefont{O.}~\bibnamefont{Buneman}},
  \bibinfo{journal}{Phys.Rev.} \textbf{\bibinfo{volume}{115}},
  \bibinfo{pages}{503} (\bibinfo{year}{1959}).

\bibitem[{\citenamefont{{Farley}}(1963)}]{1963PhRvL..10..279F}
\bibinfo{author}{\bibfnamefont{D.~T.} \bibnamefont{{Farley}}},
  \bibinfo{journal}{Physical Review Letters} \textbf{\bibinfo{volume}{10}},
  \bibinfo{pages}{279} (\bibinfo{year}{1963}).

\bibitem[{\citenamefont{{Anderson} et~al.}(2001)\citenamefont{{Anderson},
  {Fedele}, and {Lisak}}}]{2001AmJPh..69.1262A}
\bibinfo{author}{\bibfnamefont{D.}~\bibnamefont{{Anderson}}},
  \bibinfo{author}{\bibfnamefont{R.}~\bibnamefont{{Fedele}}}, \bibnamefont{and}
  \bibinfo{author}{\bibfnamefont{M.}~\bibnamefont{{Lisak}}},
  \bibinfo{journal}{American Journal of Physics} \textbf{\bibinfo{volume}{69}},
  \bibinfo{pages}{1262} (\bibinfo{year}{2001}).

\bibitem[{\citenamefont{Samuelsson et~al.}(2010)\citenamefont{Samuelsson,
  Lopez-Monsalvo, Andersson, and Comer}}]{Samuelsson:2009up}
\bibinfo{author}{\bibfnamefont{L.}~\bibnamefont{Samuelsson}},
  \bibinfo{author}{\bibfnamefont{C.}~\bibnamefont{Lopez-Monsalvo}},
  \bibinfo{author}{\bibfnamefont{N.}~\bibnamefont{Andersson}},
  \bibnamefont{and} \bibinfo{author}{\bibfnamefont{G.}~\bibnamefont{Comer}},
  \bibinfo{journal}{Gen.Rel.Grav.} \textbf{\bibinfo{volume}{42}},
  \bibinfo{pages}{413} (\bibinfo{year}{2010}), \eprint{0906.4002}.

\bibitem[{\citenamefont{Hawke et~al.}(2013)\citenamefont{Hawke, Comer, and
  Andersson}}]{Hawke:2013haa}
\bibinfo{author}{\bibfnamefont{I.}~\bibnamefont{Hawke}},
  \bibinfo{author}{\bibfnamefont{G.}~\bibnamefont{Comer}}, \bibnamefont{and}
  \bibinfo{author}{\bibfnamefont{N.}~\bibnamefont{Andersson}},
  \bibinfo{journal}{Class.Quant.Grav.} \textbf{\bibinfo{volume}{30}},
  \bibinfo{pages}{145007} (\bibinfo{year}{2013}), \eprint{1303.4070}.

\bibitem[{\citenamefont{Gubankova et~al.}(2006)\citenamefont{Gubankova,
  Schmitt, and Wilczek}}]{Gubankova:2006gj}
\bibinfo{author}{\bibfnamefont{E.}~\bibnamefont{Gubankova}},
  \bibinfo{author}{\bibfnamefont{A.}~\bibnamefont{Schmitt}}, \bibnamefont{and}
  \bibinfo{author}{\bibfnamefont{F.}~\bibnamefont{Wilczek}},
  \bibinfo{journal}{Phys. Rev.} \textbf{\bibinfo{volume}{B74}},
  \bibinfo{pages}{064505} (\bibinfo{year}{2006}), \eprint{cond-mat/0603603}.

\bibitem[{\citenamefont{Deng et~al.}(2007)\citenamefont{Deng, Schmitt, and
  Wang}}]{Deng:2006ed}
\bibinfo{author}{\bibfnamefont{J.}~\bibnamefont{Deng}},
  \bibinfo{author}{\bibfnamefont{A.}~\bibnamefont{Schmitt}}, \bibnamefont{and}
  \bibinfo{author}{\bibfnamefont{Q.}~\bibnamefont{Wang}},
  \bibinfo{journal}{Phys. Rev.} \textbf{\bibinfo{volume}{D76}},
  \bibinfo{pages}{034013} (\bibinfo{year}{2007}), \eprint{nucl-th/0611097}.

\bibitem[{\citenamefont{Huang et~al.}(2007)\citenamefont{Huang, Hao, and
  Zhuang}}]{Huang:2006kr}
\bibinfo{author}{\bibfnamefont{X.}~\bibnamefont{Huang}},
  \bibinfo{author}{\bibfnamefont{X.}~\bibnamefont{Hao}}, \bibnamefont{and}
  \bibinfo{author}{\bibfnamefont{P.}~\bibnamefont{Zhuang}},
  \bibinfo{journal}{New J. Phys.} \textbf{\bibinfo{volume}{9}},
  \bibinfo{pages}{375} (\bibinfo{year}{2007}), \eprint{cond-mat/0610610}.

\bibitem[{\citenamefont{{Yukalov} and {Yukalova}}(2004)}]{2004LaPhL...1...50Y}
\bibinfo{author}{\bibfnamefont{V.~I.} \bibnamefont{{Yukalov}}}
  \bibnamefont{and} \bibinfo{author}{\bibfnamefont{E.~P.}
  \bibnamefont{{Yukalova}}}, \bibinfo{journal}{Laser Physics Letters}
  \textbf{\bibinfo{volume}{1}}, \bibinfo{pages}{50} (\bibinfo{year}{2004}),
  \eprint{cond-mat/0401234}.

\bibitem[{\citenamefont{{Melnikovsky}}(2009)}]{2009JPhCS.150c2057M}
\bibinfo{author}{\bibfnamefont{L.~A.} \bibnamefont{{Melnikovsky}}},
  \bibinfo{journal}{Journal of Physics Conference Series}
  \textbf{\bibinfo{volume}{150}}, \bibinfo{eid}{032057} (\bibinfo{year}{2009}).

\bibitem[{\citenamefont{Landea}(2014)}]{Landea:2014naa}
\bibinfo{author}{\bibfnamefont{I.~S.} \bibnamefont{Landea}}
  (\bibinfo{year}{2014}), \eprint{1410.7865}.

\bibitem[{\citenamefont{Rees}(1922)}]{1922}
\bibinfo{author}{\bibfnamefont{E.~L.} \bibnamefont{Rees}},
  \bibinfo{journal}{The American Mathematical Monthly}
  \textbf{\bibinfo{volume}{29}}, \bibinfo{pages}{pp. 51}
  (\bibinfo{year}{1922}), ISSN \bibinfo{issn}{00029890}.

\end{thebibliography}

\end{document}